\documentclass{aa}
\usepackage{graphicx}
\usepackage{txfonts}
\usepackage{natbib}
\bibpunct{(}{)}{;}{a}{}{,}
\newcommand{\msol}{\mbox{M}_{\sun}}

\newcommand{\teff}{T_\mathrm{eff}}
\newcommand{\logg}{\log\left(g\right)}
\newcommand{\logy}{\log\left(y\right)}
\newcommand{\kel}{\mbox{K}}
\newcommand{\dex}{\mbox{dex}}
\newcommand{\mgi}{\ion{Mg}{i}}
\newcommand{\hal}{\mbox{H}\alpha}

\newcommand{\hdel}{\mbox{H}\delta}
\newcommand{\ace}{\alpha_\mathrm{CE}}
\newcommand{\ath}{\alpha_\mathrm{th}}

\newcommand{\qcrit}{q_\mathrm{crit}}
\begin{document}
  \title{Hot subdwarfs from the ESO Supernova Ia Progenitor
  Survey\thanks{Based on observations collected at the Paranal
  Observatory of the European
  Southern Observatory for program No.~165.H-0588(A) and 167.D-0407(A).}}

  \subtitle{I. Atmospheric parameters and cool companions of sdB stars}

 \titlerunning{Hot subdwarfs from SPY -- I. Atmospheric parameters and
 cool companions of sdB stars}

  \author{T. Lisker\inst{1,2}
    \and
    U. Heber\inst{1}
    \and
    R. Napiwotzki\inst{1,3}
    \and
    N. Christlieb\inst{4}
    \and
    Z. Han\inst{5}
    \and
    D. Homeier\inst{6}
    \and
    D. Reimers\inst{4}}

  \offprints{T. Lisker\\
    \email{lisker@sternwarte.uni-erlangen.de}}

  \institute{Dr.-Remeis-Sternwarte, Astronomisches Institut der
    Universit\"at Erlangen-N\"urnberg, Sternwartstr.~7, 96049
    Bamberg, Germany, lisker@sternwarte.uni-erlangen.de
    \and
    Institute of Astronomy, ETH Z\"urich, Department of Physics, HPF
    D8, ETH H\"onggerberg,
    8093 Z\"urich, Switzerland 
    \and
    Department of Physics \& Astronomy, University of Leicester, University
    Road, Leicester LE1 7RH, UK
    \and
    Hamburger Sternwarte, Universit\"at Hamburg, Gojenbergsweg 112,
    21029 Hamburg, Germany
    \and
    The Yunnan Observatory, Academia Sinica, Kunming, 650011, P.R. China
    \and
    Department of Physics and Astronomy, University of Georgia,
    Athens, GA 30602-2451, USA}

  \date{Received date / Accepted date}

  \abstract{
    We present the analysis of a high-resolution, high-quality sample
    of optical spectra for 76 subdwarf B (sdB) stars from the ESO Supernova
    Ia Progenitor Survey (SPY, \citealt{spy0}). Effective temperature, surface gravity,
    and photospheric helium abundance are determined simultaneously by
    fitting the profiles of hydrogen and helium
    lines using synthetic spectra calculated from LTE and NLTE model
    atmospheres. We perform a detailed comparison of our
    measurements with theoretical calculations, both for single star
    evolution and for binary population synthesis models of close
    binary evolution. The 
    luminosity evolution given by the
    standard EHB evolutionary tracks from \citet{dor93} shows an
    overall agreement in shape with our observations, although a
    constant offset in 
    luminosity exists. The various simulation sets for binary
    formation channels of sdB stars calculated by \citet{han03} are
    compared individually to our data for testing our current understanding of sdB
    formation processes and the physical effects involved. The
    best-matching sets manage to reproduce the observed sdB
    distribution in the temperature-gravity-plane well. However, they
    do not match the observed cumulative luminosity function,
    indicating that theoretical improvement is necessary. We also investigate
    composite-spectrum objects showing clear signatures of a cool
    companion with optical and infrared photometry. These stars have cool main sequence companions of
    spectral types F to K. Typical helium abundances of composite and
    non-composite sdB stars do not differ.
  }

  \maketitle


  \section{Introduction \label{sec_int}}

  \defcitealias{han03}{HPMM}

  Early studies provided evidence that subdwarf B (sdB) stars are core helium burning stars with a
  canonical mass of $M \approx 0.5\,\msol$, and a very thin hydrogen envelope
  ($M_\mathrm{env} < 0.01\,\msol$), placing them on the very hot end of the
  horizontal branch (HB), the so-called extreme horizontal branch
  (EHB). Unlike the typical post-HB evolution, these objects do not
  ascend the asymptotic giant branch (AGB) after core helium
  exhaustion, since hydrogen is not burned continuously in a shell due
  to the very low envelope mass. Instead, they evolve more or less
  directly to the 
  white dwarf stage. Proposed formation scenarios are based on
  a late core helium flash in single star evolution
  \citep[e.g.][]{dcr96} or on mass transfer in close binary systems \citep[e.g.][]{men76};
  furthermore, the merger of two helium white dwarfs (He-WDs) could be the
  origin of sdB stars \citep{ibe90}.
  
  Today, more than 200 sdB stars have been analyzed for atmospheric
  parameters \citep[e.g.][]{saf94,max01,ede03}, and this interpretation still
  seems to be valid in general: effective temperature and surface gravity of the
  \emph{observed} sdBs mostly lie within the theoretically
  determined start and end points for core helium burning on the EHB,
  zero-age EHB (ZAEHB) and terminal-age EHB (TAEHB). However, the extensive
  theoretical study of binary formation mechanisms of sdB stars by \citet[ hereafter HPMM]{han03}
  shows that a huge number of sdB stars may be missing in observational surveys
  due to selection effects. These are primarily caused by
  main sequence companions that outshine the sdBs (main sequence
  spectral type
  A and earlier) or appear as composite spectrum objects (main sequence type F to
  K). The latter have mostly been set aside because their spectral
  analysis is rendered difficult by the companion spectrum.

  More than ten years after the first quantitative
  estimate of the contribution of different binary channels to
  the population of sdB stars \citep{tut90}, it is clear from the observations of \citet{max01} 
  that close binary evolution is indeed of
  great importance to sdB formation processes, since they find that two
  thirds of their observed sdB stars are members of close
  binaries. The variety of physical parameters involved in those
  processes led HPMM to produce twelve simulation sets for sdB stars
  that formed in several evolutionary channels, i.e.~by stable Roche lobe
  overflow (RLOF), common envelope
  (CE) ejection, or merging of two He-WDs.
  In comparing them with an observational sample providing good
  quality in atmospheric parameters, it should be possible to constrain the
  simulations' parameter range and to draw conclusions about the
  relative importance of each formation channel.
  
  The ESO Supernova Ia Progenitor Survey (SPY, \citealt{spy0}) provides such a
  sample, since the ambitious project obtained high-resolution optical
  spectra of over
  1000 white dwarf candidates, containing some 140 previously
  misclassified hot subdwarfs of various types.
  Earlier analyses of sdBs \citep[e.g.][]{saf94,max01,ede03} were
  limited by the lower resolution of their spectra, sometimes drawn
  from inhomogeneous data sets. In many cases the wavelength
  coverage was incomplete as well, e.g.~not all Balmer lines were included. Here we present the quantitative analysis
  of an unprecedented homogeneous set of high-resolution, high-quality
  spectra with large wavelength coverage, all of which makes it an
  excellent
  means for testing and increasing the knowledge of sdB formation
  and evolution.
  
  Our report on the analysis of 76 sdB stars from the SPY sample is
  outlined as follows. In Sect.~2, we
  describe the observations and data reduction of our
  objects. Section 3 contains information on the model atmospheres
  and the fit procedure used to determine atmospheric parameters, as
  well as on the examination of the $\hal$ line and of spectral lines
  from cool companions. In
  Sect.~4, we analyze the sdB stars showing no spectral signatures of a
  cool companion and compare them to earlier studies. The
  composite-spectrum objects are investigated in Sect.~5,
  where we attempt to draw conclusions about the nature and spectral types
  of the companions. Finally, in Sect.~6, a detailed
  comparison of our observations with calculations for sdB formation
  and evolution is presented.


  \section{Observations and data reduction \label{sec_obs}}

  Observations were obtained at the ESO Very Large Telescope with
  UT2 (Kueyen) equipped with the UV-Visual Echelle Spectrograph
  \citep[UVES,][]{uves}. A slit width of
  $2\farcs1$ was used, resulting in a spectral resolution of $18\,500$
  ($0.36\,\AA$ at $\hal$) or better. Wavelength coverage of
  $3300-6650\,\AA$ is achieved, with gaps at $4500-4600\,\AA$ and
  $5600-5700\,\AA$. For most of the stars, two exposures in different
  nights were taken, since SPY was originally intended to search for
  RV-variable objects. The spectra
  were then reduced with a procedure developed by Karl (in prep.) using the ESO MIDAS
  software package, partly based on the UVES online reduction
  facility. After accounting for cosmic ray
  hits and bad CCD pixels, bias and interorder background were
  subtracted, followed by order extraction for sky background,
  flatfield, and object. For each extracted one dimensional order, sky
  background subtraction and flatfielding were performed, and finally the
  orders were merged, resulting in three partial spectra separated by
  the gaps mentioned above. Those spectral parts are hereafter referred to as blue, lower red,
  and upper red part, respectively. Finally, each of them was divided by a
  smoothed spectrum of a DC white dwarf, which by definition shows no spectral features at all and
  therefore provides an excellent means of correcting for the
  instrumental response.

  The spectra were then convolved with a Gaussian of $1.0\,\AA$ FWHM and
  rebinned to $0.4\,\AA$ stepsize.
  A signal-to-noise ratio ($S/N$) was defined and calculated for the blue parts of
  the spectra, since they contain most of the lines used for fitting, as the
  ratio of mean flux to standard deviation in a continuum area. The
  median value is 178, all except one have values of
  $S/N \ge 10$, nine spectra even have $S/N > 1000$.

  Photometric data are available for all of our objects.
  $B$ magnitudes of most of the stars are photographic measurements provided by the Hamburg/ESO
  Survey \citep[HES,][]{he0}, the Hamburg Quasar Survey \citep[HQS,][]{hs0}, and
  \citet{mccook}. For nine HES stars, $B$ and $V$ values were taken
  from the CCD photometry of \citet{alt04}. In some cases, only $V$
  or Str\"omgren-$y$ magnitudes were available. $J$ magnitudes have been drawn from the Two
  Micron All Sky Survey (2MASS) archive.


  \section{Spectral analysis and classification\label{sec_spec}}

  \subsection{Model atmospheres and line profile fitting \label{sec_fit}}

  \begin{figure}
    \resizebox{\hsize}{!}{\includegraphics{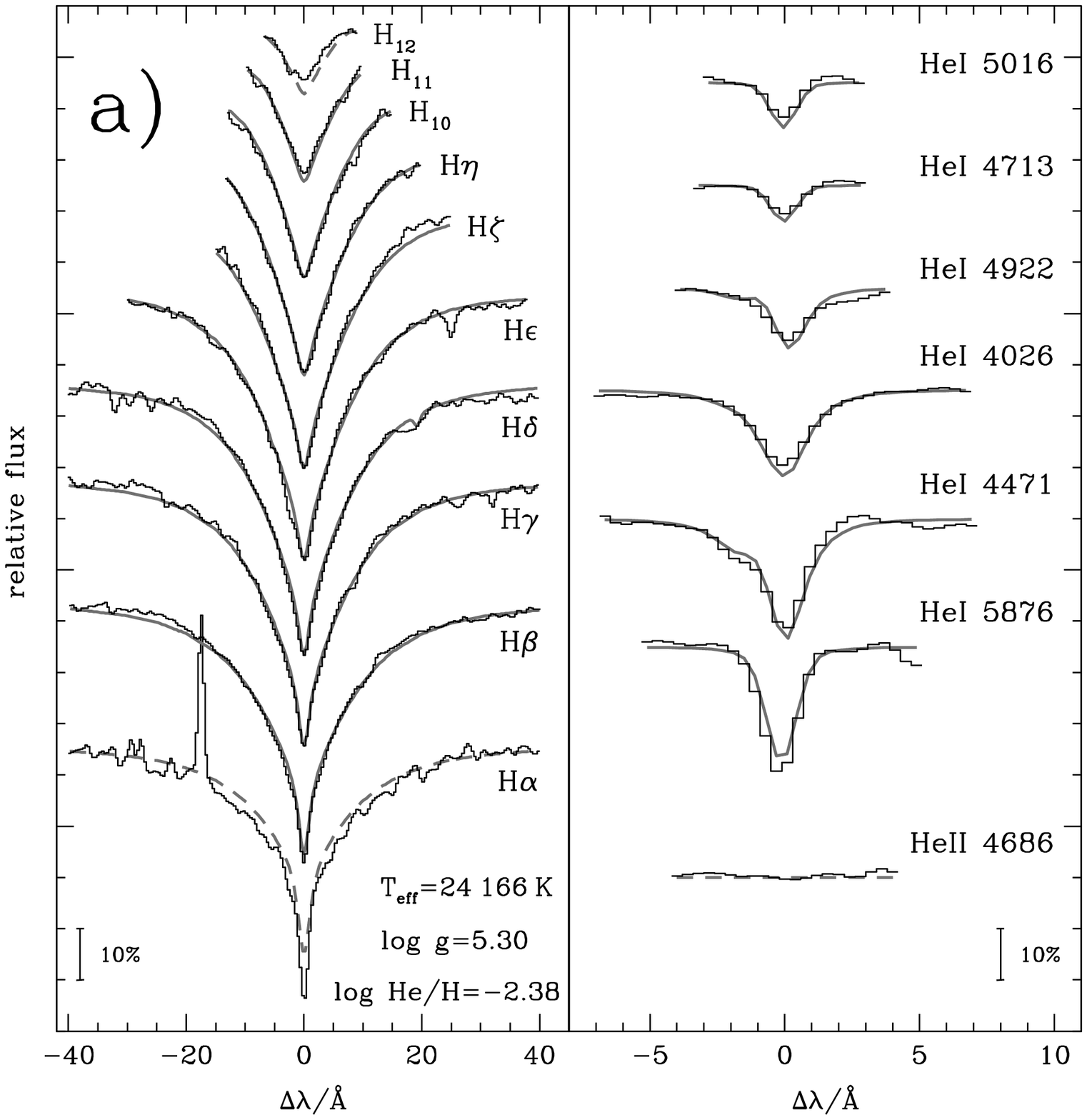}}
    \resizebox{\hsize}{!}{\includegraphics{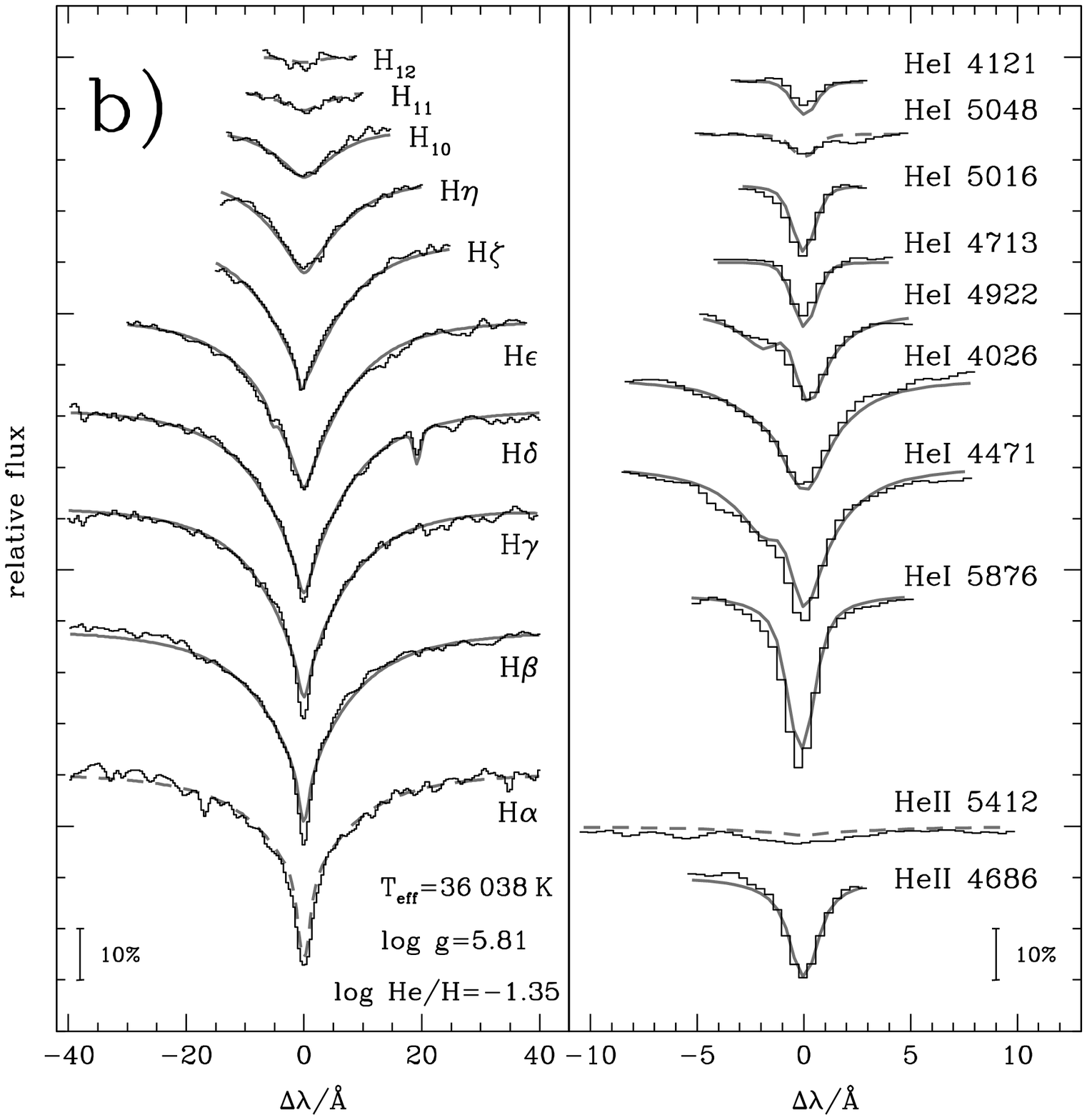}}
    \caption{{\bf ~a)} Sample fit of a relatively cool sdB
    (\object{GD 687}) using synthetic spectra calculated from LTE model
    atmospheres.
    {\bf b)} Sample fit of a relatively hot sdB (\object{GD 619})
    using synthetic spectra calculated from NLTE model atmospheres.
    In both panels, the observed spectra are shown as histograms. When a
    spectral line was used for parameter determination, model spectra
    are plotted as solid grey lines, otherwise as dashed grey lines. The continua of
    observed and model spectrum were scaled to the same level
    on both sides of the line. The resulting atmospheric parameter
    values are displayed in the figure.}
    \label{typfit}
  \end{figure}

  Effective temperatures ($\teff$), surface gravities ($\logg$), and helium
  abundances ($y = N_\mathrm{He}/N_\mathrm{H}$) were determined by fitting
  simultaneously each hydrogen
  and helium line to synthetic model spectra, using a procedure
  developed by R.~Napiwotzki \citep{napfit} based on \citet{saf94}.
  Three different sets of models were used, where the last two make use of PRO2 \citep{pro2}:
  \begin{enumerate}
    \item
      A grid of metal line-blanketed LTE model atmospheres with solar abundance
      \citep{heb00} for stars with $\teff < 32\,000\,\kel$.\smallskip
    \item
      A grid of partially line-blanketed NLTE model atmospheres
      \citep{napatm} for stars with $\teff > 32\,000\,\kel$.\smallskip
    \item
      A grid of partially line-blanketed NLTE model atmospheres for
      helium rich objects, used for two stars with $\logy > -1$
      \citep[updated version of][]{dre90}.\smallskip
  \end{enumerate}

  A typical LTE and NLTE fit is shown in
  Fig.~\ref{typfit}. For each line, the continuum level is
  determined and normalized to 1, in order to compare it to
  the synthetic spectrum. $\hal$ was never included in the
  parameter determination itself, but was kept in the final plot for
  examining possible deviations from the model $\hal$ line. The core
  of $\hal$ is sensitive to NLTE effects because it is formed in the
  very outer parts of the atmosphere. Therefore, LTE models often fail
  to reproduce the observations. This is evident in the example given
  for a LTE analysis in Fig.~\ref{typfit}a. NLTE analyses, however,
  usually are able to reproduce the $\hal$ profiles of sdB stars very
  well \citep{hebwind}, as exemplified in Fig.~\ref{typfit}b. We will
  discuss the $\hal$ lines of the programme stars in more detail in
  Sect.~\ref{sec_hal1}.   

  \subsection{Spectral classification \label{sec_clas}}

  \begin{figure}
    \resizebox{\hsize}{!}{\includegraphics{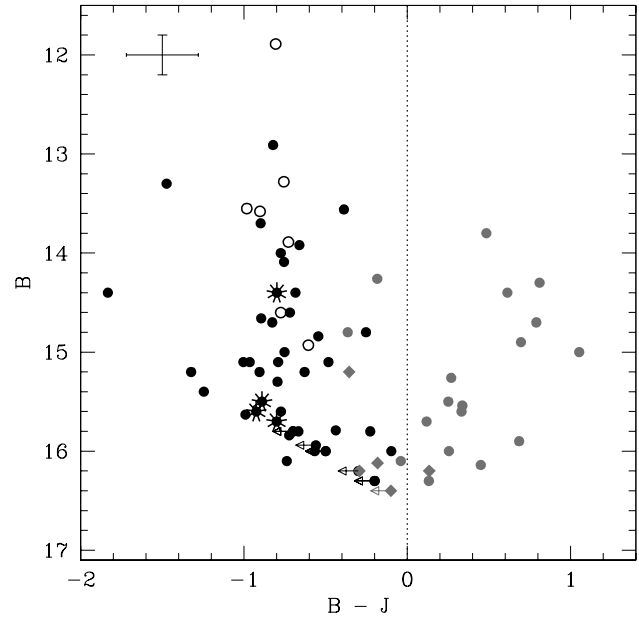}}
    \caption{Colour-magnitude diagram of $B$ versus $B-J$ for all 76
    objects. Composite spectrum objects are plotted as grey circles if
    they are classified from the presence of the \mgi~triplet in their
    spectra, or as grey diamonds if they are classified from the
    contribution of the companion to the $\hal$ line profile (see
    text). Asterisks represent
    the four stars with peculiar $\hal$ profiles. The dotted vertical line
    marks the value $B-J = 0\fm0$. For the
    stars marked as open symbols, we adopted $B \approx V-0\fm25$ or $B
    \approx y-0\fm25$, because no $B$ measurements are available.}
    \label{bjb}
  \end{figure}
  
  The hot subdwarfs had been selected from the SPY data by
  using Balmer line widths as an indicator for surface gravity and by
  performing a line profile fit where visual inspection did not
  yield unambigous results. After having excluded objects that clearly
  have helium-dominated
  atmospheres, the presence and strength of the \ion{He}{ii} lines
  was used for distinguishing between sdB, sdOB, and sdO stars. In
  this paper we present analyses of the
  sdB and sdOB stars from SPY, subsequently referred to as sdB
  stars. This sample comprises 76 sdBs with
  effective temperatures 
  between $20\,000\,\kel < \teff
  < 38\,000\,\kel$, gravities in the range $4.8 < \logg < 6.0$, and
  helium abundances between $-4.0 \le \logy < -0.8$ (see Table
  \ref{tab_res}). 51 
  program stars are objects 
  from HES, 9 from HQS, 14 from \citet{mccook}, and 2 are from the list
  of white dwarf central stars of planetary nebulae by
  \citet{napatm}, see Table \ref{tab_res}.

  \begin{table*}[t!]
    \centering
    \caption{Results of our spectral analysis. $\teff$, $\logg$ and
      $\logy$ are mean values from
      two or more exposures, if possible. Luminosity in units of solar
      luminosity is calculated by assuming $M_\mathrm{sdB} =
      0.5\,\msol$.}
    \begin{tabular}{llllllllllll}
      \hline
      \hline
      \noalign{\smallskip}
      Object &  $\teff$ &       $\logg$ &       $\logy$ &       $\log(L)$ &
      $\log(L)$ &       $M_\mathrm{V}$ &        $d$ &   RA\,(J2000) &   DEC\,(J2000) &  $B$ &   Note\\
      & $\mathrm{K}$ &  $\mathrm{cm\ s^{-2}}$ &
      & $L_\mathrm{edd}$ &      $\mathrm{L_{\sun}}$ &   mag
      & $\mathrm{kpc}$ &        hh:mm:ss.ss &   $\pm$dd:mm:ss.s &       mag &   \\
      \noalign{\smallskip}
      \hline
      \noalign{\smallskip}
      \object{EGB 5} &  34060 & 5.85 &  $-$2.77 &       $-$2.84 &       1.37 &  4.50 &  0.7  &  08:11:12.77 &   +10:57:16.8 & 13.8$^\mathrm{\,c}$ &     7\\
      \object{HE 0007$-$2212} & 28964 & 5.68 &  $\le-$4.00 &    $-$2.95 &       1.25 &  4.42 &  1.4  \ &        00:09:45.91 &   $-$21:56:14.4 & 15.0 & 10 \\
      \object{HE 0016+0044} &   28264 & 5.38 &  $-$2.66 &       $-$2.70 &       1.51 &  3.72 &  1.7  &  00:18:43.59 &   +01:01:22.5 &   14.8  \\
      \object{HE 0019$-$5545} & 35662 & 5.86 &  $-$1.44 &       $-$2.76 &       1.44 &  4.45 &  2.2  &  00:21:27.68 &   $-$55:29:12.3 & 15.8  \\
      \object{HE 0101$-$2707} & 35568 & 5.97 &  $-$0.92 &       $-$2.88 &       1.32 &  4.73 &  1.2  &  01:03:43.73 &   $-$26:51:53.8 & 15.1 & 10 \\
      \object{HE 0123$-$3330} & 36602 & 5.87 &  $-$1.49 &       $-$2.73 &       1.47 &  4.45 &  1.6  &  01:25:22.46 &   $-$33:15:11.4 & 15.2 & 10 \\
      \object{HE 0135$-$6150} & 27020 & 5.59 &  $-$2.47 &       $-$2.99 &       1.22 &  4.34 &  2.8  &  01:37:30.08 &   $-$61:34:57.8 & 16.3  \\
      \object{HE 0136$-$2758} & 28202 & 5.47 &  $\le-$4.00 &    $-$2.79 &       1.41 &  3.95 &  2.8  &  01:39:14.46 &   $-$27:43:21.7 & 15.94 & 5,10\\
      \object{HE 0151$-$3919} & 20841 & 4.83 &  $-$2.07 &       $-$2.68 &       1.53 &  2.94 &  1.9  &  01:53:11.18 &   $-$39:04:18.2 & 14.09 & 5,9\\
      \object{HE 0207+0030} &   31414 & 5.83 &  $-$1.83 &       $-$2.95 &       1.25 &  4.63 &  1.4  &  02:10:14.97 &   +00:45:02.1 &   15.1 & 8  \\
      \object{HE 0230$-$4323} & 31552 & 5.60 &  $-$2.58 &       $-$2.72 &       1.48 &  4.04 &  0.9  &  02:32:54.66 &   $-$43:10:27.9 & 13.56 & 5\\
      \object{HE 0306$-$0309} & 26710 & 5.79 &  $\le-$4.00 &    $-$3.20 &       1.01 &  4.86 &  1.9  &  03:08:40.80 &   $-$02:58:02.8 & 16.1  \\
      \object{HE 0321$-$0918} & 25114 & 5.67 &  $-$3.02 &       $-$3.18 &       1.02 &  4.69 &  1.3  &  03:23:45.85 &   $-$09:08:15.6 & 15.2  \\
      \object{HE 0415$-$2417} & 32768 & 5.12 &  $-$2.44 &       $-$2.18 &       2.03 &  2.76 &  5.6  &  04:17:31.49 &   $-$24:09:50.7 & 16.2  \\
      \object{HE 0513$-$2354} & 26758 & 5.50 &  $-$2.33 &       $-$2.91 &       1.29 &  4.14 &  2.4  &  05:15:15.30 &   $-$23:51:09.5 & 15.8  \\
      \object{HE 0532$-$4503} & 25710 & 5.33 &  $-$3.07 &       $-$2.81 &       1.40 &  3.79 &  2.8  &  05:33:40.51 &   $-$45:01:35.3 & 15.84 & 5\\
      \object{HE 0539$-$4246} & 23279 & 5.51 &  $-$3.91 &       $-$3.16 &       1.05 &  4.43 &  1.2  &  05:41:06.71 &   $-$42:45:31.9 & 14.60 & 5\\
      \object{HE 0929$-$0424} & 29602 & 5.69 &  $-$2.01 &       $-$2.92 &       1.28 &  4.40 &  1.9  &  09:32:02.15 &   $-$04:37:37.8 & 15.4  \\
      \object{HE 1021$-$0255} & 35494 & 5.81 &  $-$1.46 &       $-$2.73 &       1.48 &  4.33 &  1.7  &  10:24:18.06 &   $-$03:10:38.8 & 15.3  & 1,8\\ 
      \object{HE 1033$-$2353} & 36204 & 5.76 &  $-$1.46 &       $-$2.64 &       1.56 &  4.19 &  2.6  &  10:36:07.23 &   $-$24:08:35.4 & 16.0  & 9\\
      \object{HE 1038$-$2326} & 30573$^\mathrm{\,a}$ &  5.21$^\mathrm{\,b}$ &   $-$1.73$^\mathrm{\,b}$ &        &       &       &       &       10:40:36.97 &   $-$23:42:39.4 & 15.9  & 1,2,9\\ 
      \object{HE 1047$-$0436} & 30280 & 5.71 &  $-$2.35 &       $-$2.91 &       1.30 &  4.41 &  1.3  &  10:50:26.93 &   $-$04:52:35.8 & 14.7 & 8  \\
      \object{HE 1050$-$0630} & 34501 & 5.79 &  $-$1.40 &       $-$2.76 &       1.45 &  4.33 &  0.9  &  10:53:26.52 &   $-$06:46:15.6 & 14.0 & 8  \\
      \object{HE 1140$-$0500} & 34522$^\mathrm{\,a}$ &  4.97$^\mathrm{\,b}$ &   $-$2.62$^\mathrm{\,b}$ &        &       &       &       &       11:42:57.85 &   $-$05:17:14.1 & 14.8  & 2,8 \\ 
      \object{HE 1200$-$0931} & 33419 & 5.78 &  $-$1.88 &       &       &       &       &       12:03:21.77 &   $-$09:48:06.6 & 16.2  & 3 \\ 
      \object{HE 1221$-$2618} & 32606$^\mathrm{\,a}$ &  5.51$^\mathrm{\,b}$ &   $-$1.84$^\mathrm{\,b}$ &        &       &       &       &       12:24:32.72 &   $-$26:35:17.1 & 14.7  & 2,9 \\ 
      \object{HE 1254$-$1540} & 29700 & 5.63 &  $\le-$4.00 &    &       &       &       &       12:57:19.36 &   $-$15:56:22.8 & 15.2  & 3,9 \\ 
      \object{HE 1309$-$1102} & 27109$^\mathrm{\,a}$ &  5.36$^\mathrm{\,b}$ &   $-$2.46$^\mathrm{\,b}$ &        &       &       &       &       13:12:02.39 &   $-$11:18:16.3 & 16.1  & 2 \\ 
      \object{HE 1352$-$1827} & 35674$^\mathrm{\,a}$ &  5.53$^\mathrm{\,b}$ &   $-$1.77$^\mathrm{\,b}$ &        &       &       &       &       13:55:26.67 &   $-$18:42:09.4 & 16.0  & 2,9 \\ 
      \object{HE 1407+0033} &   37309 & 5.54 &  $-$2.99 &       $-$2.38 &       1.83 &  3.61 &  2.4  &  14:10:20.73 &   +00:18:54.6 &   15.5  & 4,8\\ 
      \object{HE 1415$-$0309} & 29520 & 5.56 &  $-$2.87 &       $-$2.80 &       1.41 &  4.09 &  3.0  &  14:18:20.93 &   $-$03:22:54.1 & 16.3  \\
      \object{HE 1419$-$1205} & 34171 & 5.71 &  $-$1.65 &       &       &       &       &       14:22:02.17 &   $-$12:19:30.9 & 16.2  & 3 \\ 
      \object{HE 1421$-$1206} & 29570 & 5.53 &  $\le-$4.00 &    $-$2.76 &       1.45 &  4.01 &  1.9  &  14:24:08.81 &   $-$12:20:21.5 & 15.1 & 8  \\
      \object{HE 1422$-$1851} & 33896$^\mathrm{\,a}$ &  5.19$^\mathrm{\,b}$ &   $-$3.07$^\mathrm{\,b}$ &        &       &       &       &       14:24:48.65 &   $-$19:05:01.3 & 16.3  & 1,2,9\\ 
      \object{HE 1441$-$0558} & 36396$^\mathrm{\,a}$ &  5.79$^\mathrm{\,b}$ &   $-$1.63$^\mathrm{\,b}$ &        &       &       &       &       14:44:12.11 &   $-$06:10:44.7 & 14.4  & 2 \\ 
      \object{HE 1448$-$0510} & 34760 & 5.53 &  $-$3.41 &       $-$2.48 &       1.72 &  3.66 &  1.5  &  14:51:13.13 &   $-$05:23:16.9 & 14.4  & 4,8\\ 
      \object{HE 1450$-$0957} & 34563 & 5.79 &  $-$1.29 &       $-$2.75 &       1.45 &  4.32 &  1.7  &  14:53:24.19 &   $-$10:09:21.9 & 15.1 & 9 \\
      \object{HE 1459$-$0234} & &       &       &       &       &       &       &       15:02:12.29 &   $-$02:46:00.9 & 14.9  & 2 \\
      \object{HE 1519$-$0708} & 34498 & 5.73 &  $-$1.52 &       $-$2.70 &       1.50 &  4.18 &  2.3  &  15:21:53.20 &   $-$07:19:23.6 & 15.6 & 8  \\
      \object{HE 2135$-$3749} & 29924 & 5.87 &  $-$2.45 &       $-$3.08 &       1.13 &  4.83 &  0.7  &  21:38:44.18 &   $-$37:36:15.1 & 13.70 & 5\\
      \object{HE 2150$-$0238} & 29846 & 5.90 &  $-$2.36 &       $-$3.12 &       1.09 &  4.91 &  1.7  &  21:52:35.81 &   $-$02:24:31.6 & 15.8  \\
      \object{HE 2151$-$1001} & 34984 & 5.70 &  $-$1.60 &       $-$2.64 &       1.57 &  4.07 &  2.2  &  21:54:31.49 &   $-$09:47:30.5 & 15.6  & 4\\ 
      \object{HE 2156$-$3927} & 27995$^\mathrm{\,a}$ &  5.50$^\mathrm{\,b}$ &   $-$2.35$^\mathrm{\,b}$ &        &       &       &       &       21:59:35.53 &   $-$39:13:15.3 & 14.26 & 2,5,11 \\ 
      \object{HE 2201$-$0001} & 27062 & 5.51 &  $-$3.29 &       $-$2.90 &       1.31 &  4.14 &  2.7  &  22:04:18.27 &   +00:12:36.7 &   16.0  \\ 
      \object{HE 2208+0126} &   24277 & 5.67 &  $-$2.98 &       $-$3.25 &       0.96 &  4.75 &  1.4  &  22:10:45.47 &   +01:41:35.4 &   15.2  & 1,8\\ 
      \object{HE 2222$-$3738} & 30248 & 5.69 &  $-$3.65 &       $-$2.88 &       1.32 &  4.36 &  1.3  &  22:24:56.50 &   $-$37:23:30.7 & 14.66 & 5\\ 
      \object{HE 2237+0150} &   25606 & 5.38 &  $-$1.92 &       $-$2.86 &       1.34 &  3.92 &  2.7  &  22:40:14.38 &   +02:06:31.3 &   15.8  & 8 \\ 
      \object{HE 2238$-$1455} & 30393 & 5.47 &  $-$2.37 &       $-$2.66 &       1.54 &  3.80 &  3.0  &  22:41:38.27 &   $-$14:39:39.5 & 16.0  \\ 
      \object{HE 2307$-$0340} & 23260 & 5.51 &  $-$3.65 &       $-$3.16 &       1.04 &  4.44 &  2.1  &  23:10:24.09 &   $-$03:24:02.3 & 15.8  \\ 
      \object{HE 2322$-$0617} & 28106$^\mathrm{\,a}$ &  5.50$^\mathrm{\,b}$ &   $-$1.93$^\mathrm{\,b}$ &        &       &       &       &       23:25:31.91 &   $-$06:01:12.1 & 15.7  & 2 \\ 
      \object{HE 2322$-$4559} & 25512$^\mathrm{\,a}$ &  5.30$^\mathrm{\,b}$ &   $-$2.47$^\mathrm{\,b}$ &        &       &       &       &       23:25:09.05 &   $-$45:43:06.5 & 15.5  & 2 \\ 
      \object{HE 2349$-$3135} & 28520 & 5.44 &  $-$3.84 &       $-$2.74 &       1.47 &  3.86 &  2.6  &  23:51:43.63 &   $-$31:18:52.9 & 15.63 & 5,10\\ 
      \object{HS 1530+0542} &   &       &       &       &       &       &       &       15:33:10.74 &   +05:32:26.8 &   14.3 &  2,8 \\
      \object{HS 1536+0944} &   35114$^\mathrm{\,a}$ &  5.83$^\mathrm{\,b}$ &   $-$0.82$^\mathrm{\,b}$ &        &       &       &       &       15:38:42.88 &   +09:34:42.8 &   15.6 &  1,2,8\\ 
      \noalign{\smallskip}
      \hline
      \multicolumn{10}{l}{\textit{continued on next page}}\\
      \hline
      \label{tab_res}
    \end{tabular}
  \end{table*}
  
  \begin{table*}
    \centering
    \begin{tabular}{llllllllllll}
      \hline
      \multicolumn{10}{l}{\textit{continued from previous page}}\\
      \hline
      \noalign{\smallskip}
      Object &  $\teff$ &       $\logg$ &       $\logy$ &       $\log(L)$ &
      $\log(L)$ &       $M_\mathrm{V}$ &        $d$ &   RA\,(J2000) &   DEC\,(J2000) &  $B$ &   Note\\
      & $\mathrm{K}$ &  $\mathrm{cm\,s^{-2}}$ &
      & $L_\mathrm{edd}$ &      $\mathrm{L_{\sun}}$ &   mag
      & $\mathrm{kpc}$ &        hh:mm:ss.ss &   $\pm$dd:mm:ss.s &       mag\\
      \noalign{\smallskip}
      \hline
      \noalign{\smallskip}
      \object{HS 1710+1614} &   34826 & 5.72 &  $-$1.64 &       $-$2.68 &       1.53 &  4.13 &  2.4  &  17:13:03.18 &   +16:10:42.8 &   15.7 &  4\\ 
      \object{HS 2033+0821} &   32706 & 5.87 &  $-$1.56 &       $-$2.93 &       1.28 &  4.64 &  1.0  &  20:35:29.34 &   +08:31:51.7 &   14.4 \\ 
      \object{HS 2043+0615} &   26157 & 5.28 &  $-$2.38 &       $-$2.73 &       1.48 &  3.63 &  3.4  &  20:46:20.86 &   +06:26:24.4 &   16.0 \\ 
      \object{HS 2125+1105} &   32542 & 5.76 &  $-$1.86 &       &       &       &       &       21:27:32.17 &   +11:18:17.1 &   16.4 &  3 \\ 
      \object{HS 2216+1833} &   34361$^\mathrm{\,a}$ &  5.51$^\mathrm{\,b}$ &   $-$1.70$^\mathrm{\,b}$ &        &       &       &       &       22:18:30.60 &   +18:48:09.4 &   13.8 &  2 \\ 
      \object{HS 2357+2201} &   27629 & 5.55 &  $-$2.54 &       $-$2.90 &       1.30 &  4.20 &  0.8  &  00:00:18.41 &   +22:18:03.0 &   13.3 &  1 \\ 
      \object{HS 2359+1942} &   31434 & 5.56 &  $-$3.58 &       $-$2.69 &       1.52 &  3.95 &  1.4  &  00:02:08.46 &   +19:59:12.8 &   14.4 &  1,8 \\ 
      \object{PHL 932} &        33644 & 5.74 &  $-$1.64 &       $-$2.75 &       1.46 &  4.25 &  0.4  &  00:59:56.65 &   +15:44:13.6 &   12.1$^\mathrm{\,c}$  &  6,8\\ 
      \noalign{\smallskip}
      \hline
      \multicolumn{10}{l}{\textit{sdB stars misclassified in the
          \citet{mccook} catalog of white dwarfs:}}\\
      \hline
      \noalign{\smallskip}
      \object{CBS 275} &        29262 & 5.72 &  $-$2.46 &      $-$2.97 &        1.23 &  4.50 &  1.3  &  14:36:07.33 &   $-$27:13:14.3 & 14.8 & 9 \\ 
      \object{GD 617} & &       &       &       &       &       &       &       00:31:13.06 &   $-$27:12:54.4 & 15.2$^\mathrm{\,d}$ &   2,10 \\
      \object{GD 619} & 36097 & 5.82 &  $-$1.33 &       $-$2.71 &       1.49 &  4.34 &  0.9  &  00:33:53.88 &   $-$27:08:23.6 & 13.9   & 10        \\ 
      \object{GD 687} & 24350 & 5.32 &  $-$2.38 &       $-$2.90 &       1.31 &  3.87 &  1.1  &  01:10:18.46 &   $-$34:00:26.4 & 14.1$^\mathrm{\,d}$ & 10\\ 
      \object{GD 1237} &        &       &       &       &       &       &       &       23:45:26.69 &   $-$15:28:38.6 & 15.5 &  2 \\
      \object{KUV 01542$-$0710} & 27760 & 5.44 &  $-$2.91 &       &       &       &       &       01:56:42.42 &   $-$06:55:40.2 & 16.3$^\mathrm{\,c}$ &   3 \\ 
      \object{PG 0258+184} &    28092$^\mathrm{\,a}$ &  5.52$^\mathrm{\,b}$ &   $-$3.03$^\mathrm{\,b}$ &        &       &       &       &       03:01:12.87 &   +18:40:54.0 &   15.3  & 2 \\ 
      \object{PG 1207$-$032} &  35693 & 5.82 &  $-$1.48 &       $-$2.73 &       1.48 &  4.35 &  0.7  &  12:09:36.04 &   $-$03:33:08.0 & 13.5$^\mathrm{\,d}$   & 1 \\ 
      \object{PG 1549$-$001} &  28252 & 5.49 &  $-$2.66 &       $-$2.80 &       1.40 &  4.00 &  1.7  &  15:52:02.77 &   $-$00:04:39.4 & 15.2$^\mathrm{\,c}$ \\ 
      \object{PG 2122+157} &    26015$^\mathrm{\,a}$ &  5.22$^\mathrm{\,b}$ &   $-$2.69$^\mathrm{\,b}$ &        &       &       &       &       21:24:54.89 &   +15:59:03.6 &   15.0 &  2 \\ 
      \object{PHL 555} &        34126 & 5.77 &  $-$1.36 &       $-$2.76 &       1.45 &  4.30 &  0.8  &  23:31:49.97 &   $-$28:52:53.1 & 13.8$^\mathrm{\,c}$ &  \\ 
      \object{PHL 861} &        29668 & 5.50 &  $\le-$4.00 &    $-$2.73 &       1.47 &  3.93 &  1.5  &  00:51:03.97 &   $-$20:00:00.3 & 14.9$^\mathrm{\,d}$ & 10\\ 
      \object{SB 485} & 27738 & 5.51 &  $-$2.50 &       $-$2.85 &       1.36 &  4.09 &  0.6  &  01:12:11.65 &   $-$26:13:27.9 & 12.9 & 10 \\ 
      \object{TON S 155} &      32318$^\mathrm{\,a}$ &  5.16$^\mathrm{\,b}$ &   $-$3.04$^\mathrm{\,b}$ &        &       &       &       &       00:23:59.35 &   $-$23:09:53.5 & 16.1$^\mathrm{\,d}$ &   2,10 \\ 
      \noalign{\smallskip}
      \hline
      \hline
      \label{tab_res2}
    \end{tabular}
    \begin{list}{}{}
    \item[$^{\mathrm{a}}$] The given value is an upper limit due
      to the presence of a cool companion.
    \item[$^{\mathrm{b}}$] The given value is a lower limit due
      to the presence of a cool companion.
    \item[$^{\mathrm{c}}$] Johnson $V$ magnitude
    \item[$^{\mathrm{d}}$] Str\"omgren $y$ magnitude
    \item[1] = stars for which there was only one useful exposure.
    \item[2] = a cool companion shows \mgi~in the spectrum, and
      possibly additional features. 
    \item[3] = the presence of a companion was deduced solely from a
      flux contribution at $\hal$. Atmospheric parameters have been
      determined from Balmer and helium lines in the blue part of the
      spectrum. Unlike for the stars with earlier type companions, the
      contribution of the companion in this spectral region is 
      irrelevant, and the parameters can be regarded as reliable (for details see Sect.~\ref{sec_comp}).
    \item[4] = single-lined objects showing peculiar $\hal$
      profiles.
    \item[5] = independently classified as sdB by \citet{alt04}.
    \item[6] = central star of the planetary nebula \object{PN G 125.9$-$47.0}\,.
    \item[7] = central star of the planetary nebula \object{PN G 211.9+22.6}\,.
    \item[8] = also in the Palomar-Green survey \citep[PG,][]{pg}.
    \item[9] = also in the Edinburgh-Cape survey \citep[EC,][]{ec}.
    \item[10] = also in the Montreal-Cambridge-Tololo survey \citep[MCT,][]{mct}.
    \item[11] = In Table 1 of \citet{alt04}, $V$ magnitudes of this
    star and \object{HE 2156$-$1732} have to be interchanged (M.~Altmann, priv.~comm.).
    \end{list}
  \end{table*}

  \subsection{Spectroscopic and photometric signatures of cool companions \label{sec_mgi}}

  \begin{figure*}[t]
    \centering
    \includegraphics[width=5.5cm]{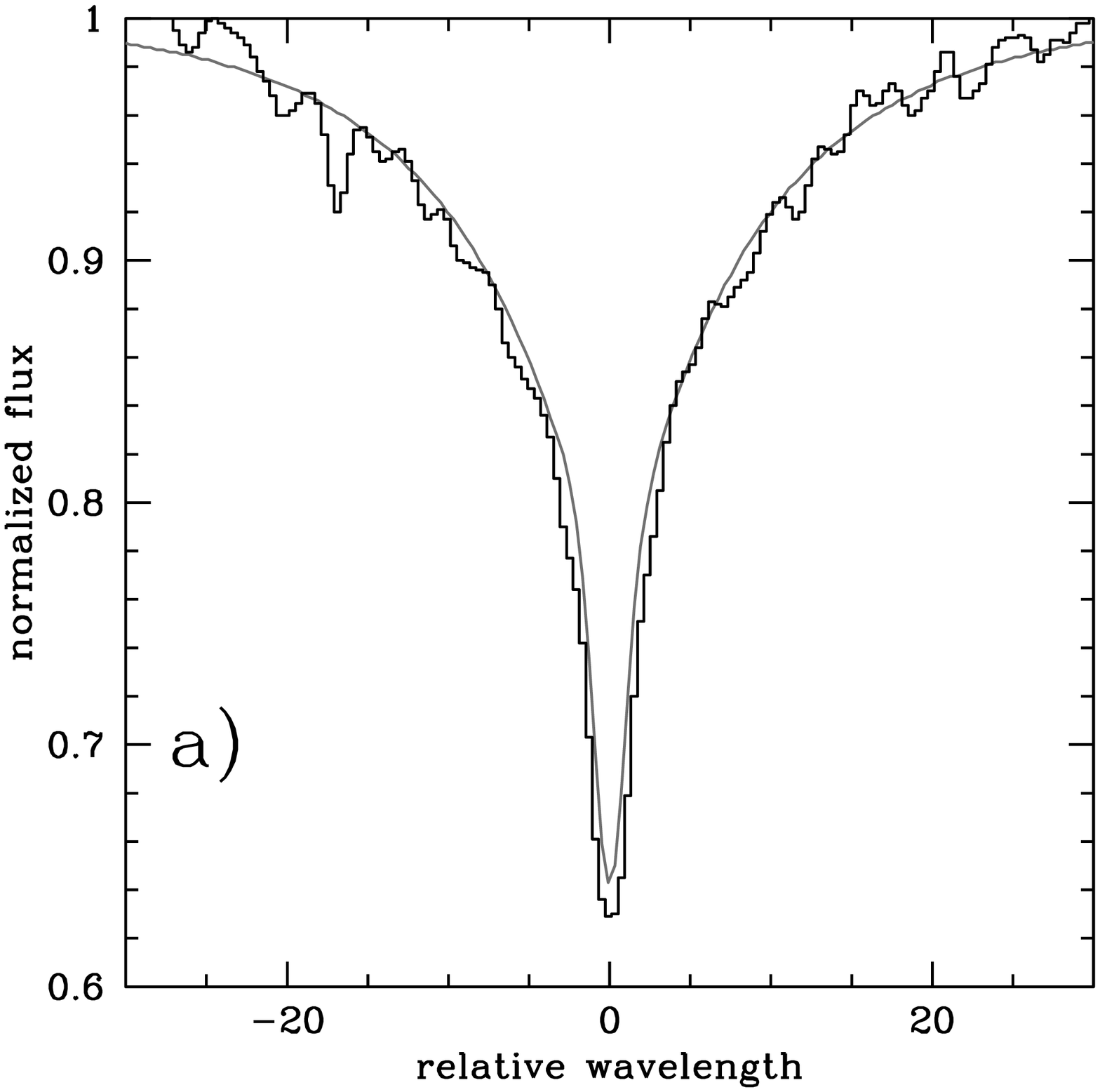}\includegraphics[width=5.5cm]{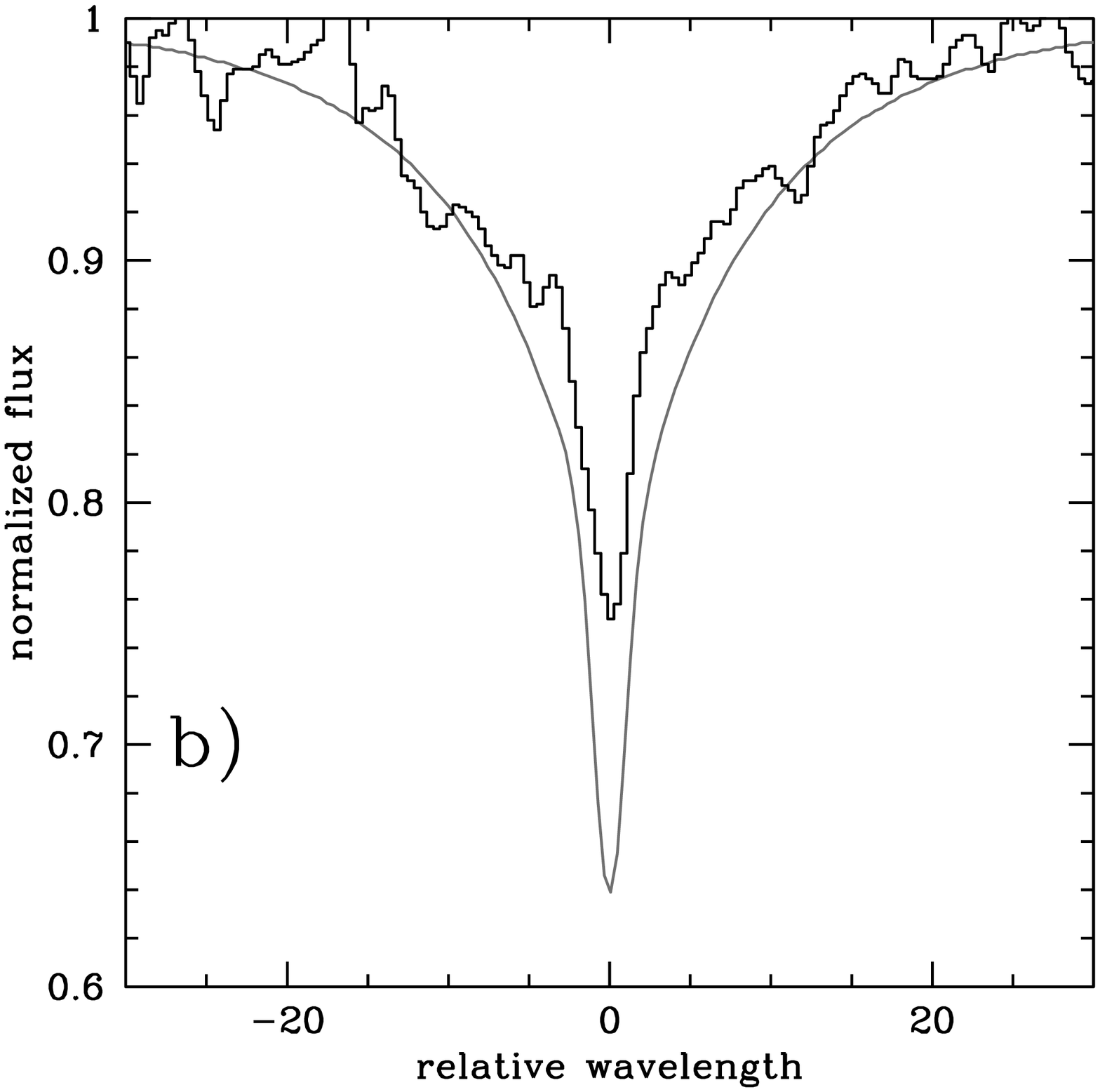}\includegraphics[width=5.5cm]{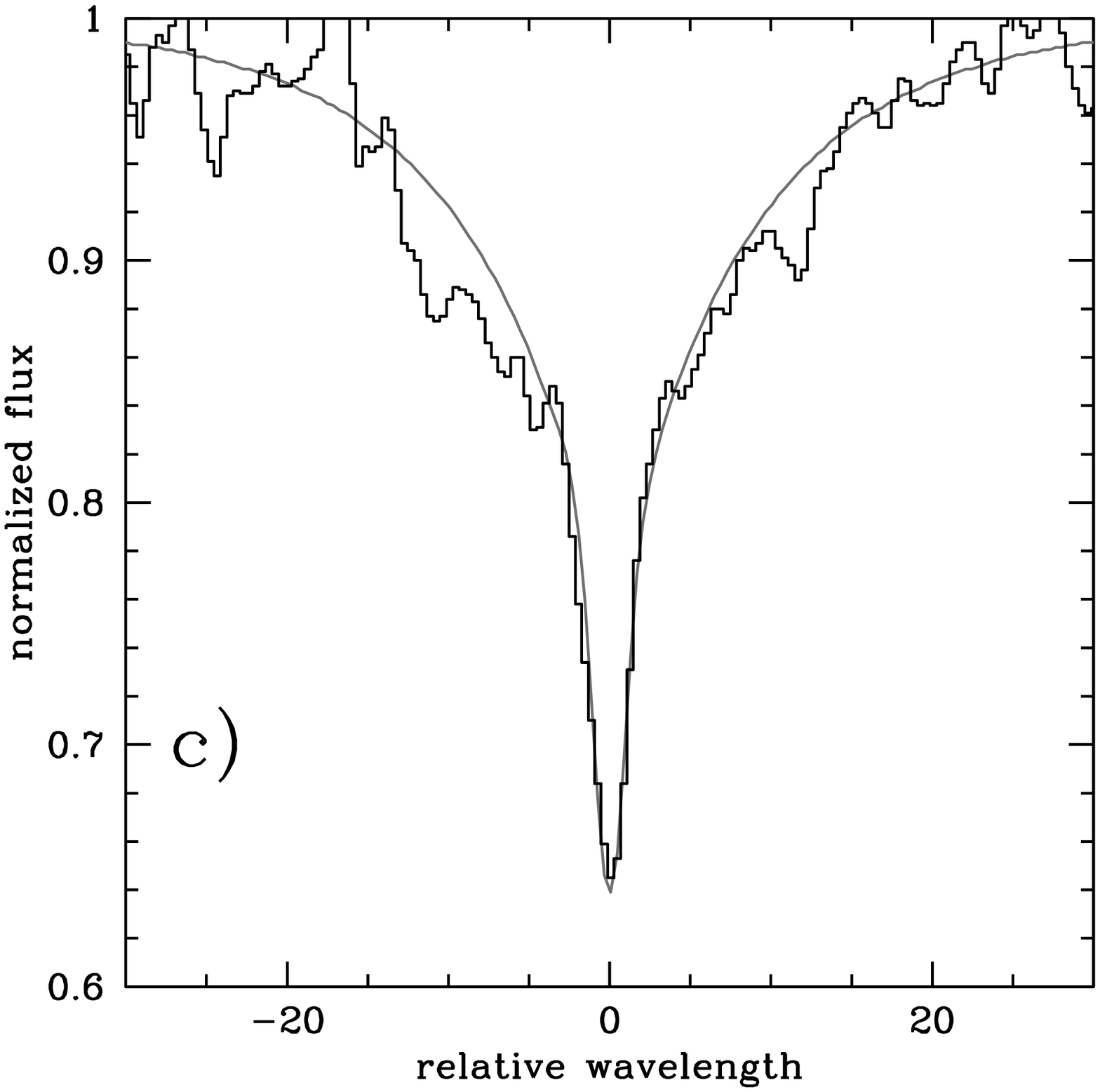}
    \caption{Comparison of synthetic $\hal$ line profiles calculated
      from NLTE model atmospheres to observed ones (see
      text). {\bf ~a)} \object{GD 619}: a typical sdB, {\bf b)}
      \object{HE 1441$-$0558}: a composite sdB star, {\bf c)}
      \object{HE 1441$-$0558}: after correction for continuum light
      from the cool companion.}
    \label{ha_comp}
  \end{figure*}

  \begin{figure}
    \centering
    \includegraphics[width=5.5cm]{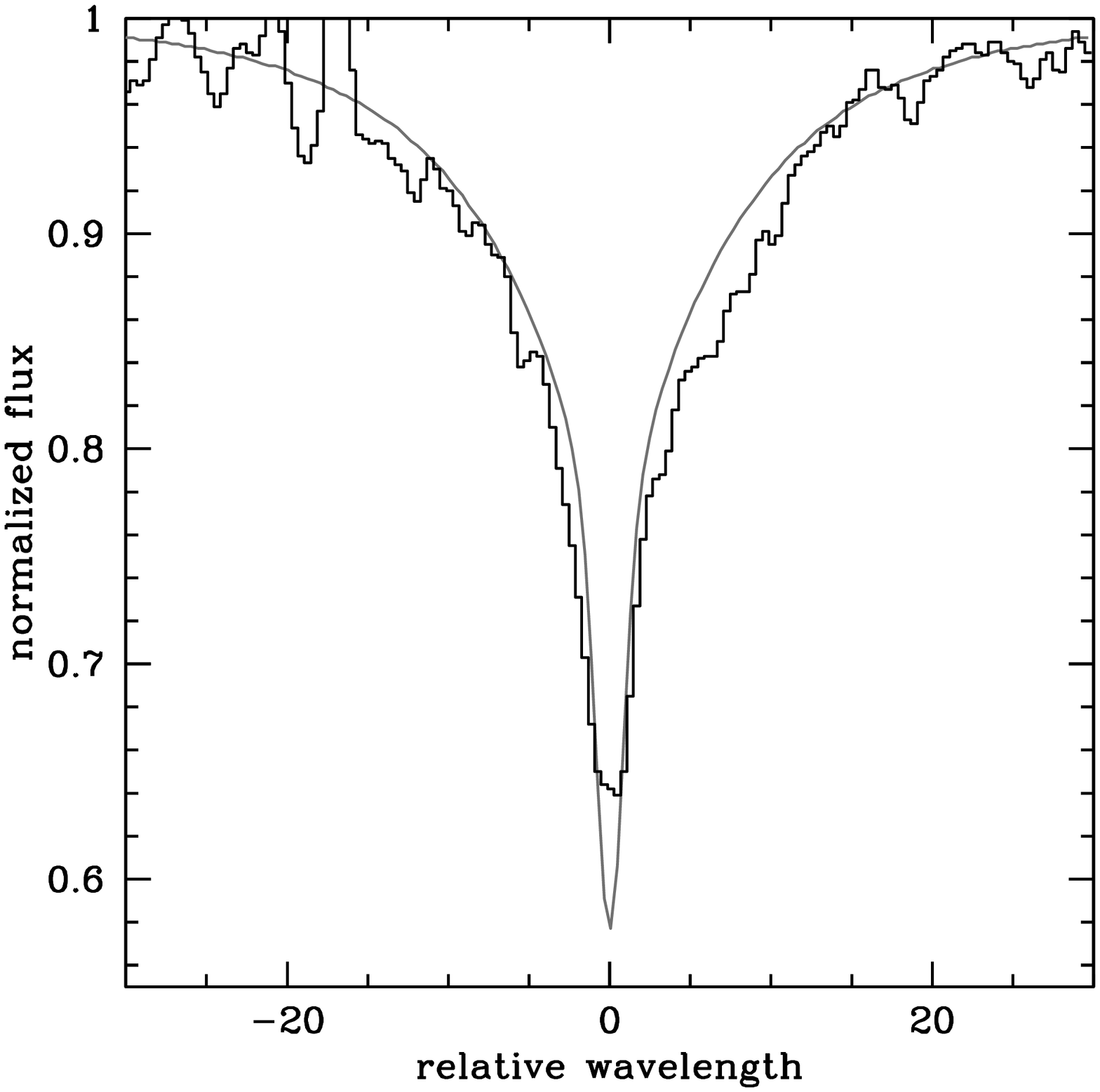}
    \caption{~Same as Fig.~\ref{ha_comp}, but here we show the second kind of
   $\hal$ profile peculiarity we observed, in the
   spectrum of \object{HE 1448$-$0510} (shown here) and four other
   stars (for details see text).}
    \label{halpha}
  \end{figure}

  Our sample of sdB stars includes many objects where signatures of a cool
  companion can be seen in the spectrum. In some of them, the
  companion shows a large contribution to the total flux even in the blue
  part of the spectrum, causing a broad Ca K line ($3933\,\AA$) as
  opposed to the frequently occurring narrow interstellar Ca K line. In
  addition, G-band absorption was taken as a clear companion signature. For most
  objects though, the contribution of the
  companion is best visible in the lower red part of the spectrum,
  where the \mgi~triplet at $5167\,\AA, 5173\,\AA$, and $5184\,\AA$ is
  the most striking
  feature. Since the latter is not known to occur in sdB
  atmospheres nor in the interstellar medium, we used it as a tracer
  to find even very faint companions in sdB spectra that appeared to
  be non-composite at first glance. These examinations yielded
  altogether 19 sdB stars with a cool
  companion contributing to the spectrum (see Table
  \ref{tab_res}). They are treated separately
  (cf.~Sect.~\ref{sec_doub})
  from the non-composite objects, because the
  results from line profile fitting cannot be considered reliable.

  Since our spectra have not been flux calibrated, it was not possible
  to verify these findings by making use of the flux
  distribution in the upper red part, which should otherwise clearly show the
  contribution of the companion. Instead, we retrieved
  infrared photometric data for our stars from the 2MASS database to examine the
  quality of the \mgi~triplet as tracer for cool 
  companions.  
  67 of our objects show a detection at least in $J$ and are
  unambiguously identified, whereas 9 show
  no detection at all and cannot be mistaken with nearby
  sources. In the latter cases, we derived $J < 16\fm5$
  as a conservative
  estimate of the detection limit from the fluxes of nearby
  sources in the field of view.

 The 2MASS observations are
  complete\footnote{http://www.ipac.caltech.edu/2mass/overview/about2mass.html}
  down to $J \le 15\fm8$, $H \le 15\fm1$ and $K \le 14\fm3$.
  Since more than half of our sample is fainter than $J = 15\fm5$,
  their $H$ and $K$ fluxes are of little use and were not
  considered. We based our analysis on $B-J$ colours, because
  $V$-band measurements are unavailable for most
  stars. In Fig.~\ref{bjb} we plot $B$ vs.~$B-J$ for all
  76 sdB stars. The apparently single sdB stars separate from objects
  with a cool 
  companion at about $B-J \approx 0$, which nicely agrees with the
  detection of \mgi~in their spectra (grey circles in
  Fig.~\ref{bjb}).

  For companion spectral types later than K0, the difference
  in colours between an sdB+main sequence system and an sdB
  alone decreases and finally almost vanishes, because of the
  decreasing contribution of the companion to the total flux
  (cf.~Sect.~\ref{sec_comp}). Hence, there is a limit in spectral type
  above which an sdB+main sequence system may lie at $B-J < 0$, and therefore cannot
  be identified as composite. Three stars with $B-J < 0$ show \mgi~as well,
  confirming the quality of the latter as an even better
  tracer for cool companions than colour values are.

  \subsection{Peculiar $\hal$ line profiles \label{sec_hal1}}

  A close examination of the $\hal$ line of the apparently non-composite
  objects yielded ten
  stars with peculiar $\hal$ profiles. They can be divided into two
  groups: five of them show a line core which is shallower than that
  of the corresponding model line, but has a normal line core
  shape. This peculiarity is seen in seven composite objects as well,
  suggesting that this is another hint of a cool companion. It is
  demonstrated in Fig.~\ref{ha_comp} where
  we compare synthetic $\hal$ line profiles calculated from NLTE model
  atmospheres to observed ones for a typical sdB (Fig.~\ref{ha_comp}a)
  and for a representative composite one showing this peculiarity (Fig.~\ref{ha_comp}b). Atmospheric parameters
  have been derived for each star from other Balmer and helium lines
  in the blue part of its spectrum. It must
  be stressed that we did not \emph{fit} the $\hal$ line profile by a
  synthetic spectrum. As can be clearly seen, the $\hal$ line profile
  in a typical sdB is quite well matched, whereas in the peculiar star
  it is too shallow. Since the cool companion adds some continuum
  flux, it dilutes the $\hal$ profile of the sdB. Hence,
  \emph{after} normalization, the $\hal$ line depth is lower than it would be
  without the companion contribution, and therefore lower than that of
  the synthetic profile. This is exactly what we see
  in those stars. We subtracted an appropriate constant from the
  spectrum around
  $\hal$. Now the corrected $\hal$ profile is perfectly matched by the
  synthetic profile (see Fig.~\ref{ha_comp}c). Since this is shown for
  a star that we already classified as composite, by subtracting a
  \emph{constant} we automatically demonstrated that the contribution of the
  companion $\hal$ line is negligible. Applying this treatment
  to the five non-composite stars leads to the same result, which we take as
  evidence for the presence of a cool companion
  (see Table \ref{tab_res}), although other spectral features
  (i.e.~\mgi) are not
  detected in these stars.

  Since the companion contribution decreases with
  decreasing wavelength, it is understandable why in the blue and
  lower red part of the spectrum no cool star features can be
  seen. The companions are therefore expected to be of very late type,
  consistent with the colour-magnitude-diagram ($B$, $B-J$)
  (Fig.~\ref{bjb}): one star lies among
  the composite objects, the others lie among the single sdBs, but
  are very close to $B-J = 0$, which we regard as the photometric
  detection limit for composite objects. As
  discussed in the previous subsection, this
  is exactly what one expects for very late type companions, thus supporting our conclusion.
  We summarize that 24 out of 76 sdB stars show signatures of a cool
  companion, hereafter termed double-lined objects, even if the stars
  just discussed do not show any obvious companion \emph{lines}. As mentioned
  above, they are treated separately from the 52 non-composite
  (single-lined) sdBs.

  A second kind of peculiarity of the $\hal$ profile is found in five
  other objects of our sample: the core is flat, and the inner wing is
  broader than predicted by the model, as seen in
  Fig.~\ref{halpha}. One of these stars, \object{EGB 5}, is the central
  star of a planetary nebula. The $\hal$ emission of its nebula may
  not have been correctly subtracted in the semi-automatic data
  reduction process. Close
  inspection of sdB spectra that we regard as normal
  (Fig.~\ref{ha_comp}a) reveals that slight mismatches of the
  $\hal$ line core occur even in those stars. However, these effects are so small that they
  could be caused by errors in the data reduction and/or the
  atmospheric models and are therefore considered as
  insignificant. Nevertheless, we are left with four stars with peculiar
  $\hal$ profiles.

  The colour values of this group of peculiar stars
  (Fig.~\ref{bjb}, asterisks) clearly show that they cannot be composite objects.
 There is no
  significant line profile variation observed from one to the next exposure of the
  star. Time intervals range from $3\,\mathrm{d}$ to $1\,\mathrm{yr}$.
  Although the observed line broadening could possibly be caused by
  rotation, convolution of the synthetic spectra with rotational
  profiles does
  not lead to acceptable fits for any hydrogen or helium line. Even for $\hal$ itself, the
  match is still not satisfying, implying that other physical effects
  are present.

  Since at this point there is no interpretation for this
  $\hal$ profile peculiarity, we proceed with our analysis, and will distinguish
  those stars from the rest of single-lined objects wherever applicable.
 

  \section{Single-lined objects \label{sec_sing}}

  \subsection{Error determination \label{sec_err}}

  \begin{table}
    \centering
    \caption{Errors derived from the Gaussians in
    Fig.~\ref{myerrors}. For stars with one exposure (last column),
    the errors equal the Gaussian widths. For objects with two
    exposures or more, those values are divided by $\sqrt{2}$ (middle column).}
         \begin{tabular}{lll}
            \hline
            \noalign{\smallskip}
            Quantity & Errors & Errors$\,^{\mathrm{a}}$\\
            \noalign{\smallskip}
            \hline
            \noalign{\smallskip}
            $\teff$ & $374\,\kel$ & $529\,\kel$\\
            $\logg$ & $0.049\,\dex$ & $0.069\,\dex$\\
            $\logy$ & $0.044\,\dex$ & $0.062\,\dex$\\
            $\log(L)\,^{\mathrm{b}}$ & $0.038\,\dex$ & $0.053\,\dex$\\
            \noalign{\smallskip}
            \hline
            \label{tab_err}
         \end{tabular}
         \begin{list}{}{}
         \item[$^{\mathrm{a}}$] For stars with only one useful exposure.
         \item[$^{\mathrm{b}}$] Luminosity $L$.
         \end{list}
  \end{table}

  \begin{figure}
    \resizebox{\hsize}{!}{\includegraphics{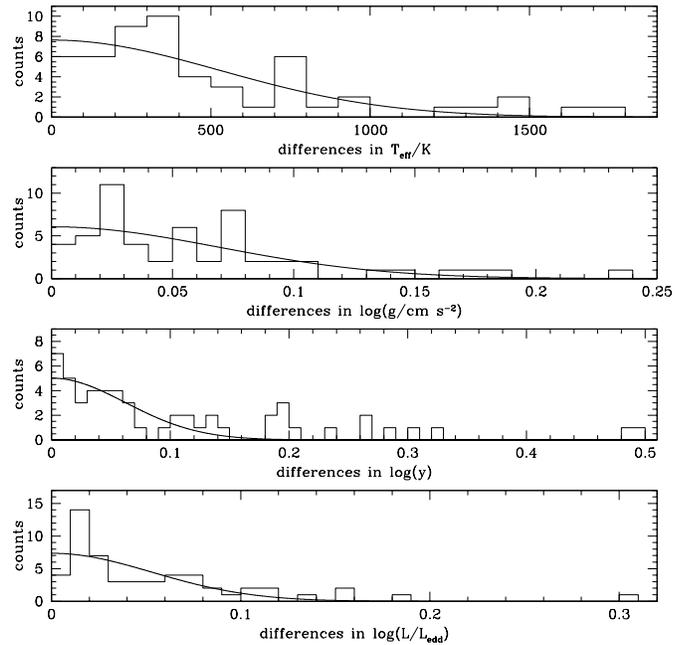}}
    \caption{Distributions of differences in atmospheric parameters between
    two exposures of non-composite stars. Solid lines show the Gaussians fitted to the data. From
    top to bottom the distributions for $\teff$, $\logg$, $\logy$ and
    luminosity $L$ with respect to the Eddington luminosity are shown.}
    \label{myerrors}
  \end{figure}

  The statistical 1-$\sigma$-errors from the fit procedure are
  typically lower than $100\,\kel$, $0.02\,\dex$, and $0.04\,\dex$ for
  $\teff$, $\logg$, and $\logy$, respectively. However,
  despite the high resolution and the low noise level of our data, we do not
  believe these to be the ``true errors''. To obtain a good estimate
  for the latter, we take advantage of the SPY observing strategy: as
  mentioned above, at least two exposures were taken for most of the
  stars.
  Hence, for determining reliable errors we used the distribution of
  differences in the fit results of
  individual exposures of each object
  (Fig.~\ref{myerrors}). Fitting this distribution
  by a Gaussian, we adopt the width of the latter as the error of the fit
  results of individual exposures. This is a quite conservative treatment
  intended to avoid underestimating the errors: since it is
  usually assumed that the \emph{individual} results show a
  Gaussian error distribution as well, the width of the latter
  would be lower by a factor of $\sqrt{2}$ than the width we derived from our
  distribution of \emph{differences} of the results of two
  exposures. However, we do not think that this standard assumption
  would be
  well justified here, since in many cases the differences are much larger than
  the typical statistical errors mentioned above. 

  Furthermore, when we take the mean value of two exposures, the
  errors are automatically reduced by a factor of $\sqrt{2}$, increasing
  the accuracy of the derived parameters. For stars with two or more
  spectra, those reduced Gaussian widths are $\Delta\teff = 360\,\kel$
  and $\Delta\logg = 0.05\,\dex$, which we regard as our best error
  estimates. More details are given in Table \ref{tab_err}.

An uncertainty is caused by the unknown metallicity of the program stars.
We used solar metallicity, fully line-blanketed LTE models (see Sect.~\ref{sec_fit}). 
A comparison shows that adopting a much lower metallicity ($[m/H]=-2$) in the LTE models has a negligible effect at
$\teff = 32\,000\,\kel$ \citep[see also][]{heb00} but increases $\teff$ by 
up to $800\,\kel$ at lower temperatures. Surface gravity is only marginally 
affected ($0.02\,\dex$) while the helium abundance remains almost unchanged.
Spectroscopic analyses of high resolution optical spectra of about two dozen 
bright sdB stars found near-solar iron abundances \citep{heb00,hebkeele},
corroborating our choice of metallicity.
Hence the uncertainties introduced by
the unknown metallicity are smaller than the errors of the 
spectroscopic analyses quoted above for the majority of our objects.

Another source of systematic error lies with the usage of both LTE and NLTE
model atmospheres. \citet{hebkeele} have shown that a systematic 
offset in gravity exists: analysis with
our grid of NLTE models leads to a gravity lower by
$0.06\,\dex$ than derived from our LTE models, while the resulting effective 
temperatures and helium abundances are the same. 

We find that neither of the effects above affects any of our 
conclusions (see Sect.~\ref{sec_evol}).

  \subsection{Effective temperature and surface gravity \label{sec_ehb}}

  \begin{figure}
    \resizebox{\hsize}{!}{\includegraphics{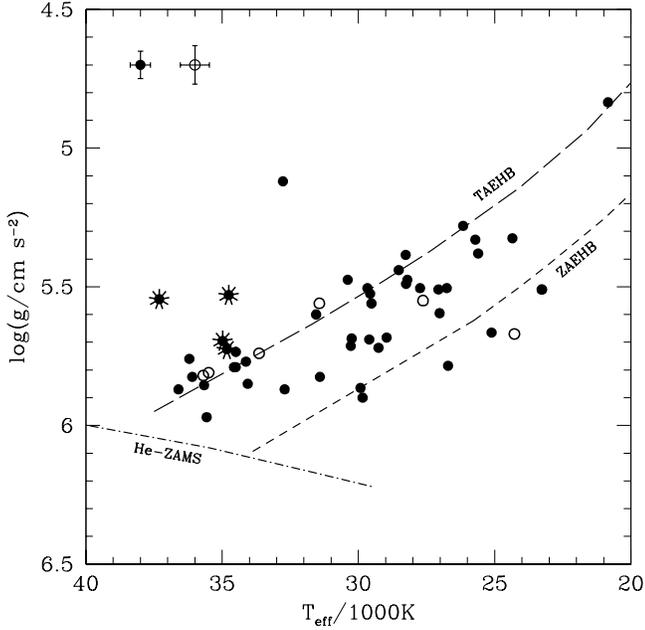}}
    \caption{Distribution of our sdB sample in the
      $\teff$-$\logg$-plane. Results derived from
      only one exposure are
      shown as open circles, whereas filled circles represent mean
      values from two or more spectra. Errors are quoted separately for 
      the former and the latter. Asterisks represent
    the four stars with peculiar $\hal$ profiles. The ZAEHB (short-dashed line), TAEHB
      (long-dashed line) and He-ZAMS (dashed-dotted line) are shown
      (see text for details).}
    \label{myehb}
  \end{figure}

    The fit results are listed in Table \ref{tab_res}. They were
  calculated as the mean value of the results from individual
  exposures of each star, using only spectra with $S/N \ge 10$. For
  nine stars, only one (useful) exposure was available (see Table \ref{tab_res}).

  Figure \ref{myehb} shows the position of our sdB stars in the $\teff$-$\logg$-plane,
  along with the ZAEHB and TAEHB for solar metallicity \citep{dor93}
  and the theoretical zero-age main
  sequence for pure helium stars \citep[He-ZAMS,][]{pac71}. Up to
  effective temperatures of $\teff \approx 32\,000\,\kel$, most of our sdB stars lie
  in the so-called EHB strip, defined by ZAEHB and TAEHB. For higher $\teff$, almost all of
  the stars lie near the TAEHB or above it, including the four objects
  with peculiar $\hal$ profiles. Six stars fall slightly below the
  ZAEHB. Possible explanations
  could be selection effects and/or evolutionary effects, which
  will be examined in Sects.~\ref{sec_othe} and \ref{sec_evol},
  respectively.

  From $\teff$ and $\logg$, the luminosity could be calculated when
  a mass for the sdB is assumed. However, we
  prefer to use luminosity in units of
  the Eddington luminosity, thereby eliminating the mass. It is
  calculated from\smallskip\\
  $\log\left(\frac{L}{L_\mathrm{edd}}\right) =
  4\times\log\left(\teff/\kel\right) -
  \log\left(g/\mathrm{cm\,s^{-2}}\right) - 15.118$\ ,\smallskip\\
  where pure electron scattering in a fully ionized
hydrogen atmosphere is assumed.

  In order to estimate distances we derive the absolute
    visual magnitude $M_\mathrm{V}$ from 
    the surface size and flux. To calculate the former, we need the
    surface gravity, and an assumption for the mass, taken to be
    $M_\mathrm{sdB} = 0.5\,\msol$. The surface flux is
    calculated from Kurucz model atmospheres
    \citep{napflux}.
    Subsequently, the distance is derived from
    $M_\mathrm{V}$ and the apparent visual brightness
    $m_\mathrm{V}$. For most stars, though, only $m_\mathrm{B}$ is
    available. Therefore, we adopted $B-V = -0\fm25$ for
    these objects, which is the typical value for non-composite sdB
    stars \citep[ see also Fig.~\ref{bjb}]{sta03}. We assumed
    Str\"omgren-$y$ to be identical to $V$. All these additional
    quantities are given in Table \ref{tab_res}.

  \subsection{Helium abundance \label{sec_nhe}}

  \begin{figure}
    \resizebox{\hsize}{!}{\includegraphics{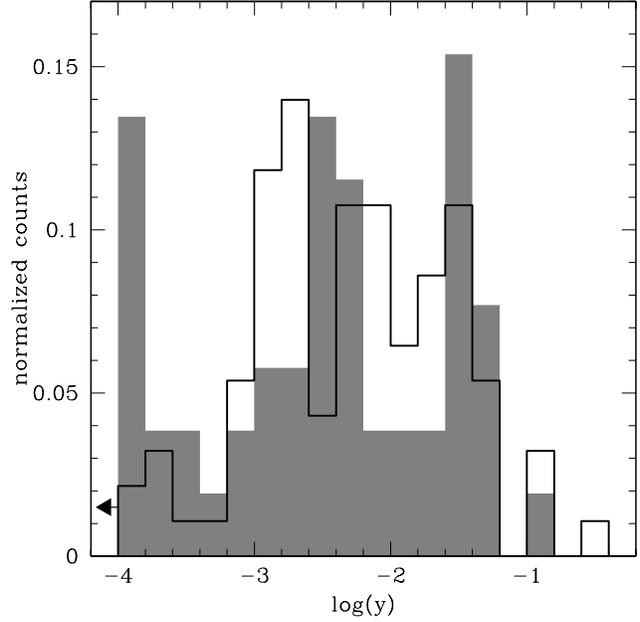}}
    \caption{\emph{Shaded histogram:} Distribution of photospheric
    helium abundance of our sdB stars. \emph{Open histogram:}
    Distribution from \citet{ede03}. The
    value $-4$ is the lowest helium abundance in the model
    atmospheres of both studies. Since no extrapolation was done, it has to be
    regarded as an upper limit for the true abundance.}
    \label{hehist}
  \end{figure}

  \begin{figure}
    \resizebox{\hsize}{!}{\includegraphics{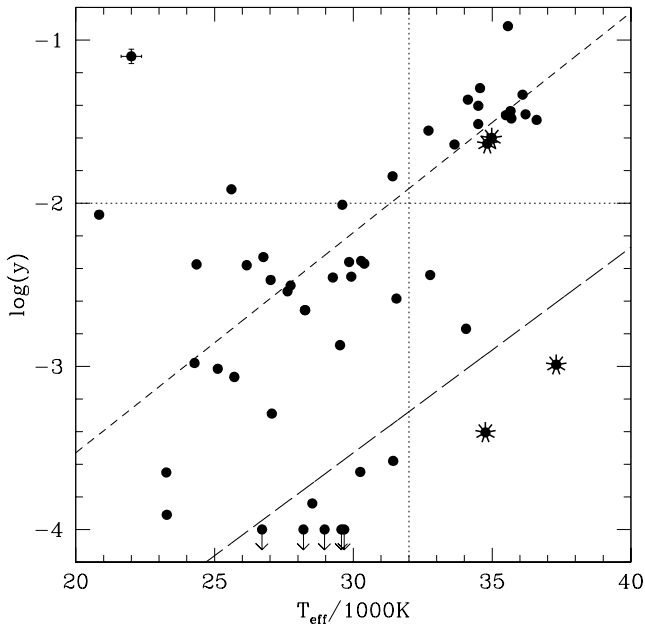}}
    \caption{Helium abundance vs.~effective temperature. Asterisks represent
    the four stars with peculiar $\hal$ profiles. Arrows denote upper limits for helium
    abundance. The dotted horizontal line marks a value of $\logy =
    -2$, the vertical line $\teff = 32\,000\ K$. The dashed lines are
    from \citet{ede03}, see text for details.}
    \label{teffnhe}
  \end{figure}

  \begin{figure}
    \resizebox{\hsize}{!}{\includegraphics{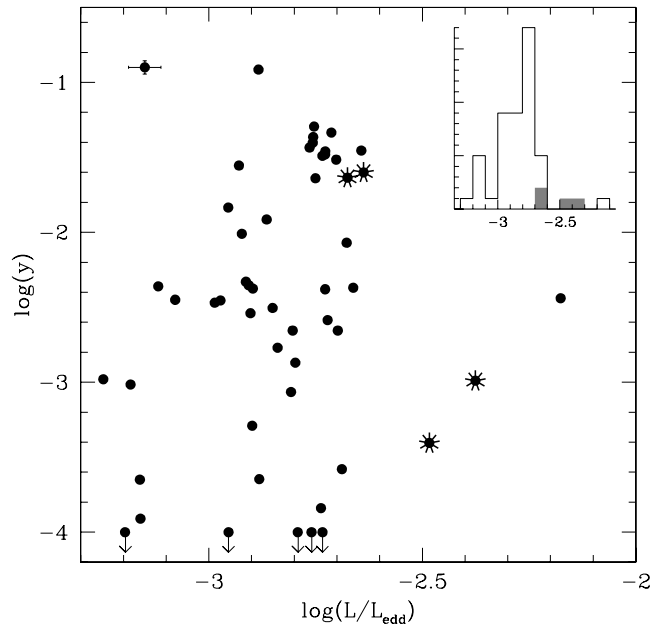}}
    \caption{Helium abundance vs.~luminosity with respect to the Eddington
    luminosity. Asterisks represent
    the four stars with peculiar $\hal$ profiles. Arrows denote upper
    limits for helium abundance. The inset shows the luminosity
    distribution as histogram for the four stars with peculiar $\hal$
    profiles (shaded) and the other objects (open).}
    \label{lumnhe}
  \end{figure}

  Our distribution of helium abundances (Fig.~\ref{hehist}) shows a wide spread
  from the lower boundary of our model grids, $\logy = -4$, with sometimes no helium visible at all, to slightly supersolar
  helium abundance $\logy = -0.92$. There is overall agreement with
  the results from the sample of \citet{ede03} (see Fig.~\ref{hehist}),
  but we find a higher fraction
  of stars with very low helium abundances. Possible correlations with other quantities are
  presented in Figs.~\ref{teffnhe} and \ref{lumnhe}, where helium
  abundance is compared to effective temperature and luminosity,
  respectively. The comparison with $\teff$ shows that for $\logy > -2$, almost all of
  the sdB stars are hotter than $\teff > 32\,000\,\kel$ (see the dotted lines in
  Fig.~\ref{teffnhe}). At these high temperatures, only two objects have a very low
  helium abundance $\logy < -3.0$, whereas there are twelve such
  stars at lower temperatures. Hence, at $\teff \approx 32\,000\,\kel$, our
  objects could be divided into two groups that are roughly correlated
  with spectral type sdB and sdOB, respectively, the latter having
  higher temperatures and an overall higher helium abundance.

  In addition, a slight trend with luminosity can also be observed: At
  higher $\teff$, the sdB stars are somewhat more luminous (see Table
  \ref{tab_res}). Consequently, the average
  luminosity is higher for higher helium
  abundances, as seen in Fig.~\ref{lumnhe}.

  Interestingly, the two stars which have the lowest helium
  abundance of the high-temperature objects show peculiar $\hal$
  profiles, and are more luminous than the other two stars of this
  kind. One might thus speculate about underlying physical
  explanations. However, the $\hal$ profiles of these two subgroups do
  not show any obvious differences, and from a statistical point of
  view there would be no significance in such a separation.

  \citet{ede03} found two distinct sequences of sdB stars, separated
  mainly by an offset in helium abundance. Their corresponding linear fits are also plotted in
  Fig.~\ref{teffnhe}. The sequence for higher helium abundance (short
  dashed line) matches the majority of our
  objects quite well, especially at higher temperatures. The sequence
  for lower $\logy$ (long dashed line), with which
  \citet{ede03} matched about 1/6th of their
  data, can be neither confirmed nor rejected, mainly because we can
  only derive upper limits for five stars.

  As discussed in previous work \citep[e.g.][]{fon97}, the
  diffusive equilibrium abundances of gravitational settling
  versus radiative levitation lie two orders of magnitude below the
  observed average helium abundances; hence,
  this diffusion model is too simplistic. \citet{fon97} included mass loss of
  $10^{-14}\,\msol/\mbox{yr}$ as an additional force in their
  diffusion models, and
  concluded that it is
  possible to reach equilibrium abundances as high as $\logy = -3$
  after $t = 10^{7}\,\mathrm{yr}$ on the 
  EHB. \citet{ung01} showed that mass loss rates of
  $10^{-13}\,\msol/\mbox{yr}$ can produce even higher helium
  abundances $\logy = -2$, again after $t =
  10^{7}\,\mathrm{yr}$. The large number of stars with $\logy >
  -2$ in our sample indicates that either the winds must be even
  stronger for reproducing the whole range of abundances,
  or alternative explanations have to be found.

  Since the stellar winds of sdB stars are probably radiation driven,
  their mass loss rates are expected to be correlated with their luminosities
  \citep{pau88}. However, it would be premature to use that
  correlation for interpreting our
  observed slight trend of helium abundance with
  luminosity. The metallicity plays an important role both for radiative levitation and
  the radiation-driven wind. It
  would therefore be of enormous interest to determine metallicities
  for our high-resolution sdB spectra and use them as input for
  theoretical calculations of radiative forces and wind mass
  loss, which, however, is beyond the scope of this paper.

  \subsection{Peculiar $\hal$ profiles \label{sec_hal2}}

  All four stars with peculiar
  $\hal$ profiles have a higher luminosity
  than the majority of sdB stars (see inset in Fig.~\ref{lumnhe}).
  This gives us a hint to a possible explanation:
  \citet{hebwind} observed similar $\hal$ profiles in the four most luminous
  objects of their sample, and suggested
  stellar winds as the cause. Similar to our data, their profile
  shapes show broadening around the core region, while the line
  depth is lower than in the model spectrum. The core itself is
  flattened in one case, and shows a faint emission in the
  other cases, consistent with our own observations.

  First calculations of $\hal$ lines from spherical NLTE models
  including mass loss have recently become available
  \citep{vinkkeele}. While they can explain the occurrence of small
  emissions in the core and, therefore, a weakening of the core, they
  are unable to match the observed widths of the line wings. In the
  latter respect, there is no improvement yet compared to the static,
  plane parallel NLTE models we use. Therefore the physical effects
  causing the extraordinary line broadening remain obscure. Obviously,
  there is a need for more sophisticated model atmospheres.

  In the sdB mass loss calculations mentioned above, it is possible to
  reproduce the emission with mass loss rates of
  $10^{-11}\,\msol/\mathrm{yr}$. On the one hand, such a strong mass loss agrees with our
  conclusion from the previous subsection that mass loss rates above
  $10^{-13}\,\msol/\mbox{yr}$ would be needed to explain the
  upper part of the helium abundance range of sdBs. On the other hand,
  the helium abundance of 
  the four stars discussed here lies between $-3.41 \le \logy \le
  -1.60$, leading us to recall the statement that stellar winds need
  not automatically cause a correlation of luminosity and helium
  abundance.

  \subsection{Comparison with other samples \label{sec_othe}}

  \begin{figure}
    \resizebox{\hsize}{!}{\includegraphics{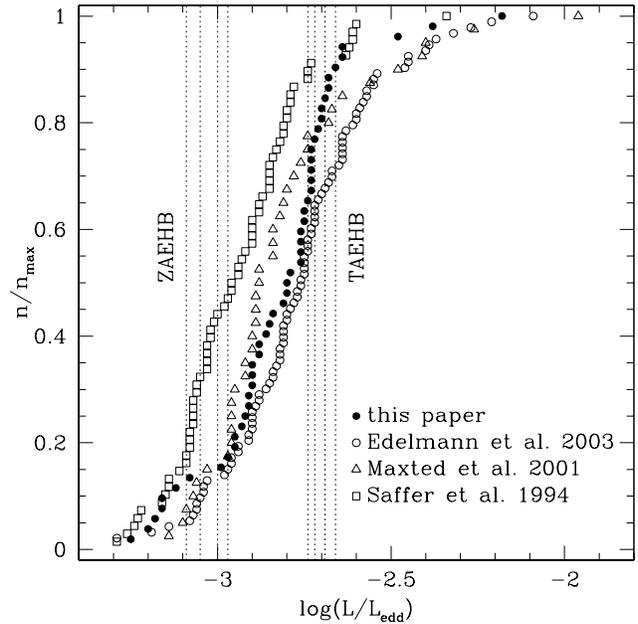}}
    \caption{Cumulative luminosity function: cumulative normalized
    counts versus luminosity in units
    of the Eddington luminosity. Shown are our
    data (filled circles) along with the measurements of \citet[ open
    circles]{ede03}, \citet[ open triangles]{max01}, and \citet[ open
    squares]{saf94}. ZAEHB and TAEHB are from \citet{dor93} for metallicities
    $[\mathrm{Fe}/\mathrm{H}] = 0.00, -0.47, -1.48, -2.26$ from left
    to right.}
    \label{cumlum}
  \end{figure}

  \begin{figure}
    \resizebox{\hsize}{!}{\includegraphics{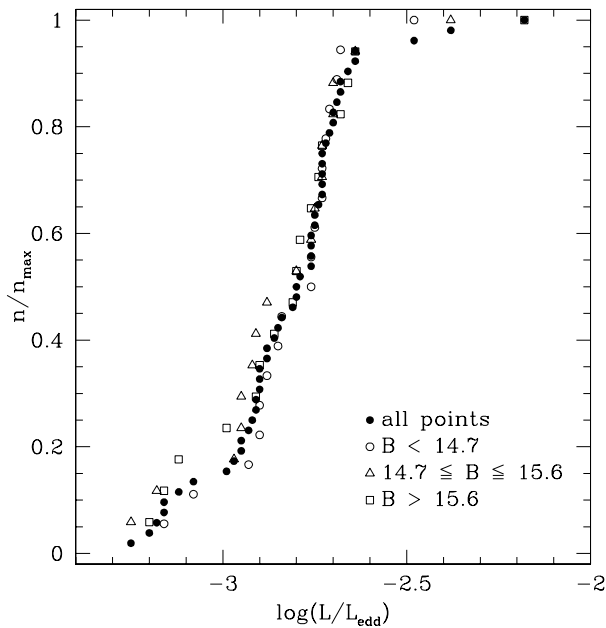}}
  \caption{Cumulative luminosity function for all our objects (filled circles, as in
    Fig.~\ref{cumlum}), as well as for three magnitude-selected subsets
    of equal number of stars: $B < 14.7$ (open circles),
    $14.7 \le B \le 15.6$ (open triangles), $B > 15.6$ (open
    squares).}
    \label{cumlum_slices}
  \end{figure}

  SdB samples of comparable size were presented in the last decade
  by \citet{saf94}, \citet{max01} and \citet{ede03}. Their main selection
  effect is the magnitude-limitation of the surveys
  that provided the targets (Palomar-Green Survey, HQS). In contrast, the selection effects of
  our sample are not well known. Since the
  SPY targets were drawn from various catalogs with criteria for selecting potential white
  dwarfs, the hot subdwarfs were included only by misclassification,
  rendering a definition of
  sdB selection criteria impossible. We now attempt to investigate whether the
  selection effects of our sample are similar to those of the samples just
  mentioned, or whether any systematic differences can be found.

  An adequate means for comparing different samples of sdB stars is
  their respective cumulative luminosity function, shown in
  Fig.~\ref{cumlum} for our sample along with the three studies
  named above. It can be seen that the SPY data agree quite well
  with the observations from \citet{max01} and \citet{ede03}, whereas
  there is a larger average offset to the data of \citet{saf94}, which has
  already been discussed in \citet{ede03}. Furthermore, several of our
  stars lie below the ZAEHB, which is plotted in Fig.~\ref{cumlum} for
  various metallicities. Interestingly, the percentage of stars below
  the solar metallicity ZAEHB (12\%) is in better agreement with the
  observations from \citet[ 15\%]{saf94} -- despite the overall offset between
  the two functions -- than with the other
  two studies (3\% and 5\%). Similarly, at the highest luminosities,
  our data again agree best with
  the sample of \citet{saf94}.

  The observed moderate differences between the samples are most
  probably the result of differences in resolution, signal-to-noise
  ratio, homogeneity, and models used for line profile fitting. As far
  as the latter are concerned, one must not only distinguish between
  the actual use of LTE or NLTE models (\citealt{saf94} only use the former), but also between
  the parameter regions in which they are applied: \citet{ede03} and \citet{max01}
  use NLTE models for objects having $\teff > 27\,000\,\kel$, our
  study for sdBs with $\teff > 32\,000\,\kel$. As another example of
  model differences, we mention that the LTE models of \citet{saf94}
  did not include metal line-blanketing, while our LTE models
  do. Differences between NLTE and LTE model atmosphere results may be
  as large as $\Delta\log(L)=0.1\,\dex$
  \citep[see][]{hebkeele}.

  We conclude from the overall comparison of the samples and the lack
  of excessive deviations that the selection effects of our study are
  mainly similar to those of previous studies -- however, as far as flux
  limitation is concerned, we attempt to make a more reliable
  statement below. A more detailed discussion in
  the context of stellar evolution is deferred to Sect.~\ref{sec_evol}. 

  \subsection{Flux limitation \label{sec_maglim}}

  Our programme stars are drawn from flux-limited samples. This may
  introduce an observational bias against objects of lower luminosity,
  although the limiting magnitudes of the surveys (e.g.~HES and HQS)
  are considerably deeper than the limit for SPY. If such a bias were
  present it would be more difficult to reconcile the lower end of the
  cumulative luminosity function with evolutionary predictions, since
  we would have to correct for low luminosity stars missing in our
  sample.

  We therefore carried out a test by comparing the cumulative
  luminosity functions for three subsets of equal number of stars to
  that of the full sample (see Fig.~\ref{cumlum_slices}). A bright
  subset ($B < 14.7$), a medium bright subset ($14.7 \le B \le
  15.6$), and a faint subset ($B > 15.6$) were defined. As can be seen
  from Fig.~\ref{cumlum_slices}, the cumulative luminosity function of
  the faint subset agrees very well with that of the bright and medium
  bright subsample, and all three agree well with that of the entire
  sample. We therefore conclude that no bias due to flux limitation is
  present. This makes the cumulative luminosity function a strong tool
  for the comparison of observation with predictions by the theory of
  stellar evolution (see Sect.~\ref{sec_evol}).


  \section{Double-lined objects \label{sec_doub}}
  
 \subsection{Companion classification \label{sec_comp}}

  For the 24 sdB stars showing spectral signatures of a cool
  companion, accurate values of the stellar parameters cannot
   be expected from a simple model spectrum fit.
 In
  particular, the surface gravity is probably underestimated in many
  cases. Nevertheless, we
  performed line profile fits of the lines $\hdel$ and bluewards for
  20 objects,
  attempting to minimize the effects of companion contribution, since
  the latter decreases with decreasing wavelength. In order to estimate the helium
  abundance, we had to include \ion{He}{i} $4471\,\AA$ in some cases,
  when no other helium line was present in the blue part of the
  spectrum. Our goal was to examine if
  the fit results can still be useful to determine absolute visual
  brightnesses $M_\mathrm{V\,sdB}$ of the sdBs, from which we can
  estimate the companion spectral type using $V$ and $J$ of the
  sdB+main sequence system.
 
  For testing purposes, we
  ``contaminated'' three of our sdB stars showing no companion
  signatures with a certain amount of light from a cool
  companion. Since the SPY stars have not been flux calibrated,
  it was first necessary to create a typical spectral energy
  distribution for the sample stars. This was achieved by using flux
  calibrated low resolution spectra of sdB stars from \citet{pav03}, which were chosen to
  have $\teff$- and $\logg$-values similar to the sample stars. All
  our spectra were convolved with a Gaussian of $2.5\,\AA$ FWHM
  and then rebinned to $1.0\,\AA$ in order to match the spectra used for the companions
(see below). After dividing the
  sample stars by the corresponding flux calibrated ones, a fit to the
  continuum of the resulting spectrum was performed manually. Finally,
  each 
  sample star spectrum was divided by the respective fit curve, resulting in a
  pseudo-flux calibration for them.

\begin{figure}
    \resizebox{\hsize}{!}{\includegraphics{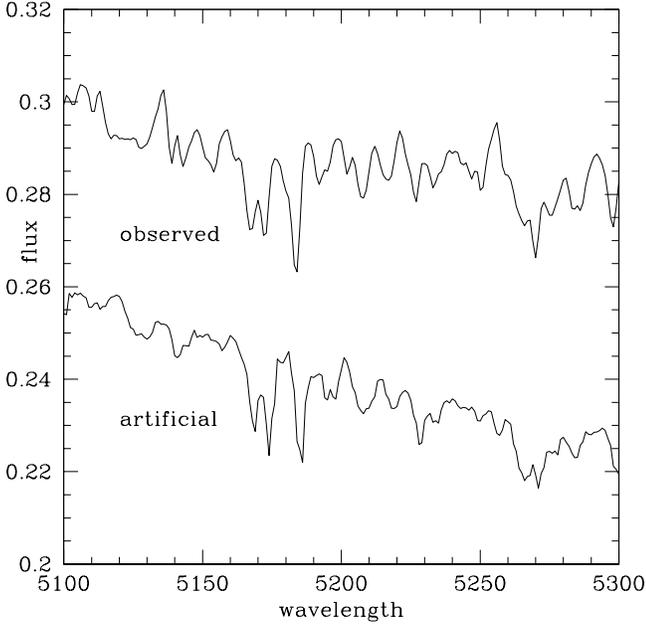}}
    \caption{Observed and artificial composite spectrum. Both spectra
    were scaled to the same flux level, and then displayed with
    an offset. \emph{Observed
    spectrum:} \object{HE 1309$-$1102}, derived companion type
    G9, corrected $M_\mathrm{V\,sdB} = 4\fm0$. \emph{Artificial
    spectrum:} \object{HE 2349$-$3135} with G4 companion,
    $M_\mathrm{V\,sdB} = 3\fm86$.}
    \label{compcomp}
  \end{figure}

  For typical main sequence stars, we used spectra from the
  STELIB\footnote{http://webast.ast.obs-mip.fr/stelib/} database \citep{stelib} for companion
  types F9, G4, K0, and K2. Their resolution is $\Delta\lambda \le 3\,\AA$, which
  matches the treatment of our spectra described above. From the absolute visual
  brightness of the sdB stars and the chosen companion spectrum, flux contribution and total
  magnitudes in $B$, $V$, and $J$ were calculated \citep{bes88,landboe,allen}, and composite spectra
  were produced (see Fig.~\ref{compcomp}). Now, Balmer line profile fitting for $\hdel$ and
  bluewards was performed, keeping the helium abundance fixed for
  simplification. $M_\mathrm{V\,sdB}$ was then calculated from the derived
  values as described in Sect.~\ref{sec_ehb}. Those values show that the effect of
  companion contribution can be quite large for F-type ($\Delta\logg =
  -0.23$ to $-0.6\,\dex$) and G-type ($\Delta\logg =
  -0.06$ to $-0.29\,\dex$) companions, which justifies our
  separate treatment of single-lined and double-lined sdB stars. For
  K-type companions the gravity estimate is almost unaffected ($\Delta\logg =
  -0.02$ to $-0.11\,\dex$).

  Let us now turn to the composite spectrum sdBs from SPY. We do not
  attempt to deconvolve the spectra, but analyze their photometric
  data ($B$, $J$). For
  obtaining apparent brightnesses $m_\mathrm{V\,tot}$ for the system
  and $m_\mathrm{V\,sdB}$ for
  the sdB alone, we begin with adopting $(B-V)_\mathrm{tot} = -0\fm25$\,\footnote{Although the Hamburg/ESO
  Survey provides $B-V$ colours \citep{chr01}, they have been
    calibrated only for main sequence stars with $B-V > -0\fm1$ to an
    accuracy of $\pm0\fm1$. Since an even better accuracy would be necessary for our
    purposes, a calibration for sdB stars would have to be carried
    out beforehand, which is beyond the scope of this paper.} to
  the system, the typical value for a single sdB star
  \citep{sta03}, and assume $m_\mathrm{B\,tot} =
  m_\mathrm{B\,sdB}$. After having performed the calculation of companion type
  once, as described below, we then
  correct $(B-V)_\mathrm{tot}$ according to the calculated type and start
  over again with a new $m_\mathrm{V\,tot}$.

  \begin{table*}
    \centering
    \caption{Determination of companion types, the results of which
    are listed in the second column. The first 15 objects
    show \mgi~in the spectrum, the last five stars only show a
    continuum contribution at $\hal$.}
         \begin{tabular}{llllllllllllll}
            \hline
            \hline
            \noalign{\smallskip}
            Object & Type &
            $m_\mathrm{B\,tot}$ & $m_\mathrm{J\,tot}\ ^{\mathrm{a}}$ & $m_\mathrm{V\,sdB}$ & $m_\mathrm{V\,comp}$ &
            $(B-V)_\mathrm{tot}$ & $M_\mathrm{V\,sdB}^\mathrm{\,out}$ & Remark\\
             & & mag & mag & mag & mag & mag & mag &\\
            \noalign{\smallskip}
            \hline
            \noalign{\smallskip}
            \object{HE 1038$-$2326} & F7 & 15.9        & 15.22  & 16.7  & 16.4  &  +0.11   & 4.0  & F7 or earlier\\
            \object{HE 1140$-$0500} & G1 & 14.8        & 15.17  & 15.1  & 17.3  &  $-$0.18 & 2.4  &    \\
            \object{HE 1221$-$2618} & G0 & 14.7        & 13.91  & 15.4  & 15.2  &  +0.13   & 4.6  &    \\
            \object{HE 1309$-$1102} & G9 & 16.1        & 16.14  & 16.4  & 18.2  &  $-$0.13 & 4.0  &    \\
            \object{HE 1352$-$1827} & G3 & 16.0        & 15.75  & 16.5  & 17.2  &  $-$0.03 & 4.1  &    \\
            \object{HE 1422$-$1851} & F9 & 16.3        & 16.17  & 16.7  & 17.7  &  $-$0.07 & 3.2  &    \\
            \object{HE 1441$-$0558} & G5 & 14.4        & 13.79  & 15.0  & 15.2  &  +0.08   & 4.9  &    \\
            \object{HE 2156$-$3927} & K3 & 14.26       & 14.44  & 14.5  & 17.2  &  $-$0.18 & 4.1  & \citet{alt04}: $(B-V)_\mathrm{tot} = -0.19$   \\
            \object{HE 2322$-$0617} & G9 & 15.7        & 15.58  & 16.1  & 17.4  &  $-$0.09 & 4.3  &    \\
            \object{HE 2322$-$4559} & G4 & 15.5        & 15.25  & 16.0  & 16.8  &  $-$0.03 & 4.2  &    \\
            \object{HS 1536+0944}   & K0 & 15.6        & 15.27  & 16.0  & 17.2  &  $-$0.07 & 4.7  &    \\
            \object{HS 2216+1833}   & G1 & 13.8        & 13.32  & 14.4  & 14.7  &  +0.04   & 4.2  &    \\
            \object{PG 0258+184}    & G8 & 15.3       & 14.99  & 15.7  & 16.7  & $-$0.06 & 4.4  &    \\
            \object{PG 2122+157}    & F7 & 15.0       & 13.95  & 16.4  & 15.0  & +0.28   & 5.1  & F7 or earlier\\
            \object{TON S 155}  & F7 & 16.1$\ ^{\mathrm{c}}$ & 15.69  & 16.8  & 17.0  & +0.02   & 3.5  & F7 or earlier\\
            \noalign{\smallskip}
            \hline                                                                                                                             
            \noalign{\smallskip}
            \object{HE 1200$-$0931} & K1 & 16.2        & 16.07  & 16.5  & 18.1  & $-$0.11 & 4.6  & \\
            \object{HE 1254$-$1540} & K7 & 15.2        & 15.56  & 15.5  & 19.0  & $-$0.22 & 4.3  & \\
            \object{HE 1419$-$1205} & K5 & 16.2        & 16.49  & 16.5  & 19.6  & $-$0.21 & 4.2  & \\
            \object{HS 2125+1105}   & K4 & 16.4        &$\ge16.5$& 16.7  & 19.2  & $-$0.18 & 4.5  & K4 or later\\
            \object{KUV 01542$-$0710}  & K2 & 16.3$\ ^{\mathrm{b}}$ & 16.30  & 16.4  & 18.8  & $-$0.18 & 4.0  & \\
            \noalign{\smallskip}
            \hline
            \hline
            \label{tab_comp}
         \end{tabular}
         \begin{list}{}{}
         \item[$^{\mathrm{a}}$] Data from 2MASS, obtained by usage of the
         VizieR database \citep{vizier}.
         \item[$^{\mathrm{b}}$] Johnson $V$ magnitude
         \item[$^{\mathrm{c}}$] Str\"omgren $y$ magnitude
         \end{list}
  \end{table*}

  The iteration itself is done in the following way:
  The distance modulus of the system is calculated from
  $M_\mathrm{V\,sdB}$ and $m_\mathrm{V\,sdB}$.
  The $J$ magnitude is determined by subtracting
  the typical value of $V-J = -0\fm6$ for single sdB stars
  \citep[ consistent with our own findings]{sta03}. Then, by comparison
  with the measured $m_\mathrm{J\,tot}$, $m_\mathrm{J\,comp}$ is
  calculated, leading to $M_\mathrm{J\,comp}$ when combined with the
  distance modulus. This automatically yields a companion type when
  compared to literature \citep{bes88,landboe,allen}, providing us with a value
  for $M_\mathrm{V\,comp}$ and subsequently $m_\mathrm{V\,comp}$. Now,
  a more reliable $m_\mathrm{V\,sdB}$ was calculated from the
  latter and $m_\mathrm{V\,tot}$, and the next iteration step was
  performed. The iteration is stopped when the spectral subtype of the companion
  stays the same from one step to the next, reflecting
  our maximum achievable accuracy.

 The iteration leads to good
  estimates of the companion types for all our artificial composite
  sdB stars, even when the companion contribution to the total
  flux is quite high. The difference of input and output type
  lies between 0 and 2 subtypes. This clearly shows that the method can indeed be
  applied to our observed sample of composite sdB stars. However, we
  must point out three sources of error:
  \begin{enumerate}
    \item
      We adopt a typical value of $(V-J)_\mathrm{sdB} = -0\fm6$. Since
      the observed scatter is quite large (see Fig.~\ref{bjb}
      and \citealt{sta03}), this could produce errors in our companion
      types. The estimated error for $B$, from which $V$ is
      determined (see below), is $0\fm2$, $J$ errors from 2MASS lie at
      about $0\fm1$. For $\Delta(V-J)_\mathrm{sdB} =
      \pm0\fm2$, the difference is two subtypes or less for low
      companion contribution (i.e.~K types), and even smaller for brighter
      companion stars (i.e.~earlier types).
    \item
      Similarly, we adopt a typical value of $(B-V)_\mathrm{sdB} =
      -0\fm25$, which in reality also shows some small scatter of the
      order of $0\fm1$ \citep{alt04}.
    \item
      For deriving companion types and absolute visual brightnesses,
      we use typical values given in the literature, again
      neglecting the natural scatter contained in the true values.
  \end{enumerate}

  We therefore conclude that the error limits of our results are a few
  subtypes. We do not account for extinction
  because of the relatively high galactic latitudes of our
  stars. The nine objects which are also in the sample of \citet[
  see Table \ref{tab_res}]{alt04} have $0.004 \le E_{B-V} \le
  0.035$,
  justifying that approach. However, extinction might be important for
  \object{PG 0258+184} and \object{PG 2122+157}, where $E_{B-V} \ge
  0.1$ \citep{sch98}. It follows that for these objects the sdB is somewhat brighter
  than what is derived by our iteration, and hence the companion
  is slightly fainter, and of later type.

  Our derived companion types for the
  observed composite sdB stars are shown in Table \ref{tab_comp}. They
  range from F7, which is an early-type limit in our iteration, to K3
  for the stars showing \mgi~in the spectrum. As expected, all of the
  five companions that were only deduced
  from a contribution at $\hal$ are of relatively late spectral type
  (K1 to K7). The derived atmospheric parameters for the corresponding
  sdBs are not affected by companion spectral contribution, since the
  latter is negligible for the lines $\hdel$ and bluewards, where line profile
  fitting was performed.
All composite
  sdB stars will be investigated in more detail in the
  future, when we will attempt to deconvolve their spectra as described in
  \citet{azn01, azn02}.

  \subsection{Main sequence or subgiant companions \label{sec_sub}}

  The luminosity class of cool companions to sdBs has been discussed
  extensively. For example, \citet{all94} and
  \citet{jef98} claimed that the
  companions to their sdB stars are mostly overluminous as compared to
  main sequence stars. In contrast, \citet{azn01, azn02} found their companion
  stars to be mostly main sequence objects. Recently, \citet{sta03}
  again discussed
  the nature of companions to sdB stars, pointing out that optical and
  infrared photometry alone is not able to distinguish between the
  main sequence and the subgiant scenario.

  \begin{figure}
    \resizebox{\hsize}{!}{\includegraphics{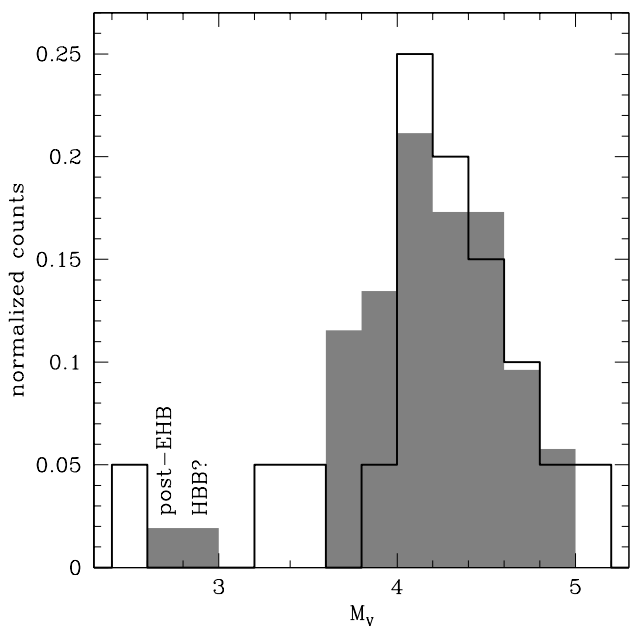}}
    \caption{\emph{Shaded histogram:} Distribution of absolute visual brightness of
    non-composite objects. The two leftmost counts come from the
    post-EHB star \object{HE 0415$-$2417} and
    the potential HBB star \object{HE 0151$-$3919}.
    \emph{Open histogram:} Distribution of 20 out of 24 sdBs with cool companions,
    as derived from the iteration described in Sect.~\ref{sec_comp}.}
    \label{mvhist}
  \end{figure}

  Figure \ref{mvhist} shows the absolute visual brightness distribution
  of our non-composite sdB stars. The values range from $3\fm6$ to $5\fm0$, with
  two extreme values at $2\fm76$ and $2\fm94$ from one
  post-EHB and one potential HBB object, respectively. In the previous
  section, our iterative calculations of
  companion types yielded as a byproduct absolute magnitudes of the
  sdB component (see Table \ref{tab_comp}), which are also shown
  in Fig.~\ref{mvhist}. These values are in very good agreement with the
  ones derived for the non-composite sdB stars, and are mostly much fainter than
  the typical values of $M_\mathrm{V} \approx 3$ for subgiant
  stars (\citealt{landboe}; also see Reid,
  http://www-int.stsci.edu/$\sim$inr/cmd.html, and references
  therein). This clearly proves that the majority of our companion stars are indeed
  main sequence stars.

  In addition, our values of $M_\mathrm{V\,sdB}$ are
  supported by observations of twelve EHB stars in the
  globular cluster \object{NGC 6752}. Its known distance allows a
  direct determination of their absolute brightnesses, which lie in
  the range $3\fm80 \le M_\mathrm{V\,sdB} \le 4\fm66$,
  with a median value of $M_\mathrm{V\,sdB} = 4\fm21$
  \citep{moe97}. The median value for
  our non-composite objects is $M_\mathrm{V\,sdB} = 4\fm20$, and
  $M_\mathrm{V\,sdB} = 4\fm2$ for 
  the sdB component of the composites.

  We therefore conclude that the vast majority of cool
  companions to observed sdB stars are main sequence stars, and
  consider the question of companion luminosity class to be
  settled. However, it follows automatically that sdB+subgiant systems
  can hardly be detected because of the subgiant being brighter than the
  typical sdB. Since ``our goal should be to describe Nature, not to confirm
  our own favorite prejudices'' (R.~A.~Wade), it has to be pointed out
  that such systems may
  exist, yet mostly remain undetected.

  \subsection{Helium abundance \label{sec_heco}}

  \begin{figure}
    \resizebox{\hsize}{!}{\includegraphics{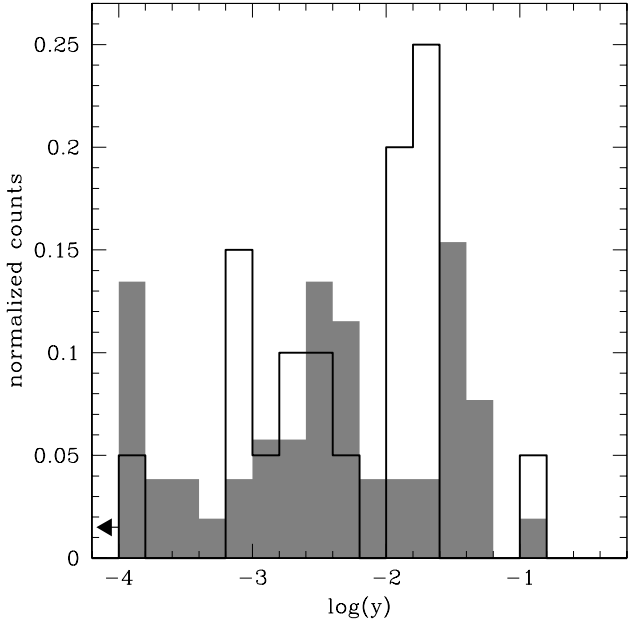}}
    \caption{\emph{Shaded histogram:} Distribution of photospheric
    helium abundance of non-composite sdB stars, as in
    Fig.~\ref{hehist}. \emph{Open histogram:} Distribution of 20 out
    of 24 objects with cool companions.}
    \label{hecomp}
  \end{figure}

  \citet{azn02} find that the composite-spectrum sdB stars show lower
  photospheric helium abundances than the non-composite objects,
  the former having $\logy < -2$. Our derived helium abundances for
  the 20 composite sdB stars for which fitting was performed can be
  seen in Table \ref{tab_res}. They are expected to be
  lower limits, since the presence of a cool companion contributing to
  the total flux dilutes the helium lines and therefore decreases the
  helium abundance derived from spectral fitting.
  A similar test as described in Sect.~\ref{sec_comp}, using
  \object{HE 2237+0150}, shows that the helium abundance is
  underestimated by $0.17\,\dex$ for an F9 companion and $0.03\,\dex$
  for a K2 companion.

  Figure \ref{hecomp} contains the derived values as a
  histogram together with the results for non-composite sdB stars. It is clear that the helium abundances for
  composite-spectrum sdB stars cover the whole range of abundances
  from $\logy \le -4$ to $\logy > -1$, with $50\%$ of the stars having
  $\logy > -2$, in obvious disagreement with the
  observations of \citet{azn02}.

  The stars discussed
  here have formed in binaries with relatively large orbital separations and long
  periods (1st RLOF channel, see Sect.~\ref{sec_evol}). \citet{azn02} suggested that the
  different strengths of tidal effects in short-period and long-period
  sdB binaries may cause differences in
  the atmospheric helium content of these two groups. This argument was taken up by \citet{ede03}
  as a possible explanation for
  their two distinct sdB sequences in the $\teff$-$\logy$-plane (see
  Fig.~\ref{teffnhe}). The absence of any differences between
  single-lined and double-lined sdBs in our sample rules out
  such a correlation. Furthermore, a large number of the
  non-composite sdB stars may be single objects, which is not taken into consideration in
  these discussions.


  \section{Observation versus theory \label{sec_evol}}

  In order to draw conclusions about potential formation scenarios and the
  evolutionary status of our objects, we consider two theoretical
  studies. \citet{dor93} computed evolutionary tracks for the
  canonical core mass established by the helium flash in the core of a
  red giant star. The core mass depends on metallicity, since the
  elemental abundances determine the core mass necessary for helium
  ignition on the first giant branch (FGB). \citet{dor93} used various
  metallicities and envelope masses, defining the
  standard ZAEHB and TAEHB, as well as EHB and post-EHB 
  evolution, which we have already used in Sect.~\ref{sec_othe}. We
  will take those calculations as representative for scenarios of
  single star sdB evolution, which we define in such a way that any
  companion star that might be present does not affect the evolution
  of the sdB.

  In contrast, the binary
  population synthesis calculations of HPMM aim to distinguish between
  several binary formation
  channels of sdB stars and then simulate the evolution of the
  resulting subsamples. Those calculations are the most recent and
  most extensive simulations for binary formation scenarios of the sdB
  population. They can be divided into three channels:
  \begin{description}
    \item[(i) \emph{The common envelope ejection channel.}]
      Dynamically unstable mass transfer in a binary system with one star being on the FGB results
      in the formation of a CE. Subsequently, the
      spiraling-in of both stars and finally the ejection of the CE can
      lead to an sdB in a very close binary system, if the giant's core is
      still able to ignite helium. The second member of the system can
      either be a main sequence star (``1st CE ejection channel'') or already a white
      dwarf (``2nd CE ejection channel'').\smallskip
    \item[(ii) \emph{The stable Roche lobe overflow channel.}]
      This channel is analogous to
      the CE ejection channel, except for the mass transfer being
      stable. The value of $\qcrit$ decides which of the two channels
      applies to a system: only if the ratio of red giant mass
      to companion mass is below $\qcrit$ (adopted to be either $1.2$
      or $1.5$ in the simulations) is mass transfer stable. This results in binary systems with a much larger
      separation and therefore much longer periods, since there is no
      spiraling-in phase affecting the orbital parameters. Again, one has
      to distinguish between the ``1st RLOF channel'' and the ``2nd RLOF
      channel''. However, as shown in HPMM, the 2nd RLOF channel
      (stable mass transfer onto a white dwarf) is negligible because
      white dwarfs of sufficiently high masses are extremely rare.\smallskip
    \item[(iii) \emph{The merger channel.}]
      The merger of two He-WDs caused by
      loss of energy due to gravitational wave radiation can produce an sdB
      star, if the mass is sufficient for core helium burning. Unlike sdB
      stars from other channels, the objects produced here are single stars.
  \end{description}

  To examine the effects of various poorly known physical parameters
  in the scenarios just described, HPMM
  produced twelve simulation sets for different values of the following
  quantities: CE ejection efficiency $\ace$, fraction $\ath$
  of CE thermal energy used for its ejection, critical mass ratio
  $\qcrit$ above which mass transfer is stable on the RGB,
  initial mass ratio distribution of the progenitor binary
  systems, and metallicity. 
 
  HPMM compared the periods and minimum companion masses for the
  single-lined sdB binaries observed by \citet{max01} and
  \citet{mor03} to the predictions of their numerical models and
  selected a simulation set that matches those observations best (``best-fit
  model''). However, this comparison does not include stars which
  formed via the merger channel, since these are single objects. Hence, it is
  very interesting to compare HPMM's predictions directly with
  our spectroscopic results, because our sample is not biased against
  single stars.

  \begin{figure}
    \resizebox{\hsize}{!}{\includegraphics{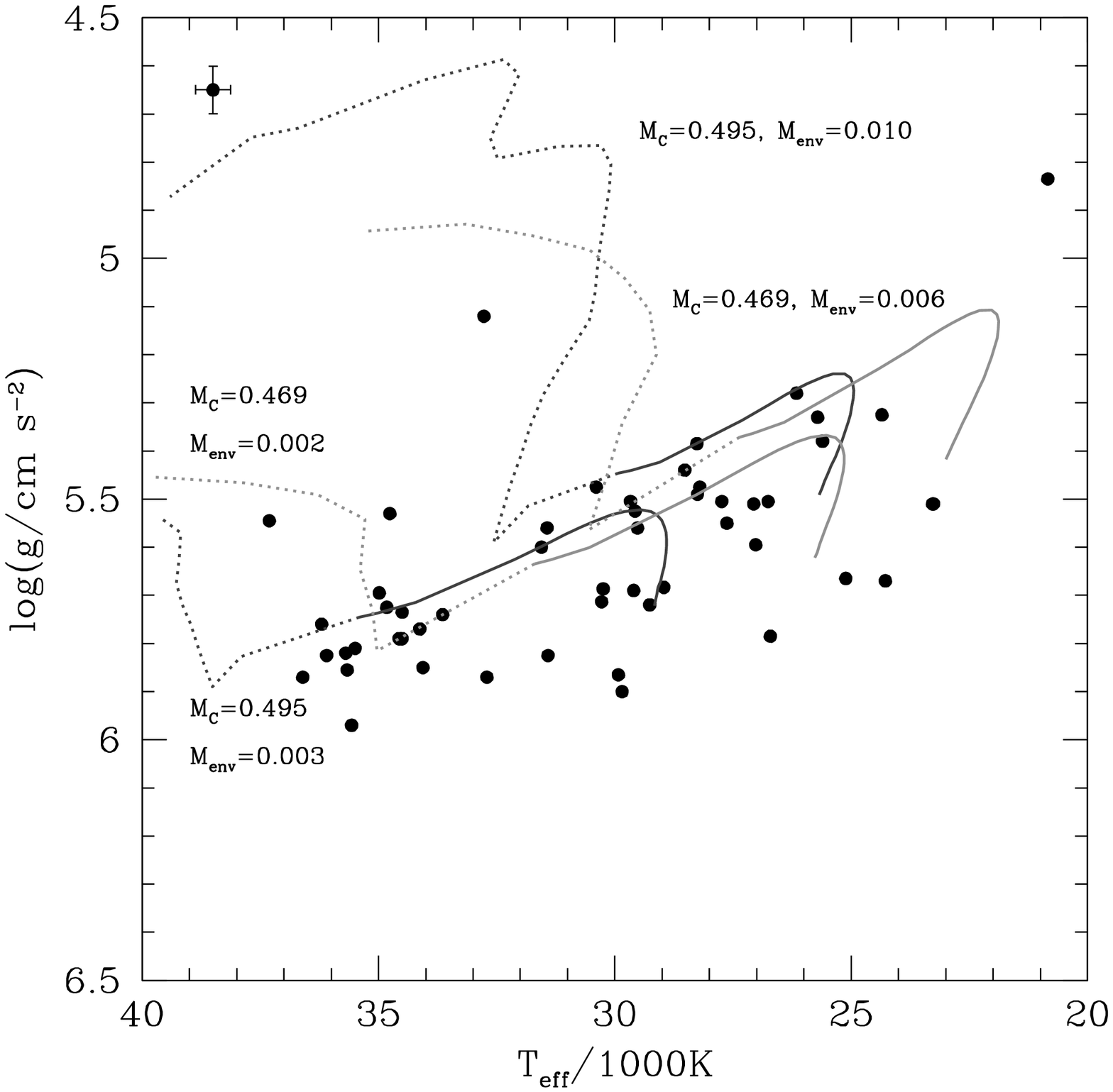}}
    \caption{$\teff$ and $\logg$ values of our sdBs as shown in
    Fig.~\ref{myehb}, along with evolutionary tracks calculated by
    \citet{dor93}. Tracks are shown for solar metallicity (helium core
    mass $M_\mathrm{core} = 0.469\,\msol$, light grey lines) as well
    as for $[\mathrm{Fe}/\mathrm{H}] = -2.26$ ($M_\mathrm{core} = 0.495\
    \msol$, dark grey lines). The respective envelope masses
    $M_\mathrm{env}$ are displayed in the figure. The solid part of
    the tracks marks the core helium burning phase, while this has
    already ceased when the star is on the dotted part of the tracks.}
    \label{dorehb}
  \end{figure}

  \begin{figure}
    \resizebox{\hsize}{!}{\includegraphics{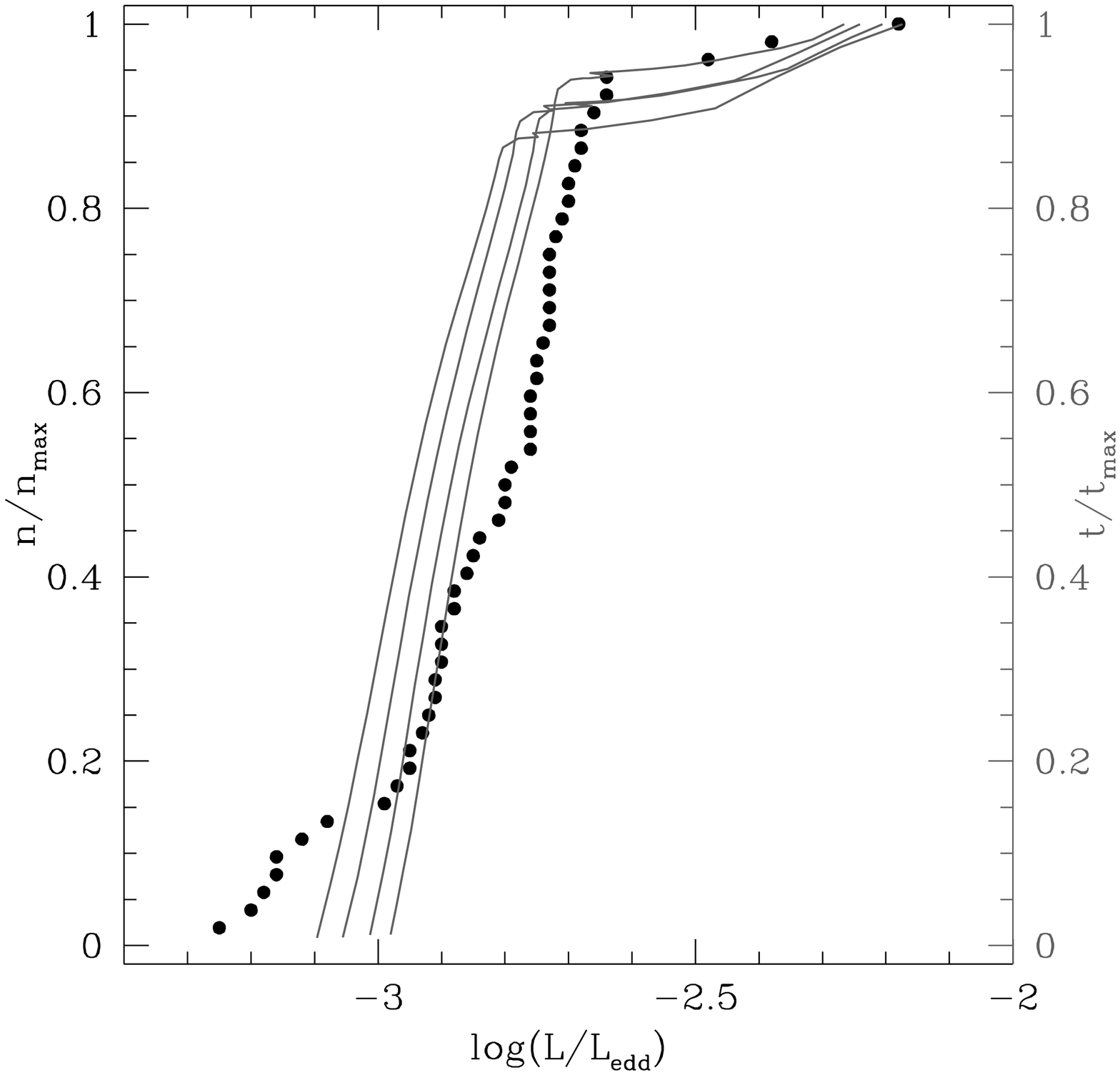}}
      \caption{Cumulative luminosity function of our sdB stars as
    shown in Fig.~\ref{cumlum}, along with sdB luminosity evolution with
    time as calculated by \citet{dor93}. The evolutionary tracks (grey
    lines) are
    followed from the ZAEHB until they leave our
    $\teff$-$\logg$-parameter space ($\teff < 40\,000\,\kel,\ \logg >
    4.5$). For these tracks, the metallicities are from left to right:
    $[\mathrm{Fe}/\mathrm{H}] =  0.00,\ -0.47,\ -1.48, -2.26$
    (corresponding to core masses of $M_\mathrm{core}~=~0.469,\
    0.475,\ 0.485,\ 0.495\,\msol$). For the leftmost track the
    envelope mass is $M_\mathrm{env} = 0.002\,\msol$, for the other
    tracks $M_\mathrm{env} = 0.003\,\msol$.}
    \label{dorcum}
  \end{figure}

  To make a comparison possible and reasonable, it is imperative to
  consider observational selection effects and apply them in the same way to
  the theoretical sample. HPMM showed that the so-called
  \emph{GK selection effect} is the most important one, since
  observational studies typically select against G, K,
  and earlier type main sequence companions. Those either outshine the sdB
  or cause a composite spectrum which renders the analysis of the sdB
  difficult or even impossible. In the simulations, HPMM apply this effect by excluding all
  sdB+main sequence systems with the companion either being brighter than the sdB
  or having $\teff > 4000\,\kel$, corresponding to M and very late K
  types. Since we excluded all composite spectrum stars carefully,
  our sample matches this approach, and allows a detailed comparison of our data with
  the theoretical predictions of HPMM.

  \subsection{Single star evolution \label{sec_dor}}

  Figure \ref{dorehb} compares our results to several evolutionary tracks
  calculated by \citet{dor93} for different core and envelope
  masses. It can be clearly seen that at
  least one of our stars (\object{HE 0415$-$2417} at $\teff =
  32768\,\kel,\ \logg = 5.12$) must be in the post-EHB stage of evolution (dotted
  part of the tracks in Fig.~\ref{dorehb}). For a more thorough
  comparison with these theoretical predictions, we plot in
  Fig.~\ref{dorcum} our cumulative luminosity function from
  Sect.~\ref{sec_othe} along with the luminosity evolution with time given by
  the described tracks, until they reach an effective temperature of
  $40\,000\,\kel$. This limit follows from our selection criteria. The
  function should reflect the ``speed of
  evolution'' across the EHB and in the post-EHB stage.

  It has been argued previously that most stars would be
  expected to lie near the ZAEHB \citep[e.g.][]{ede03}. The
  theoretical tracks, however, show a linear time-luminosity-relation
  while the star is in the EHB strip. The
  respective ZAEHB is defined by the starting point of each
  track, the TAEHB is clearly marked by the
  sharp bend in the tracks. We therefore have to expect a homogeneous
  distribution of sdB stars through the
  EHB strip, if there are no observational selection effects.

  While the discussed evolutionary tracks lie at somewhat lower average luminosity, their
  slope matches our data well, except for the six stars at luminosities lower than any of
  the plotted tracks. As shown in Sect.~\ref{sec_maglim}, any potential effects of
  magnitude limitation are negligible, and do not affect this comparison.
Furthermore, replacing NLTE models by metal line-blanketed 
LTE models for the hottest
stars would primarily lower the luminosity mainly at the top part of the 
cumulative luminosity function by about $0.06\,\dex$. Adopting a very low metallicity would increase the 
luminosities by $0.03\,\dex$ only. Hence these systematic errors are 
unimportant (see Sect.~\ref{sec_err}).

  \subsection{Close binary evolution \label{sec_han}}

  \begin{figure*}[t!]
    \centering
    \resizebox{0.0335\hsize}{!}{\includegraphics{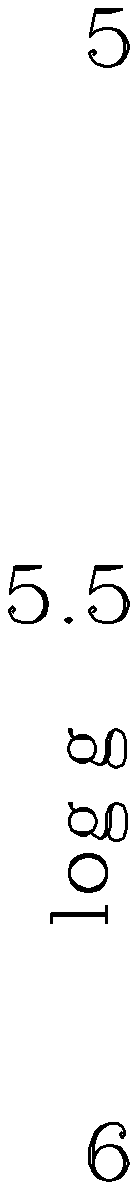}}\resizebox{0.481\hsize}{!}{\includegraphics{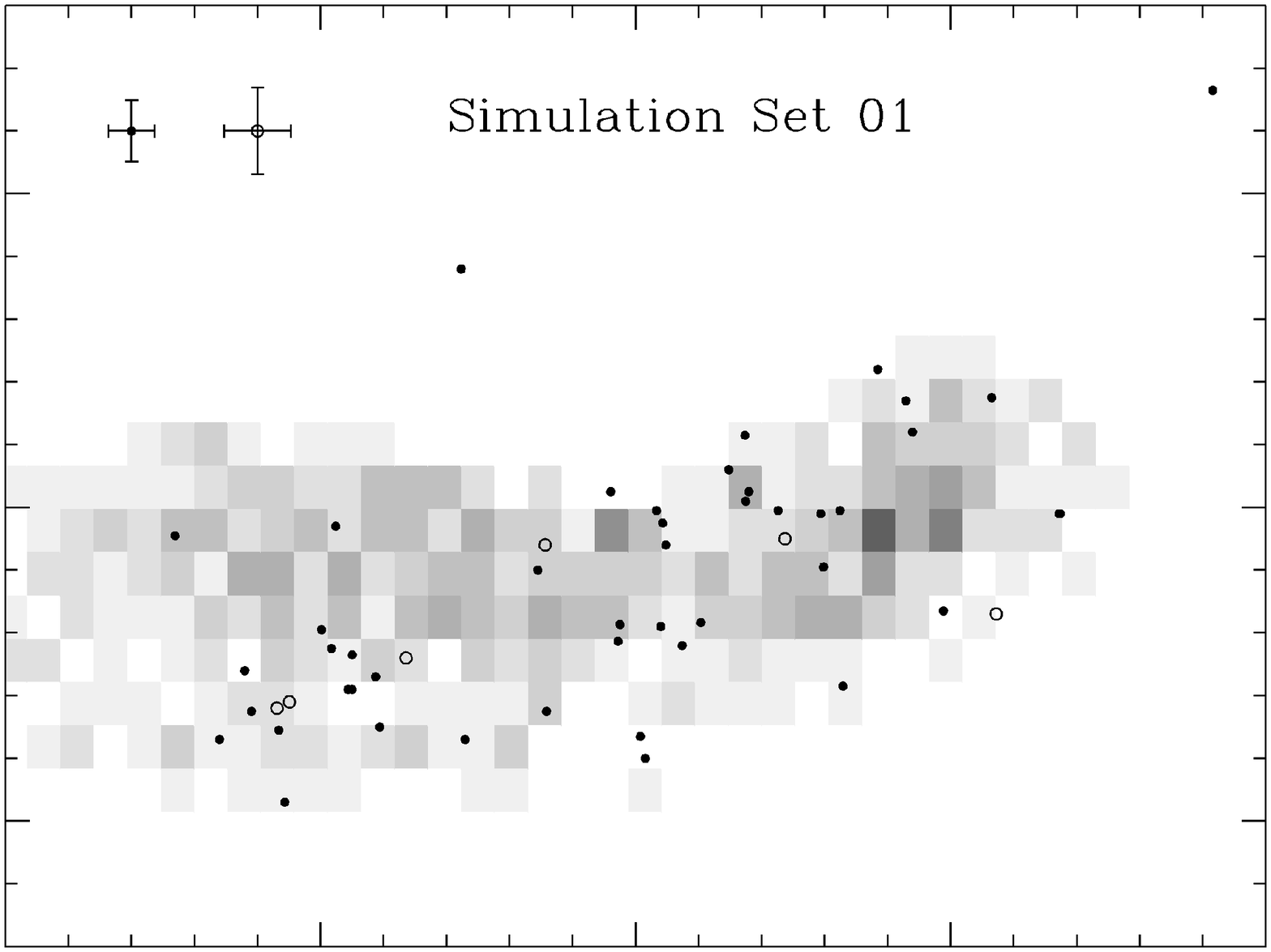}}\resizebox{0.481\hsize}{!}{\includegraphics{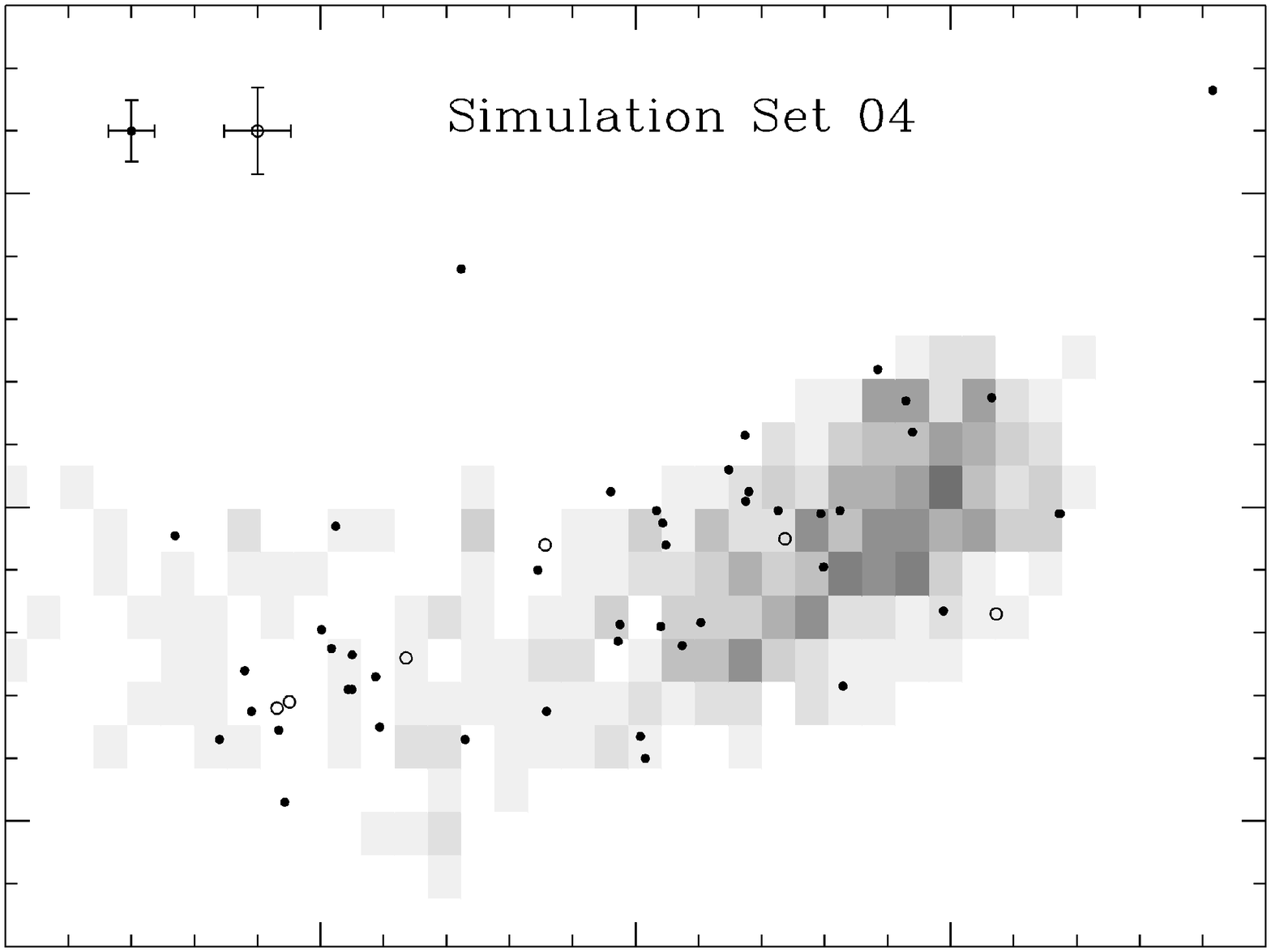}}
    \resizebox{0.0335\hsize}{!}{\includegraphics{0232f17y.eps}}\resizebox{0.481\hsize}{!}{\includegraphics{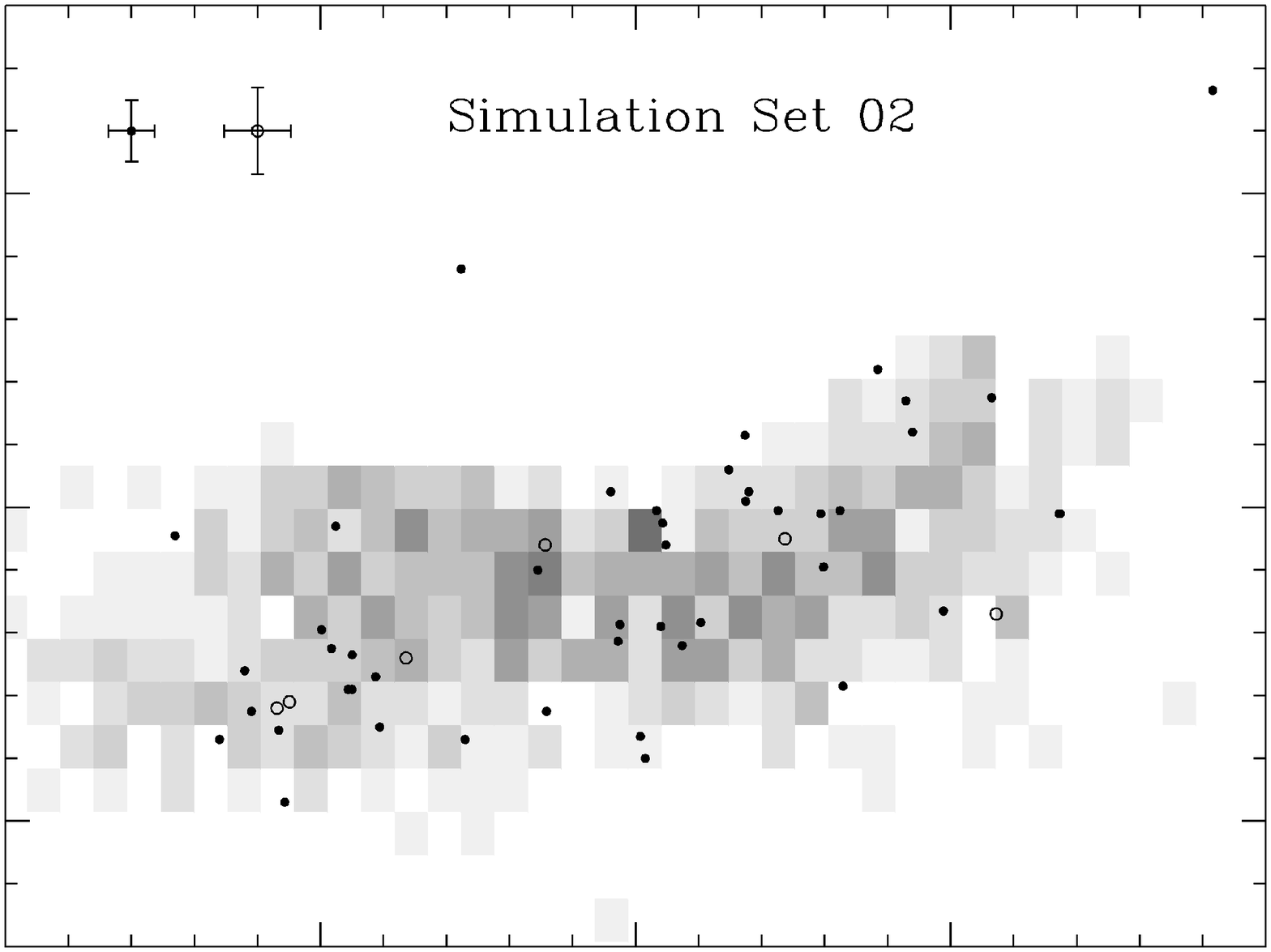}}\resizebox{0.481\hsize}{!}{\includegraphics{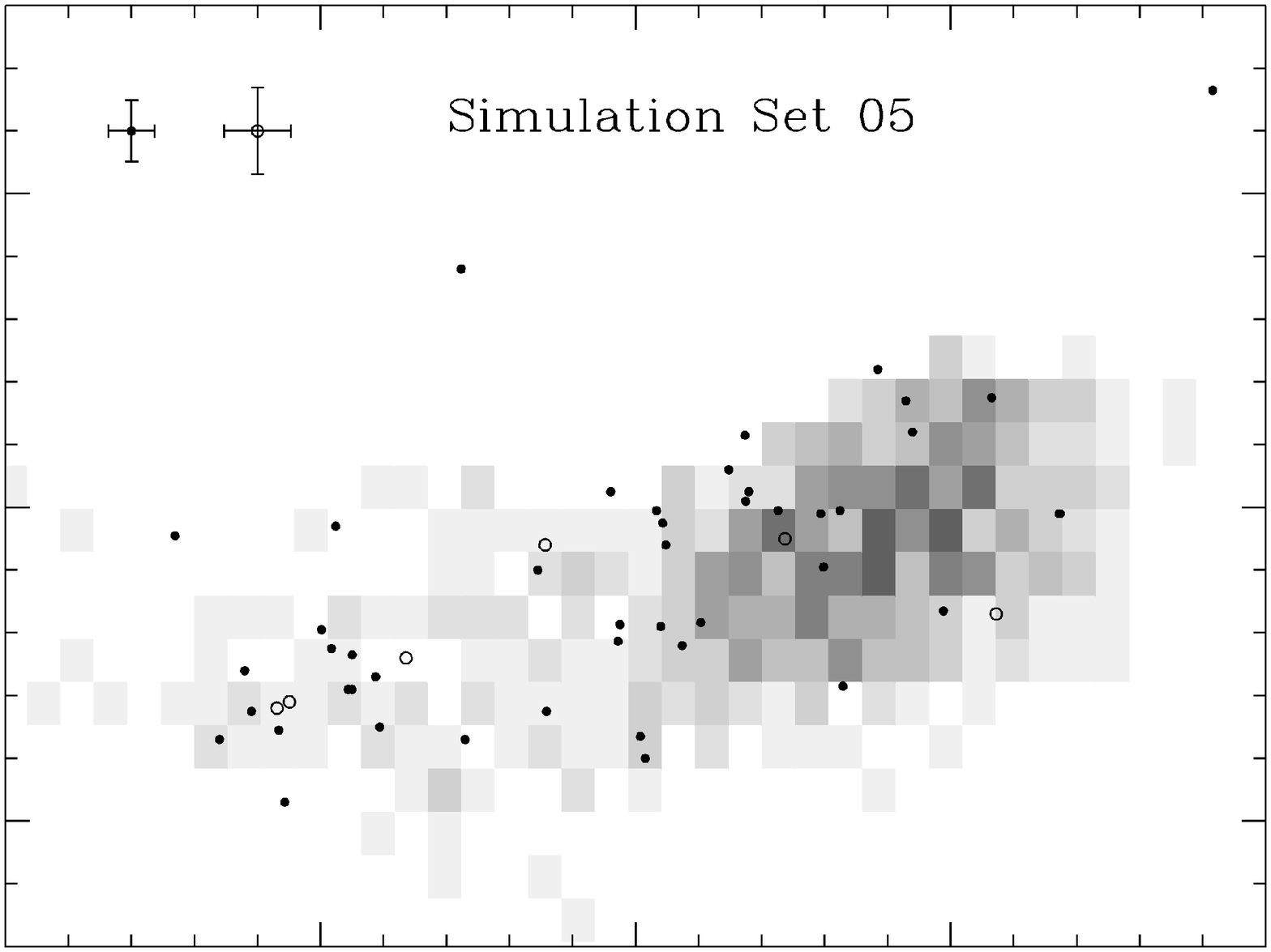}}
    \resizebox{0.0335\hsize}{!}{\includegraphics{0232f17y.eps}}\resizebox{0.481\hsize}{!}{\includegraphics{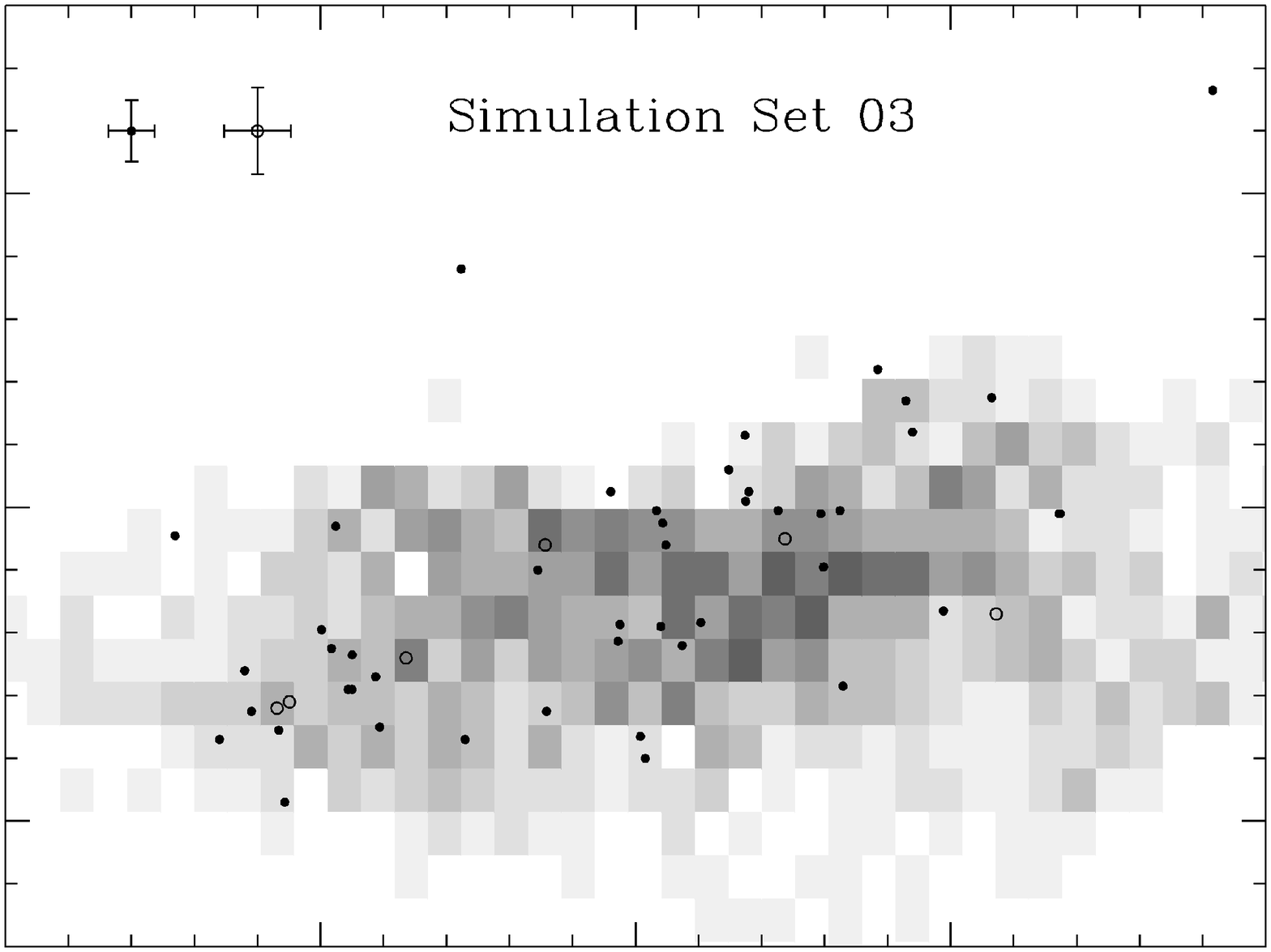}}\resizebox{0.481\hsize}{!}{\includegraphics{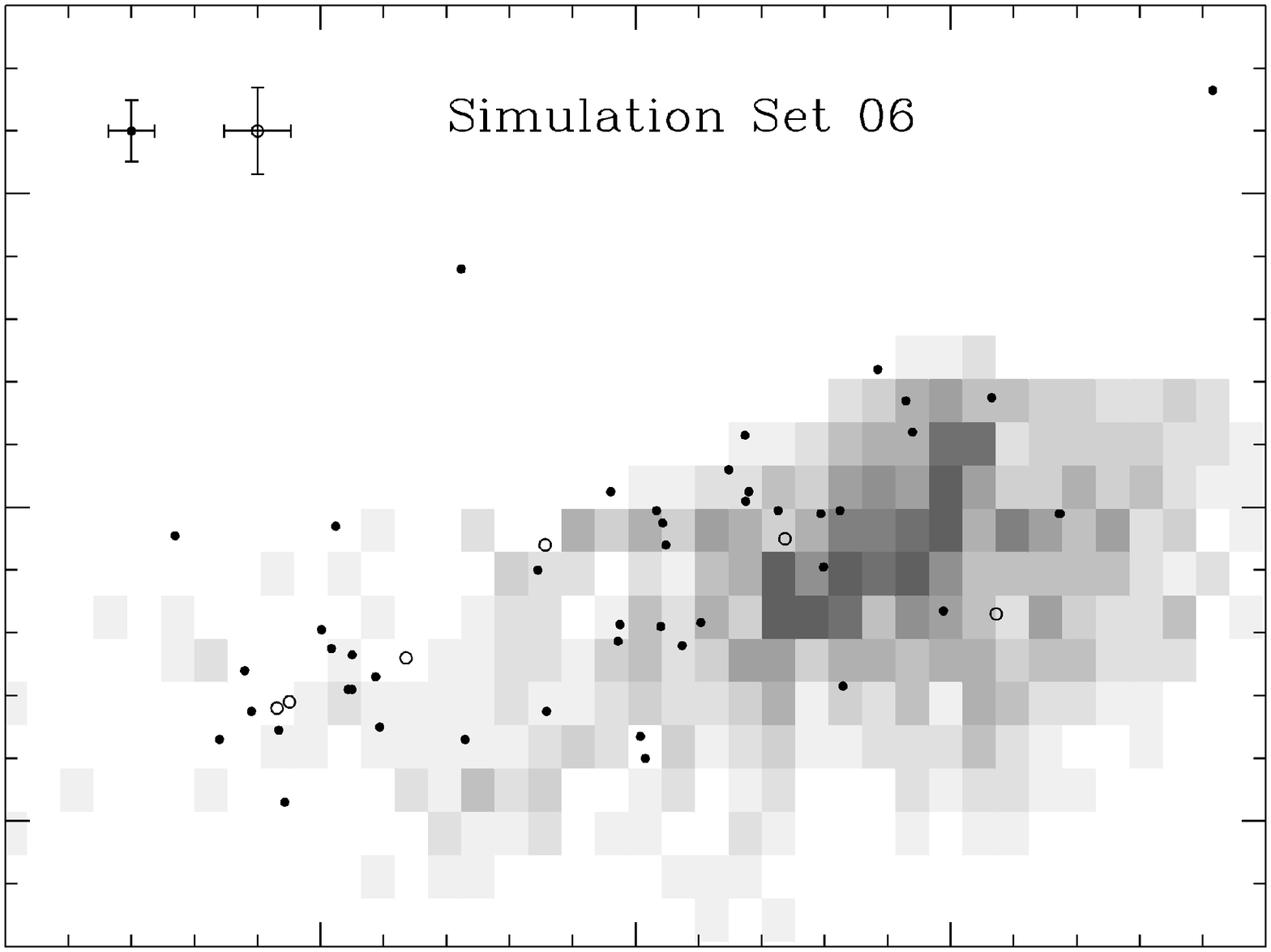}}
    \resizebox{\hsize}{!}{\includegraphics{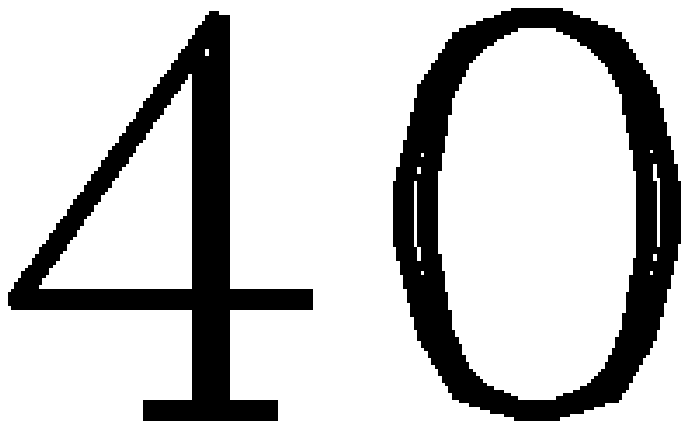}}
    \resizebox{0.9\hsize}{!}{\includegraphics{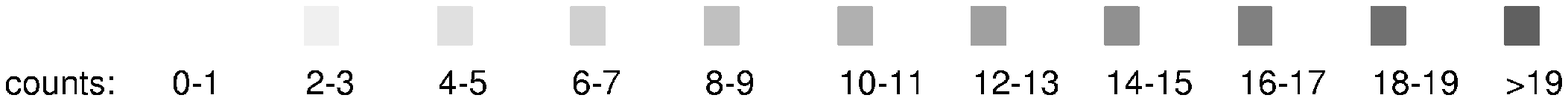}}
    \caption{Comparison of our derived values for $\teff$ and $\logg$
    (symbols and typical errors as in Fig.~\ref{myehb})
    with the simulations from HPMM. The theoretical predictions are
    shown as shaded $\teff$-$\logg$-boxes, where a higher sdB density per
    box corresponds to darker shading. The grey scale is shown below
    the figures. \emph{Continued on
    next page.}}
    \label{tg_all}
  \end{figure*}

 \begin{figure*}
    \emph{Continued from previous page.}
    \centering
    \resizebox{0.0335\hsize}{!}{\includegraphics{0232f17y.eps}}\resizebox{0.481\hsize}{!}{\includegraphics{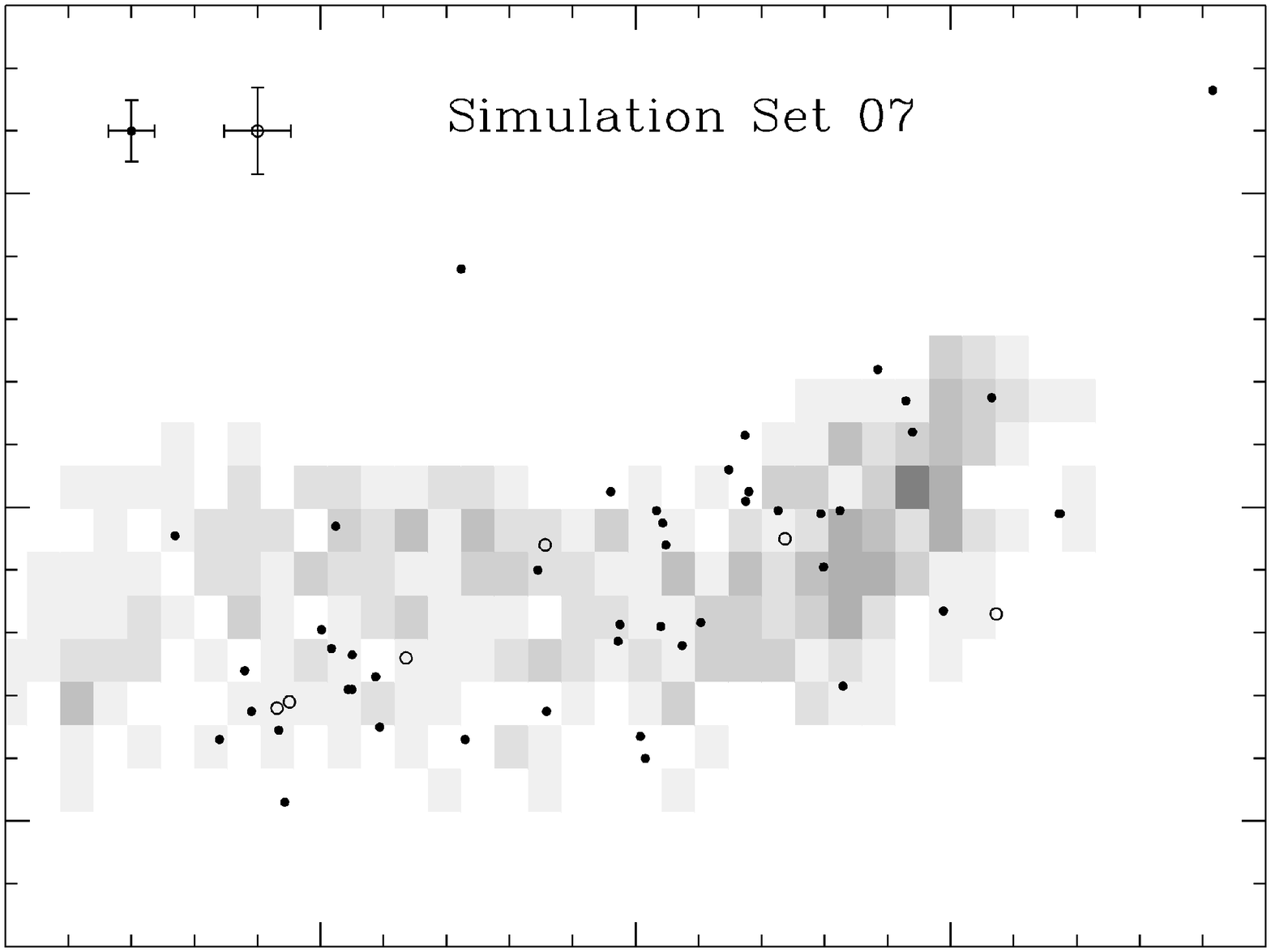}}\resizebox{0.481\hsize}{!}{\includegraphics{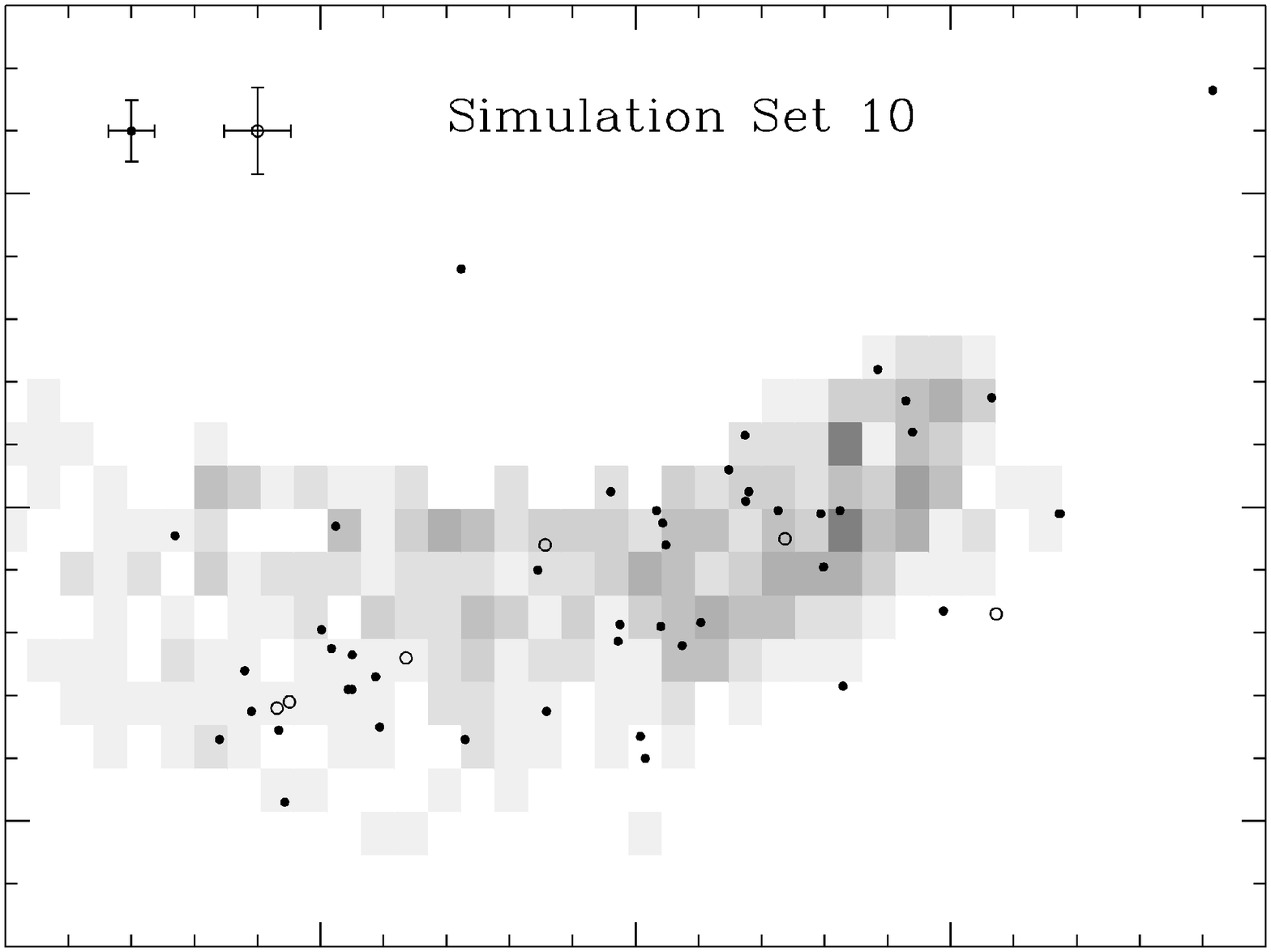}}
    \resizebox{0.0335\hsize}{!}{\includegraphics{0232f17y.eps}}\resizebox{0.481\hsize}{!}{\includegraphics{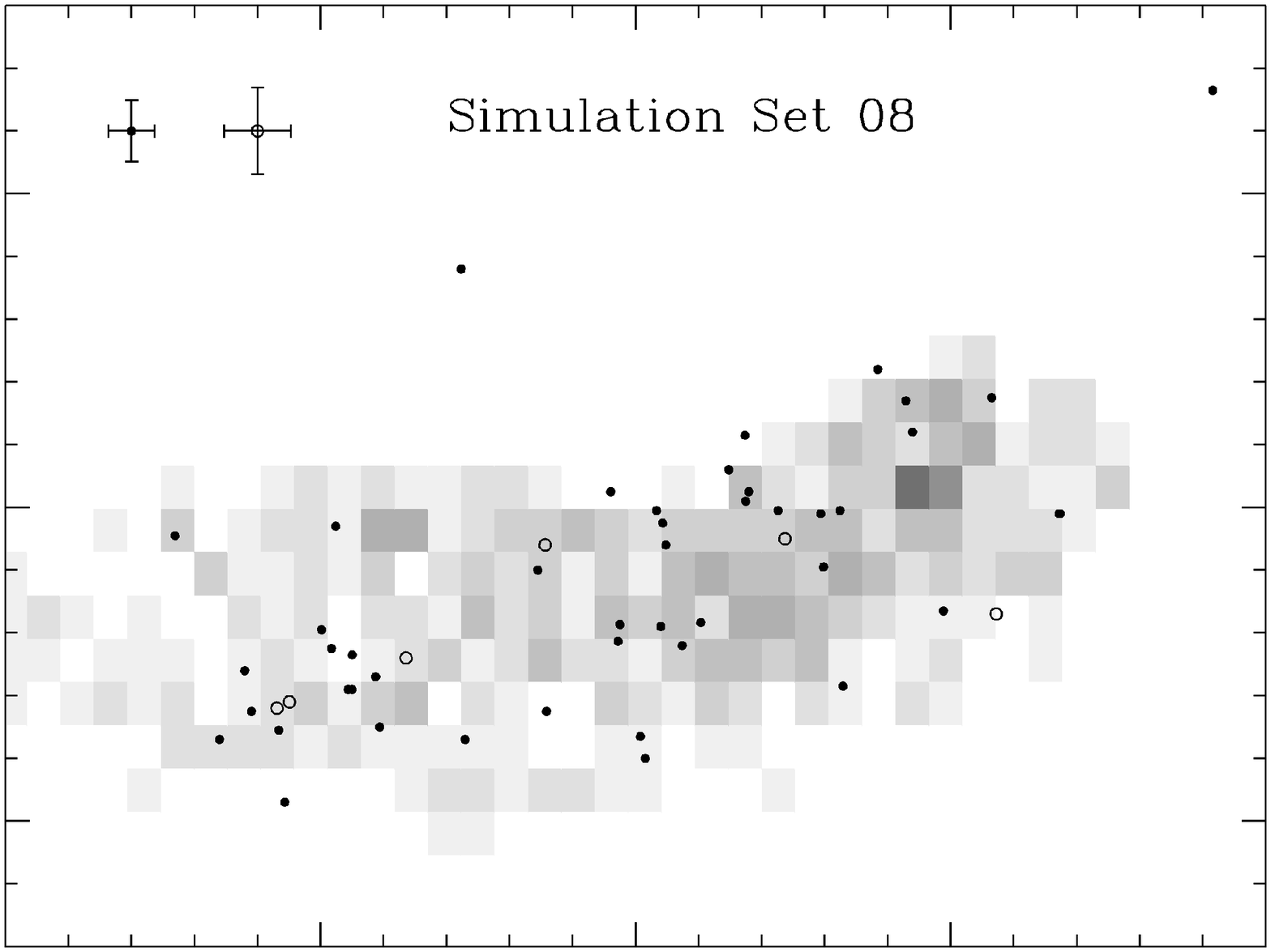}}\resizebox{0.481\hsize}{!}{\includegraphics{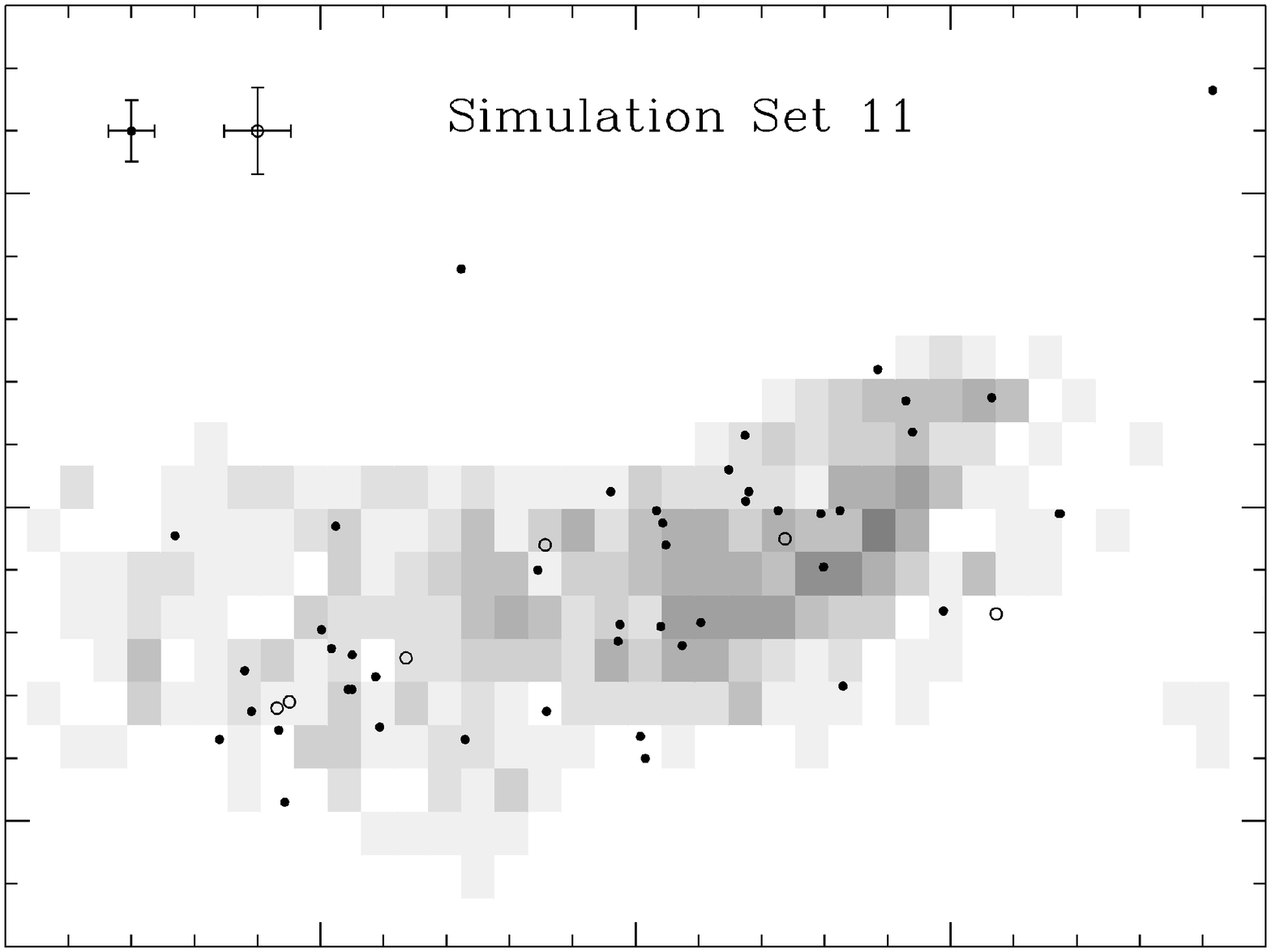}}
    \resizebox{0.0335\hsize}{!}{\includegraphics{0232f17y.eps}}\resizebox{0.481\hsize}{!}{\includegraphics{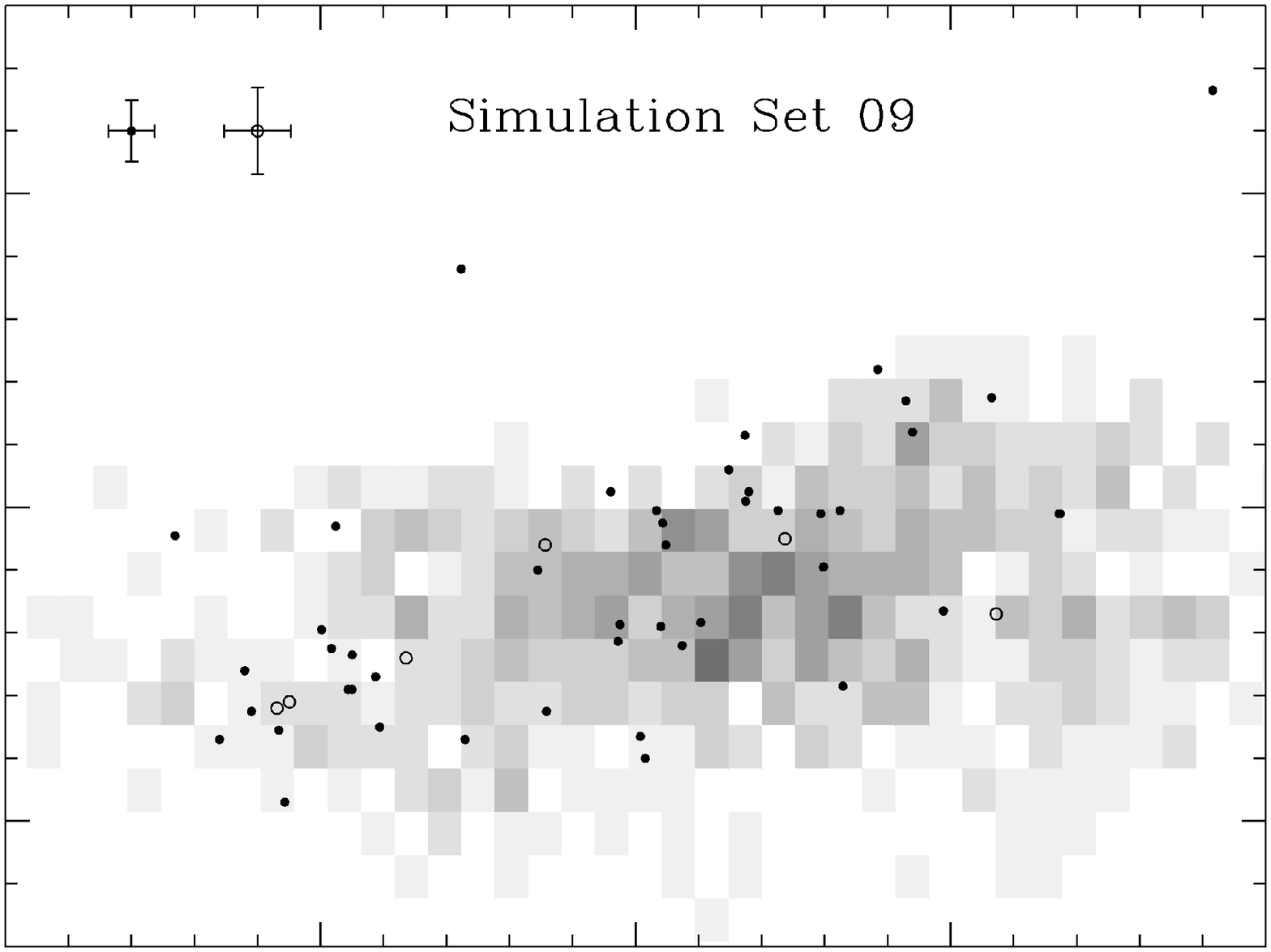}}\resizebox{0.481\hsize}{!}{\includegraphics{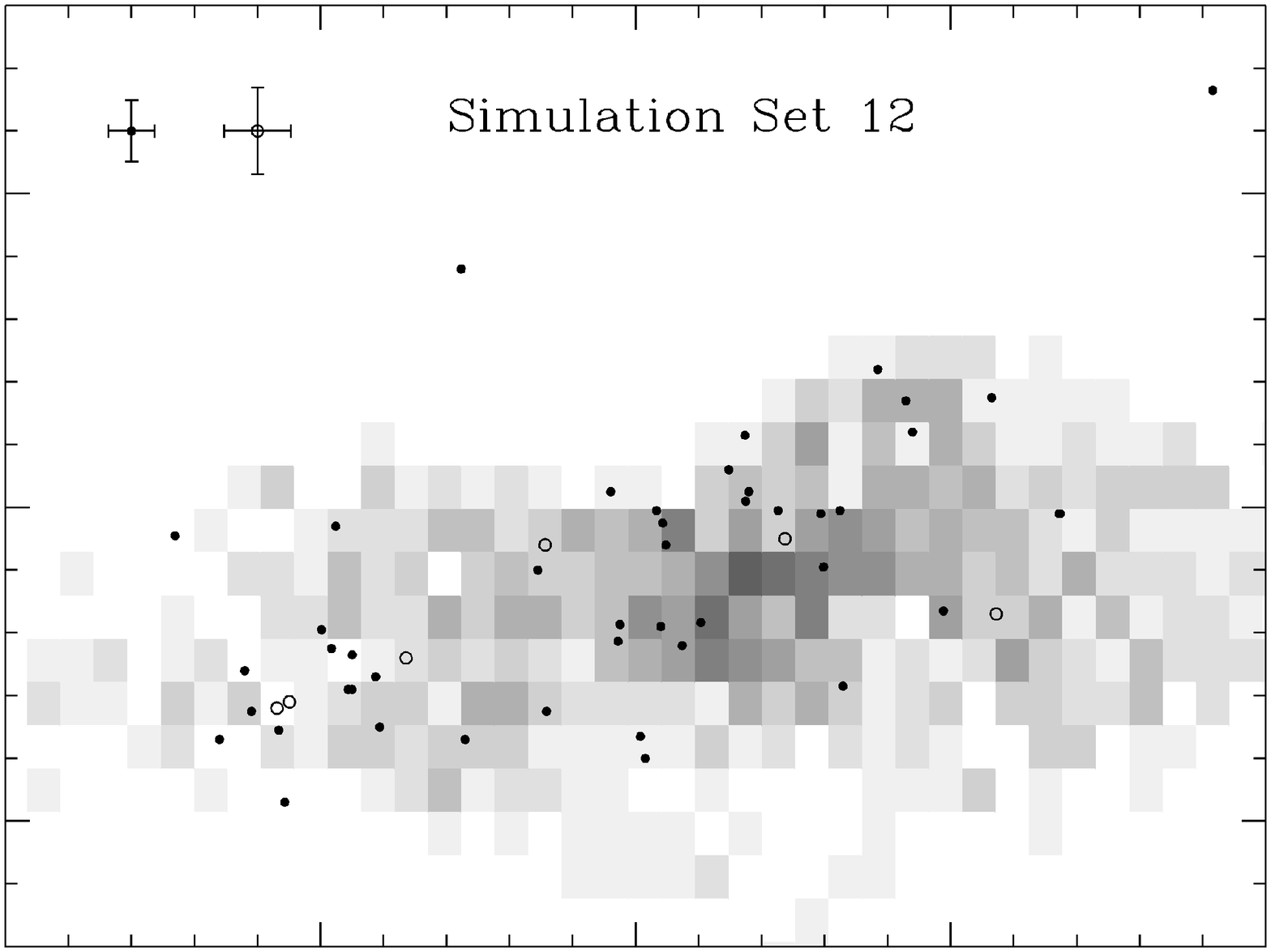}}
    \resizebox{\hsize}{!}{\includegraphics{0232f17x.eps}}
    \resizebox{0.9\hsize}{!}{\includegraphics{0232f17z.eps}}
  \end{figure*}

  We proceed with a detailed comparison of our measurements
  with all twelve simulation sets of
  HPMM.
Our goal is to exclude several sets and their respective
  parameter configurations, to test HPMM's choice of the best-fit
  model, 
and to check whether these calculations can well reproduce the
  observed sdB distribution.

  A short overview of the input parameters that
  characterize the simulation sets is given now. Three sets in a row (1+2+3, 4+5+6,
  etc.) constitute a sequence in both the common envelope ejection efficiency $\ace$
  and the thermal energy fraction $\ath$ used for ejection. The first
  set always has $\ace = \ath = 0.5$, the second $\ace = \ath = 0.75$,
  the third $\ace = \ath = 1.0$. Sets 1 to 6 have a critical mass ratio
  $q_\mathrm{crit} = 1.5$, whereas $q_\mathrm{crit} = 1.2$ is used for
  sets 7 to 12, allowing stable mass 
  transfer only for a narrower mass range than the former value. For sets 4, 5,
  and 6, an
  uncorrelated mass distribution had been adopted for the stars in sdB
  progenitor binaries, whereas in all other sets, HPMM used a
  constant distribution of the progenitor system mass ratio. Sets 10, 11,
  and 12 were
  calculated with a typical thick disk metallicity ($Z = 0.004$) while
  the other sets are for po\-pu\-la\-tion~I metallicity ($Z = 0.02$).

  \subsubsection{The $\teff$-$\logg$-plane: visual inspection \label{sec_han1}}

  Figure \ref{tg_all} shows the simulated data from HPMM
  in the $\teff$-$\logg$-plane. The number density is grey-scale coded; higher
  densities of sdB stars correspond to darker grey shading. Note that the total
  number of simulated stars is not the same for different simulation
  sets. Overlaid on each simulation data set are our measurements. The
  simulations shown here include the GK selection effect, as explained
  above, and are therefore well suited for comparison with our observations.

  Sets 4, 5, and 6 do not match our data,
  since their predicted sdB density at higher temperatures is much
  lower than for cooler stars, in obvious contrast to our
  measurements. Sets 3, 9, and 12 all show a significant number of objects at low $\teff$ and higher
  $\logg$, i.e.~in the lower right area of the diagram. These stars
  have somewhat lower luminosities, and are not seen in our
  data. Since no bias due to flux limitation is present in our sample (see Sect.~\ref{sec_maglim}), we can state a trend
  against these simulation sets. 

  \begin{table}
    \centering
    \caption{Likelihood ranking of the HPMM simulation sets. The
    logarithmic ratio of the likelihood for the best matching set 10 to
    the likelihood for each set is given.}
         \begin{tabular}{llllllllllll}
            \hline
            \noalign{\smallskip}
            Set & 10 & 8 & 11 & 2 & 1 & 3\\
            \noalign{\smallskip}
            $\log\left(\frac{\L_{best}}{\L}\right)$ & 0.0 & 1.2 & 1.5 & 2.2 & 3.5 & 3.8\\
            \noalign{\smallskip}
            \hline
            \noalign{\smallskip}
            Set &  7 & 12 & 5 & 9  & 4  & 6 \\
            \noalign{\smallskip}
            $\log\left(\frac{\L_{best}}{\L}\right)$ & 4.6 & 5.0 & 6.2 & 6.2 & 7.0 & 13.9\\
            \noalign{\smallskip}
            \hline
            \label{tab_like}
         \end{tabular}
  \end{table}

  From visual judgement only, sets 1, 2, 7, 8, 10, and 11 all match the observed sdB
  distribution well. In all sets, a trend can be seen that the average
  surface gravity at lower temperatures is somewhat higher in the simulations
  than in our data. Again, this cannot be explained with a potential
  flux limitation bias, therefore, the observed shift could
  point towards the necessity of refining some model
  details.

  Two stars are not matched by
  any simulation set: \object{HE 0151$-$3919} at $\teff =
  20841\,\kel,\ \logg = 4.83$ and \object{HE 0415$-$2417} at $\teff =
  32768\,\kel,\ \logg = 5.12$. We cannot rule out that the former is a blue
  HB star, which will ascend the AGB after core helium
  exhaustion and therefore may belong to a different evolutionary group than
  the EHB stars. The latter is believed to be in the post-EHB stage
  (see Fig.~\ref{dorehb}), which lasts only a fraction of the time
  spent on the EHB. Since there
  are no additional stars observed in this parameter region, we do not consider this to be
  a discrepancy with theory.   

  In Sect.~\ref{sec_nhe} we found that at higher $\teff$ the sdB stars are
  somewhat more luminous, and tend to have a higher helium
  abundance. The corresponding $\teff$-$\logg$-range of simulated data
  points is dominated by stars that formed in the merger channel.
  Since most of the hydrogen is expected to be burned up during the
  merging process, those objects are more likely to have a higher
  helium abundance than sdB stars from other channels
  \citep{han02}. This qualitative agreement with our results can
  be tested in a quantitative way by using radial velocity variations to
  distinguish between single objects from the merger channel and
  binaries \citep[ Napiwotzki et al. in prep.]{napkeele}.

  \subsubsection{The $\teff$-$\logg$ plane: statistical analysis \label{sec_han2}}

  To check the visual judgement, we performed a statistical test for
  the match of the different sets, using the maximum likelihood
  method. Thereby, we adopt as probability for each of our sdBs
  the fraction of simulated stars in a box around the data point,
  and then multiply the probabilities of all points. The size of the box corresponds
  to two times our errors for stars with one exposure, which was
  carefully chosen to avoid random fluctuations on the one hand, and
  too much smoothing on the other hand. In Fig.~\ref{tg_all} the
  bin size is only half of that value -- i.e.~it equals the errors --
  and one can clearly see that quite large fluctuations can occur,
  since the total number of simulated stars only lies between
  1000 and 2000, and is distributed over several hundred bins.

  \begin{figure*}
    \centering
    \includegraphics[width=5.66cm]{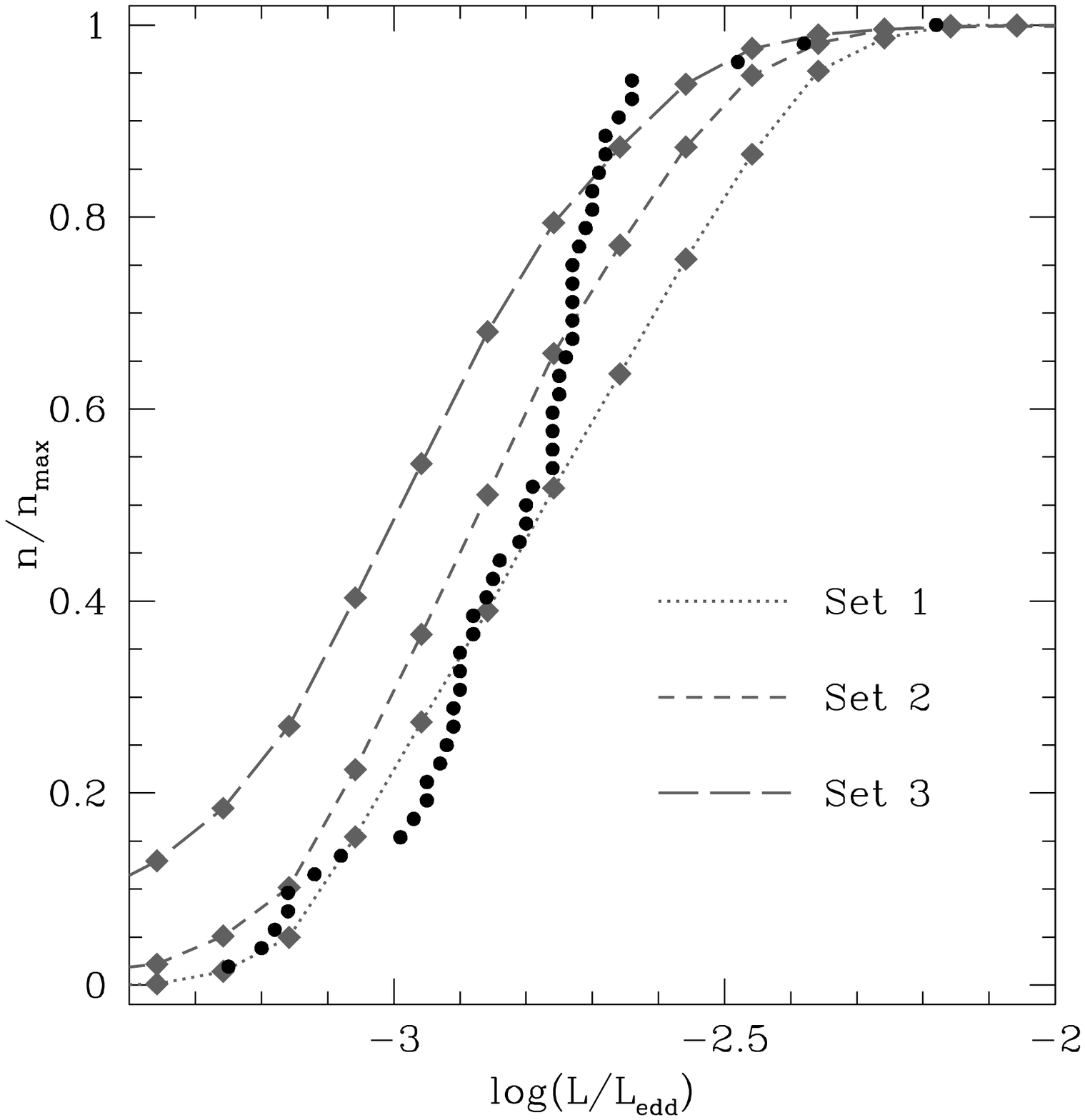}\includegraphics[width=5.66cm]{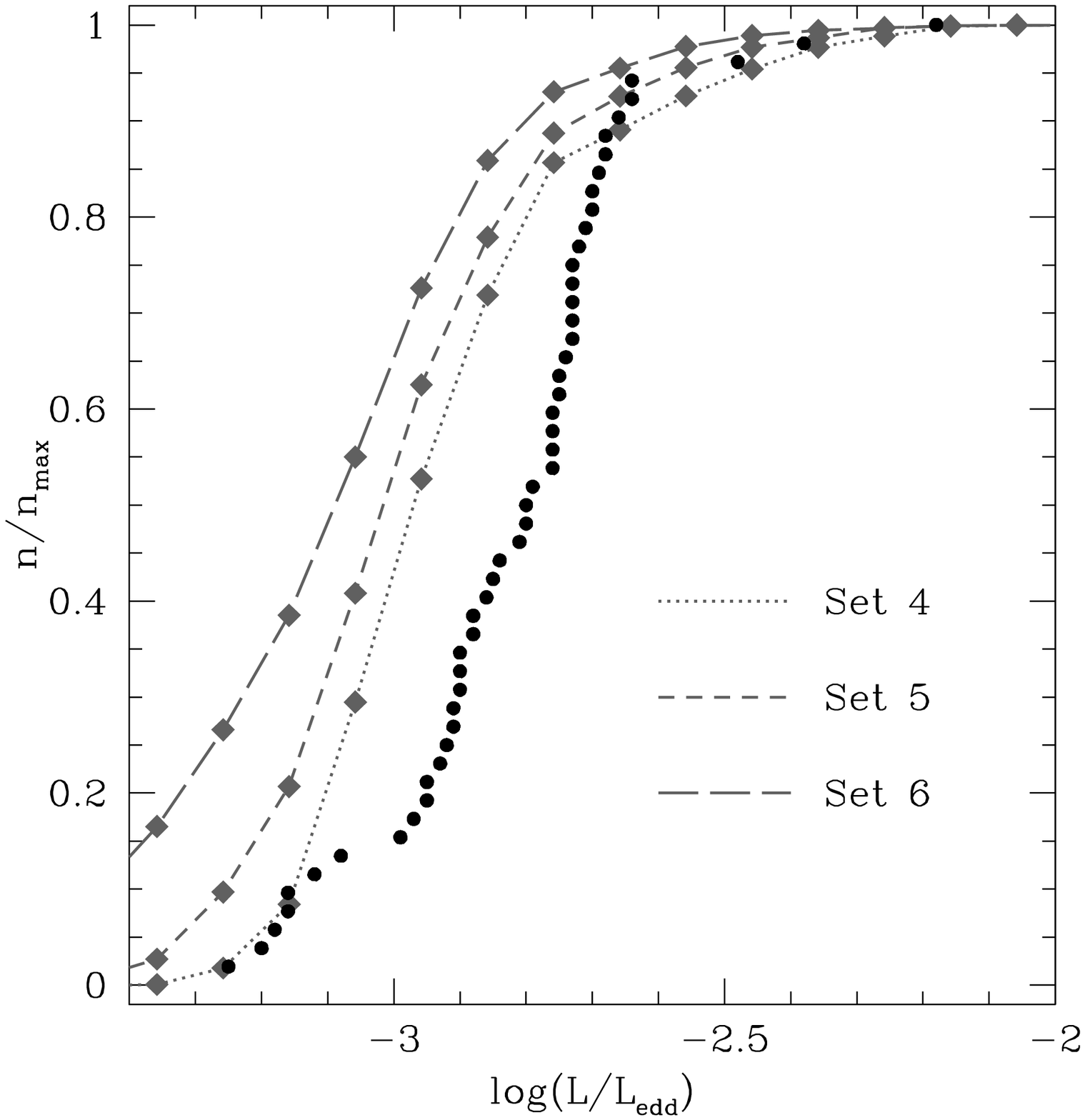}\includegraphics[width=5.66cm]{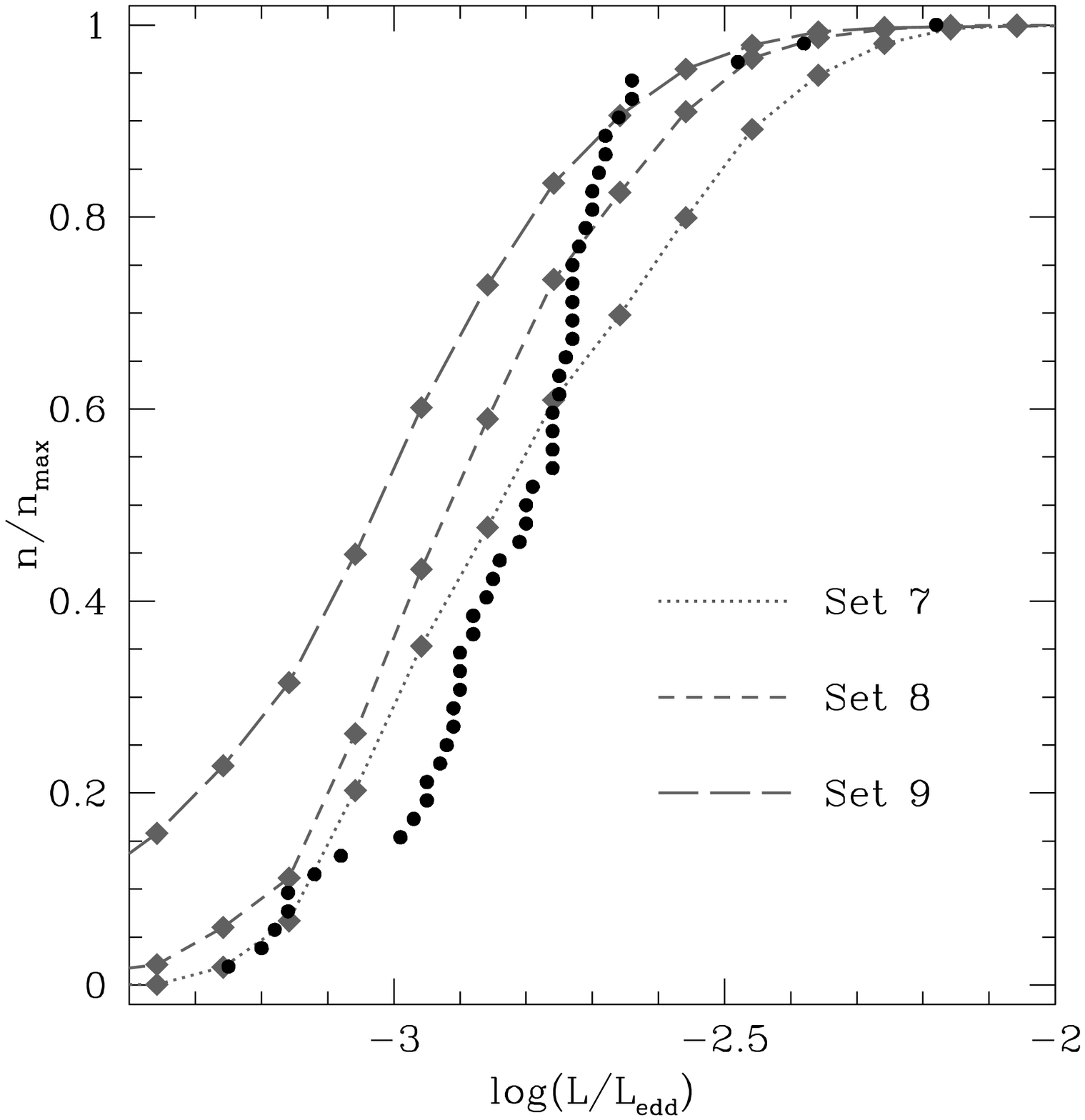}
  \end{figure*}

  The ranking derived from the likelihood of each set to produce the
  observed distribution is given in Table \ref{tab_like}, and clearly
  favors set 10, which is characterized by a low
  efficiency ($\ace = \ath = 0.5$), low metallicity ($Z = 0.004$), and a
  constant mass ratio of the progenitor binaries. We want to point out that the given
  likelihood ratios cannot be exact, since the simulation data are
  provided as discrete number counts rather than a smooth distribution
  function. Nevertheless, the overall ranking can be considered
  reliable. Sets 8, 11, and 2 form a group that still matches
  the observations well, while all of the following sets give a poor
  match to our data. Note that HPMM favored sets 2 and 8, where the
  former is their best-fit model. The rejection of sets 4, 5, and 6 is
  nicely confirmed, since they rank last, together with set
  9. Similarly, the trend that we stated above against sets 3, 9, and 12
  is supported by the statistical analysis.

  \begin{figure}
    \centering
    \includegraphics[width=5.66cm]{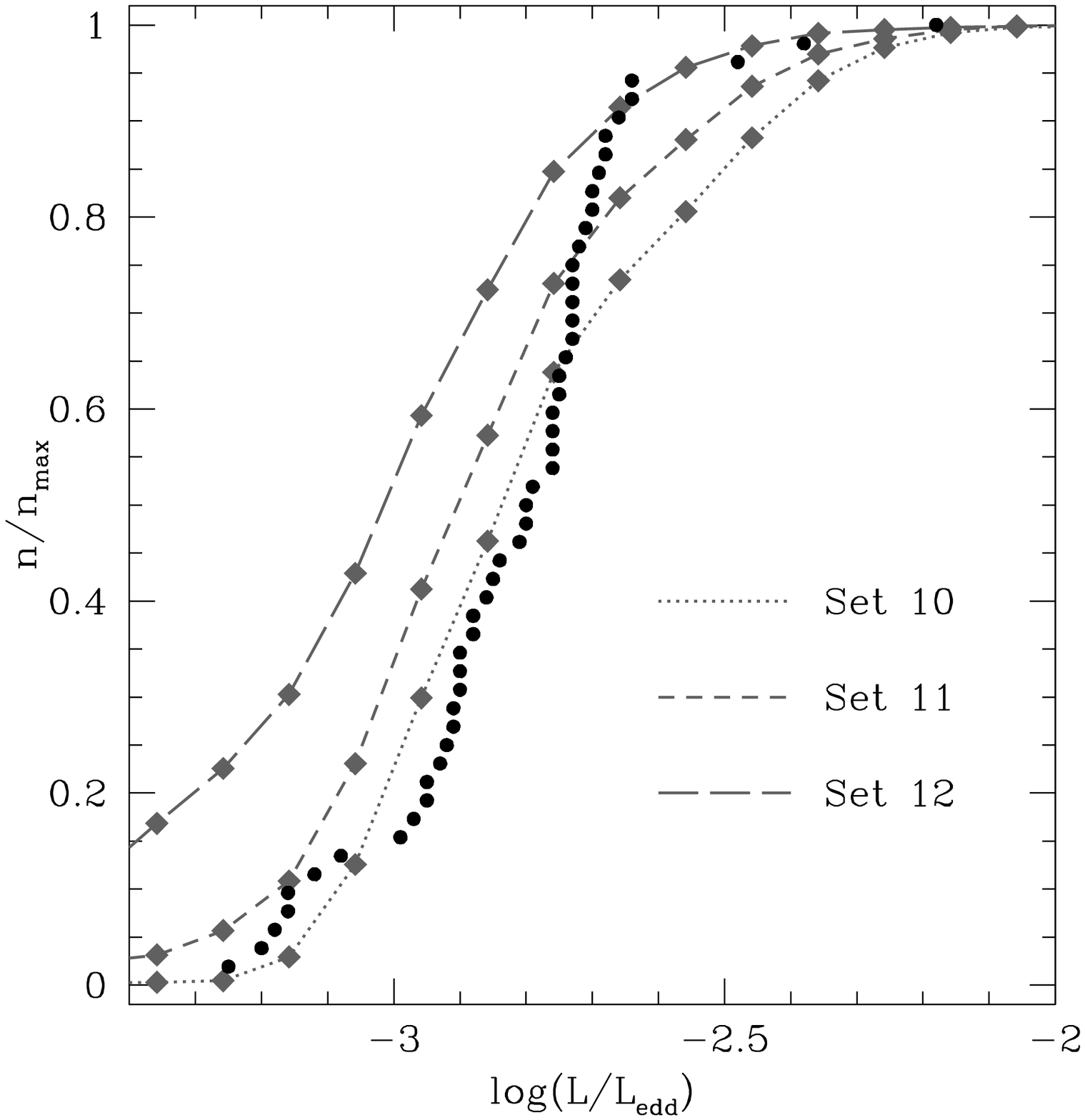}
    \caption{Cumulative luminosity function of our sdB stars as
    shown in Fig.~\ref{cumlum}, along with the functions given by the
    HPMM simulation sets.}
    \label{cumhan}
  \end{figure}

  \subsubsection{The cumulative luminosity function \label{sec_han3}}

  Fig.~\ref{cumhan} compares the cumulative luminosity function from
  all simulation sets of HPMM with our data. It is obvious that in all
  panels the respective leftmost track gives the worst match to the
  data. These tracks correspond to sets 3, 6, 9, and 12, which we have
  already found a trend against. Their
  common parameter values are 
  $\ace = \ath = 1$, leading to the conclusion that CE ejection
  processes do not reach $100\%$ efficiency. 
  
  In contrast to the luminosity functions derived from the
  \citet{dor93} models, the best matching simulation sets of HPMM --
  as derived from our statistical analysis -- are
  consistent with the observations at the low luminosity end of the
  luminosity function. It can be clearly seen, however, that they
  predict a smaller slope than we observe, and that even set 10 is far
  from reproducing the observed function.
   As mentioned before, any systematic errors due to LTE/NLTE or metallicity differences would not affect this statement.

  \subsection{Other sdB samples \label{sec_ohan}}

  We also investigated the sdB samples studied by \citet{saf94} and
  \citet{ede03}. A comparison to the theoretical luminosity function
  in the context of the single star scenario (as described for our sample
  in Sect.~\ref{sec_dor}) can be made from Figs.~\ref{cumlum} and
  \ref{dorcum}. It qualitatively shows the same overall behaviour as
  we found for our survey, i.e.~the slope of the cumulative luminosity
  function is the same, but the offsets of the luminosity scale are
  different, and the numbers of stars of too low or too high luminosity
  differ. We also compared these sdB samples 
  to the HPMM simulation sets by visual inspection of the
  $\teff$-$\logg$-plane (as described in Sect.~\ref{sec_han1}) and the cumulative
  luminosity function (as described in Sect.~\ref{sec_han3}). In
  general the result is the same as for the SPY sample. The luminosity
  function favors models with uncorrelated mass ratios (in particular
  set 4), while the $\teff$-$\logg$-plane rules them out. The latter
  favours the same models (10, 8, 2, 11) as for SPY.
  This directly leads us to a discussion of the current limits of sdB formation
  theories, and to suggestions for future improvements.


  \section{Discussion \label{sec_disc}}

  To constrain the two CE ejection parameters further
  does not seem possible nor reasonable because of poor knowledge
  of CE ejection physics. This is also reflected in the controversial
  statements by HPMM and \citet{sok03} about the treatment of CE
  ejection processes. The latter authors criticize HPMM's criterion for CE ejection, and
  suggest to account for enhanced mass-loss rates
  by binary interaction, rather than including CE ionization
  energy as HPMM did.

  To decide between a value of $1.2$ and $1.5$ for
  $q_\mathrm{crit}$ would be of enormous interest for the predicted number of sdB stars that
  are hidden from view because they are outshone by main sequence companions of
  spectral types B to F. Especially the percentage of A-type main
  sequence stars in the Galaxy that have sdB companions is highly
  sensitive to $\qcrit$, predicted to be 0.75\% if $\qcrit=1.5$, but only
  0.19\% if $\qcrit=1.2$. HPMM point out that ``the effects
  of a tidally enhanced wind can to some degree be implicitly included
  by using larger values of $q_\mathrm{crit}$''. This immediately suggests that
  various values of $q_\mathrm{crit}$ may be realized in sdB
  formation, depending on the presence and strength of tidal wind
  enhancement.

  Similarly, it cannot be doubted that various
  metallicities may play a role among the observed sdB stars,
  since their progenitors' ZAMS masses range from $M_\mathrm{ZAMS} <
  1\,\msol$ -- corresponding to main sequence lifetimes comparable to the age of the
  Galaxy -- to almost $M_\mathrm{ZAMS} = 3\,\msol$, having a much
  shorter main sequence lifetime. In addition, sdB stars exist in some
  halo globular clusters, and kinematic studies show
  that sdB stars are present in the thin disk, the thick disk, and the
  halo \citep{alt04}. Hence, a
  mixture of different parameter configurations and population
  memberships may be realized in nature
  instead of having fixed values for all physical quantities
  involved.

  These considerations require sdB formation theories to be
  even more complex than the HPMM simulations. In addition, the
  comparison of our data with the predicted cumulative luminosity
  functions of HPMM and \citet{dor93}
  in the previous section can well be interpreted with the necessity
  of \emph{combining} single star and binary formation channels to reproduce
  the observed sdB population. Although not compelling, one can see in
  Fig.~\ref{cumhan} 
  that the best-matching slope of the HPMM simulation sets is given by
  set 4 -- regardless
  of the offset -- which we had immediately excluded because it does
  not match the observed $\teff$-$\logg$-distribution.
  Since the slope of the cumulative luminosity function in the single star 
  scenario (see Fig.~\ref{dorcum}) is steeper, we could remedy the discrepancy
  in the binary scenario (HPMM) by allowing for a contribution by a single
  star evolution channel, i.e.~both scenarios contribute to the observed
  sample. An alternative possibility is that our sample is biased against 
  stars of higher luminosity, which, if added to the function, would
  flatten the steep slope and probably smooth away the sharp bend at the high
  luminosity end. Although this seems unlikely for our sdB sample, 
  we still have to investigate the relevance of the sdO stars in the SPY 
  survey for the EHB evolution, since a close connection of the
  hydrogen-rich sdO stars
  to the EHB has often been claimed. Note that sdO stars have also been
  excluded in the investigations of other published samples. 


  \section{Conclusions \label{sec_sum}}

  We have presented the results of a spectral analysis of 76 sdB stars
  from SPY. 24 objects show spectral signatures of a cool companion,
  which we investigated from optical and infrared photometry. The
  majority, if not all, of double-lined stars have
  main sequence companions of types F to K.
  When focusing on helium abundances of composite sdB stars, we find
  no difference when compared to non-composite
  objects. Of the 52 single-lined stars, four show peculiar $\hal$ profiles,
  possibly indicating stellar winds. The luminosity distribution of our
  sample is found to be in good agreement with previous studies. The
  tracks for luminosity evolution of single stars calculated by
  \citet{dor93} agree in shape with the cumulative luminosity
  function of our stars on 
  and above the EHB. They show, however, a slight offset in luminosity,
  and in addition they cannot explain
  the objects below the ZAEHB. The best-matching
  simulation sets of the binary population
  synthesis calculations by HPMM match the observed sdB
  distribution in $\teff$ and $\logg$ very well, but they cannot
  reproduce the cumulative luminosity function of our
  stars. The mismatch is most obvious for the simulation sets using
  100\% efficiency of CE ejection (sets 3, 9, and 12), which also holds for a
  comparison with the 
  luminosity function of other observational datasets \citep{saf94,ede03}. Furthermore, there is
  some evidence against an uncorrelated mass distribution of the progenitor
  systems, which follows from the trend of rejecting sets 4, 5, and
  6.

  We conclude that a combination of single star
  and binary formation channels would be necessary to achieve full
  understanding of sdB formation processes. In order to solve the problems that the latter still pose,
  future simulations are required to incorporate a more sophisticated
  description of the physics involved. In parallel, high-quality
  observational samples like SPY are necessary to enable an even better
  judgement of the simulated sdB population. A
  crucial test of theoretical calculations would be the comparison of
  predicted and observed fraction of radial
  velocity variables, i.e.~stars that formed in the CE ejection
  channels and thus are close binaries. Since SPY was initiated to
  search for RV-variations, the data are well suited for such an
  investigation. The subsequent step will then be to determine orbital
  parameters of the RV-variable sdBs from SPY by measuring their
  radial velocity curves. First results of these projects are reported
  by \citet{napkeele}, and more details will be presented in Napiwotzki et
  al.~(in prep.).

  Equally important is the completion of the analysis of hot subdwarfs
  from SPY, i.e.~the investigation of the various sdO stars
  observed. This class is much less homogeneous than the sdBs, and it
  remains to be seen whether they (or at least some of them) are
  intimately connected with the EHB.
  Increasing the sample size of sdB and sdO stars should then be the
  ultimate step. For a detailed comparison with evolutionary models,
  it is mandatory to improve number statistics. A huge number of sdB
  and sdO stars is now being discovered by sky surveys such as the
  Sloan Digital Sky Survey, and needs to be studied.


  \begin{acknowledgements}
    We express our gratitude to the ESO staff for providing
    invaluable help and conducting the service observations which
    have made this work possible.
    We are grateful to Zorica Salomon for providing us with various flux-calibrated
    sdB spectra and their atmospheric parameters. We thank Martin
    Altmann, Pierre Chayer, Victor Debattista, Christian Karl, 
     Sabine Moehler, Roy \O stensen,
    Simon
    O'Toole, Cristiano Porciani, Mike Reed, Alexander Stroeer,
    Michele Stark, Jorick Vink, and Richard Wade for fruitful
    discussions and important suggestions.
    This publication makes use of data products from the Two Micron
    All Sky Survey, which is a joint project of the University of
    Massachusetts and the Infrared Processing and Analysis
    Center/California Institute of Technology, funded by the National
    Aeronautics and Space Administration and the National Science
    Foundation. The exploitation of the
    stellar content of the Hamburg/ESO Survey has been supported by
    the DFG under grant Re 353/40.
  Work on SPY at the Dr.-Remeis-Sternwarte Bamberg is supported by the DFG under grant Na 365/2-2.
   T.L.~acknowledges support by
    the Swiss National Science Foundation.
  R.N.~acknowledges support by
    a PPARC advanced fellowship.
  \end{acknowledgements}

  \nocite{han02}
  \nocite{heb84}
  \nocite{heb86}
  \nocite{nap97}
  \nocite{sai00}
  \nocite{tut90}

  \bibliography{tlib}
  \bibliographystyle{aa}

\end{document}